\documentclass{article}
\usepackage[margin=0.8in]{geometry}
\usepackage{setspace,url,hyperref,graphicx}
\usepackage{authblk}
\usepackage{enumitem}
\usepackage{amsmath,amsthm,amssymb,amsfonts}
\usepackage{xcolor,soul}
 
\newcommand{\N}{\ensuremath{\mathbb{N}}}
\newcommand{\R}{\ensuremath{\mathbb{R}}}
\newcommand{\sub}{\mathrm{sub}}

\usepackage[sorting = none, backend = bibtex, style=numeric-comp]{biblatex}
\title{Increasing the Hardness of Posiform Planting Using Random QUBOs for Programmable Quantum Annealer Benchmarking}

\author[1]{Elijah Pelofske\thanks{Email: epelofske@lanl.gov}}
\author[2]{Georg Hahn}
\author[1,3]{Hristo N.\ Djidjev}
\affil[1]{Los Alamos National Laboratory}
\affil[2]{Harvard University, T.H.\ Chan School of Public Health}
\affil[3]{Institute of Information and Communication Technologies, Bulgarian Academy of Sciences}
\date{\vspace{-0.8cm}}

\begin{document}
\maketitle

\begin{abstract}
Posiform planting is a method for constructing QUBO problems with a single unique planted solution that can be tailored to arbitrary connectivity graphs. In this study we investigate making posiform planted QUBOs computationally harder by fusing many smaller random discrete coefficient spin-glass Ising models, whose global minimum energy is computed classically using classical binary integer programming optimization software, with posiform-planted QUBOs. The single unique ground-state solution of the resulting QUBO problem is the concatenation of (exactly one of) the ground-states of each of the smaller problems. We apply these modified posiform planted QUBOs to the task of benchmarking programmable D-Wave quantum annealers. The proposed method enables generating binary variable combinatorial optimization problems that cover the entire quantum annealing processor hardware graph, have a unique solution, are entirely hardware-graph-native, and can have tunable computational hardness. We benchmark the capabilities of three D-Wave superconducting qubit quantum annealing processors, having from $563$ up to $5627$ qubits, to sample the optimal unique planted solution of problems generated by our proposed method and compare them against simulated annealing and Gurobi. We find that the D-Wave QPU ground-state sampling success rate does not change with respect to the size of the random QUBOs we employ. Surprisingly, we find that some of these classes of QUBOs are solved at very high success rates at short annealing times compared to longer annealing times for the Zephyr connectivity graph QPUs.
\end{abstract}

\section{Introduction}
Quantum annealing is a type of analog quantum computation that aims to sample good solutions to combinatorial optimization problems \cite{PhysRevE.58.5355, farhi2000quantum, Morita_2008, das2008colloquium, Hauke_2020}. Programmable quantum annealers containing up to several thousand superconducting qubits have been manufactured by the company D-Wave \cite{johnson2011quantum, Bunyk_2014, Johnson_2010, PhysRevX.4.021041, PhysRevA.92.062328, king2022coherent, King_2023}, and have demonstrated notable compute-time and quality advantages over alternative methods on a number of different computations and simulations \cite{king2021scaling, king2024computational, tasseff2022emerging, pelofske2023shortdepth, QA_QAOA_pelofske, PRXQuantum.2.030317, bauza2024scaling}. Quantum annealing is based on the adiabatic quantum computation concept, where the system is initialized in the easy-to-prepare ground state of a Hamiltonian (typically the transverse field Hamiltonian) and then the system is slowly transitioned into a second Hamiltonian whose ground-state is not known and hard to compute \cite{farhi2000quantum, RevModPhys.90.015002}. 
In an ideal adiabatic evolution, this physical process can be used to find the ground state of Hamiltonians of interest. According to the adiabatic theorem, if certain conditions are met — such as sufficiently slow evolution — the system will remain in the ground state throughout the process \cite{born1928beweis}.

Mathematically, this adiabatic evolution is governed by a time-dependent Hamiltonian $H(t)$ that gradually transforms from an easily prepared initial state Hamiltonian (denoted as $H_{initial}$) to the problem-specific Hamiltonian of interest (denoted as $H_{ising}$), where
\begin{equation}
    H(t) = A(t)H_{initial} + B(t)H_{ising}.
    \label{equation:QA_high_level}
\end{equation}
The functions $A(t)$ and $B(t)$ in eq.~\eqref{equation:QA_high_level} define the time-dependent \emph{anneal schedules}, whereby the initial state is slowly ramped down over the course of the anneal (of time $t$) given by $A(t)$, and the magnitude of the problem of interest is increased by $B(t)$. For quantum annealing in the transverse field Ising model \cite{PhysRevE.58.5355}, $H_{initial} = \sum_{i}^n \sigma_i^x$, where $\sigma_i^x$ is the Pauli matrix acting on each qubit at index $i$. 

Combinatorial optimization problems, such as NP-hard problems, can be formulated as a Quadratic Unconstrained Binary Optimization (QUBO) \cite{Lucas2014, Boros2006, Boros2007} problem (or equivalently an Ising model) which can be written as
\begin{align}
    Q(x_1,\ldots,x_n) = \sum_{i=1}^n a_i x_i + \sum_{i<j} a_{ij} x_i x_j
    \label{eq:general_QUBO}
\end{align}
for $n \in \N$ binary variables, where $a_i \in \R$ and $a_{ij} \in \R$ are given by the user and define the optimization problem. The variables $x_i$, for $i \in \{1,\ldots,n\}$, are the decision variables whose optimal assignment—whether minimizing or maximizing the objective function—is generally unknown a priori.\footnote{Given the equivalence between spin systems and discrete combinatorial optimization problems, we will use the terms 'ground state' and 'optimal solution' interchangeably.}

In D-Wave quantum annealers, the qubits act as spins, having measured states either $+1$ or $-1$, and users can program the linear and quadratic coefficients of the hardware qubits and couplers to represent a classical Hamiltonian that represents a discrete combinatorial optimization problem. For quadratic optimization problems, the variable states can be easily converted from spins ($+1$ or $-1$) to binary states ($1$ or $0$). The problems whose decision variables are spins are called \textit{Ising models}, and the problems whose decision variables are binary are called \textit{QUBO models}. 

A reasonably large number of studies have examined various techniques for formulating and solving combinatorial optimization problems on D-Wave quantum annealers \cite{PhysRevX.5.031040, PhysRevA.91.042302, King_2016, PhysRevA.105.062406, Pelofske_2022_parallel, Albash_2018, Pelofske_2023_parallel_max_clique, Pelofske_noise_dynamics_quantum_annealers, vyskovcil2019embedding}. There are a number of practical challenges that limit the capabilities of D-Wave quantum annealers due to system constraints and sources of error, including limited qubit coherence times \cite{king2022coherent, King_2023}, having a fixed hardware graph whose connectivity is quite sparse and therefore encoding arbitrarily connected problems requires graph minor embedding \cite{PRXQuantum.2.040322, cai2014practical, lucas2019hard, choi2011minor, choi2008minor}, analog control errors \cite{pearson2019analog}, and spin bath polarization that causes anneal-correlations \cite{lanting2020probing}. For this reason, several studies have developed methods for benchmarking and mitigating different types of errors in D-Wave computations \cite{9465651, 9319535, Pelofske_noise_dynamics_quantum_annealers, grant2022benchmarking, 10.1145/3457388.3458672, vinci2015quantum, vinci2016nested, pudenz2014error}. One of the central challenges with benchmarking the capabilities of quantum annealers for solving combinatorial optimization problems is verifying the accuracy of the produced solutions (specifically if the global optimal solution could be found), since quantum annealing is a heuristic algorithm. In particular, verifying whether the optimal solution was found requires using classical deterministic exact solvers to find the optimal solution, or to construct a problem such that the optimal solution is known in advance. 

There are numerous algorithmic approaches designed to generate combinatorial optimization problems with known solutions, commonly referred to as \emph{solution planting}. Examples include frustrated loops (meaning Ising models composed of parts containing only a subset of the variables) with tunable hardness \cite{Hen2015, king2015performance}, tile-planting \cite{Perera2020}, patch-planting \cite{Wang2017}, or weighted MAX-2-SAT instances \cite{Pei2020}. However, most methods do not guarantee the uniqueness of the planted solution. Another way to generate instances is by converting sets of linear equations, called equation planting \cite{Kowalsky2022,Hen2019}, which can ensure uniqueness at the expense of not being able to generate QUBOs of a general structure. Many methods have already been made available in software implementations, for instance the \textit{Chook} toolbox \cite{chook} and the \textit{dwig} Python implementation. Benchmarks for emerging computing technologies need to be computationally hard while having known optimal solutions, yet many problem instances are easily solvable using classical algorithms. For example, it has been shown that two of the existing quantum annealing spin glass benchmarks \cite{king2017quantumannealingamidlocal, denchev2016computational} can be solved in general with a polynomial time classical algorithm because they are a type of planar spin glasses \cite{Mandr__2017}.

The focus of our study is on the previously proposed \emph{Posiform Planting} method \cite{Hahn_2023}, which allows the creation of QUBO models tailored to an arbitrary connectivity graph, with a single unique bitstring as the planted solution. Posiform planting is a very general method in that it allows many choices to be made in regards to what problem coefficients can be used to create a valid instance for a given target bitstring, and the target bitstring can be chosen arbitrarily as well. A computationally more tractable variant of posiform planting has already been proposed \cite{Isermann2024}. However, a disadvantage of posiform planted QUBOs is that they are a type of MAX-2-SAT problem, making them very tractable to solve exactly (in linear time) with classical algorithms \cite{Aspvall1979}. As shown in \cite{Hahn_2023}, this also applies to solving posiform planted QUBOs with quantum annealers. The aim of this study is to increase the computational hardness of posiform planted QUBO models by combining them with random QUBO problems, and to use the resulting problem instances to benchmark several types of D-Wave quantum annealers. In this study we design QUBO problems that are tailored to the hardware interaction graphs of several D-Wave QPUs; these graphs are known as Pegasus graphs \cite{dattani2019pegasussecondconnectivitygraph, boothby2020nextgenerationtopologydwavequantum} and Zephyr graphs \cite{zephyr}.

We expect that the proposed way of combining of random QUBO problems with posiform QUBOs produces QUBOs that are, in general, NP-hard, specifically because random QUBOs (and random Ising modes) are also in general NP-hard. However, we do not examine this in this study. It could be the case that random spin glasses defined on the sparse graphs of the underlying hardware, such as the class of QUBOs we present in this study, lend themselves to efficient classical simulations due to having a critical temperature of zero, as was shown in particular for the hardware graph of previous generations of D-Wave quantum annealers known as the Chimera graph \cite{PhysRevX.4.021008, PhysRevX.5.019901} and there is some evidence for spin glasses defined on Zephyr and Pegasus graphs possibly also having a zero critical temperature \cite{jaumà2023exploring}. However, this potential classical simulatability of hardware compatible spin glasses is not definitely known to be true for the intermediate D-Wave processor sizes currently available, and is not known in general for other classes of random Ising models defined on these relatively sparse hardware graphs \cite{jaumà2023exploring, PhysRevX.4.021008, PhysRevX.5.019901}.

The ability to control the degeneracy of ground states, which our proposed method allows, can be a valuable characteristic of a solution planting algorithm. In particular, ensuring the uniqueness of the ground state is especially relevant for random discrete-coefficient discrete decision variable optimization problems. Attenuating the degeneracy of the optimal solutions of discrete optimization problems can be a useful property since some quantum algorithms may not uniformly sample the optimal solutions \cite{Matsuda_2009, Zhu_2019, PhysRevLett.118.070502, Albash_2015, Zhang_2017, Boixo_2013, K_nz_2019}. 

This article is structured as follows. In Section~\ref{section:methods} we present the proposed improvement of the posiform planting algorithm of \cite{Hahn_2023}, and briefly describe simulated annealing of \cite{kirkpatrick1983optimization} which we use as a benchmark algorithm. Our experimental results are presented in Section~\ref{section:results}, in particular with respect to success rates for optimal solution sampling on D-Wave (Section~\ref{section:results_optimal_solution_sampling_DWave}) and simulated annealing (Section~\ref{section:results_optimal_solution_sampling_SA}), as well as Time-to-solution (TTS) results (Section~\ref{section:results_TTS}). The article concludes with a discussion in Section~\ref{section:discussion}.

\section{Methods}
\label{section:methods}
This section describes the proposed improvement of the posiform planting algorithm in Section~\ref{sec:improvement}. Section~\ref{sec:proof} gives a proof that the newly generated QUBOs preserve the unique planted solution. We also briefly introduce the Time-to-Solution metric (Section~\ref{section:methods_TTS}) and the simulated annealing algorithm (Section~\ref{section:methods_simulated_annealing}) which serves as a benchmark.

\subsection{Combining Posiform Planting with Random QUBOs}
\label{sec:improvement}
Roughly, our algorithm generates a set of smaller non-overlapping random QUBOs and one larger posiform planted QUBO that covers the hardware graph. The smaller random QUBOs are then fused with the large posiform planted QUBO covering the hardware graph. The aim of combining the larger (posiform planted) QUBO with smaller QUBOs is to alter the coefficients in such a way that the QUBO becomes harder to solve. At the same time, it is possible to add on the smaller QUBOs to the posiform planted QUBO without changing its optimal solution.

The disjoint graphs corresponding to these QUBOs are generated using the Networkx \cite{hagberg2008exploring} implementation of the Kernighan–Lin bisection algorithm \cite{6771089}. Specifically, the hardware graph is recursively bisected into equally sized partitions until a threshold is reached. The total number of partitioned subgraphs is increased starting at a maximum number of variables (for all subgraphs) of $50$. This recursive bisection technique ensures that the final partitioned subgraphs will be identically sized, unless there are any odd divisions in which case the number of variables in the partitioned subgraphs will be different from each other by at most $1$ variable (this is only true because the number of partitions is always a power of 2).

Note that the construction of these disjoint hardware subgraphs is similar to the tiled embeddings used in parallel quantum annealing \cite{Pelofske_2022_parallel, Pelofske_2023_parallel_max_clique, Pelofske_noise_dynamics_quantum_annealers, Pelofske_2022_tensor, PhysRevA.91.042314}. 
Figure~\ref{fig:hardware_graph_partitions_pegasus_4.1} in the appendix shows an example of the partitioned \texttt{Advantage\_system4.1} hardware graph and Figure~\ref{fig:hardware_graph_partitions_zephyr_prototype_1.1} shows the same for the \texttt{Advantage2\_prototype1.1} device.

The bisection partitioning algorithm is stochastic, and so different runs can produce different partitions. When generating the random QUBO instances, a new different disjoint bisection partition is constructed for every instance.

The steps used to construct the hardware-native QUBO problems are defined as follows:

\begin{enumerate}[noitemsep]
    \item Generate the bisected subgraphs that cover the entire hardware graph applying the Kernighan–Lin bisection algorithm recursively. 
    \item Choose linear and quadratic random coefficients for all of the edges and nodes within the disjoint induced subgraphs of the hardware graph computed in the previous step. Here, we generate two distinct types of random QUBOs using two different discrete coefficient sets drawn from either $\{-1, -0.9, \dots, 0.9,1\}$ (with $20$ different values) in increments of 0.1, or $\{-1, 1\}$ (with $2$ values).
    We will denote these two different types of QUBOs using the shorthand of \emph{lin$_{20}$} and \emph{lin$_{2}$}, respectively. 
    \item Use an exact classical solver to compute an optimal variable assignment for each of the random subproblems. In this case, we use CPLEX \cite{cplexv12}, which can deterministically find an optimal variable assignment (given sufficient compute time). To this end, the optimization problem is formulated as a binary Mixed Integer Quadratic Program (MIQP). 
    \item Concatenate together each of the single optimal variable assignments for each random problem found in the previous step (do this based on some consistent bit-ordering, in this case we used consistent node indexing of the hardware graph). This concatenated bitstring thus gives a single variable state for every qubit (spin) in the hardware graph. 
    \item Use the posiform planting algorithm introduced in \cite{Hahn_2023} to generate a QUBO whose planted ground-state matches exactly the concatenated bitstring computed in the previous step. There are various valid choices for generating this posiform QUBO, although some may affect the coefficient range of the resulting QUBO problem, potentially leading to suboptimal performance when encoded onto current D-Wave quantum annealers. In our case, we use the 2-SAT solver \emph{Minisat} \cite{minisat} (as in \cite{Hahn_2023}). All posiform coefficients corresponding to a 2-SAT clause are chosen as $1$. To conserve compute time, we only attempt to solve the 2-SAT instance being generated in batches. This may cause the generated 2-SAT instance to be (slightly) larger than is needed to ensure the uniqueness of the planted solution.
    \item "Glue" together all of the random QUBO problems and the posiform planted QUBO, which is done by simply summing all of the terms together. The obtained final QUBO inherits from posiform planting that it has a unique and known optimal solution. The posiform planted QUBO spans the entirety of the chip and connects together the small random QUBOs, making this final QUBO connected (and in particular, it covers a majority of the edges in the hardware graph and all of the nodes in the hardware graph). A parameter that we add into this QUBO combination step is a constant scale factor which we will refer to as the \emph{posiform scaling coefficient}. This coefficient multiplies all of the terms in the posiform QUBO prior to their addition to the random QUBO problems. We evaluated this coefficient experimentally at $0.1$ and $0.01$. The reason for this is empirical. We observed in experiments that scaling the posiform QUBO coefficients by a factor prior to combining it with the smaller QUBOs resulted in harder problems for simulated annealing.
\end{enumerate}

\begin{table}[ht!]
    \begin{center}
        \begin{tabular}{|l||l|l|l|l|}
            \hline
            D-Wave QPU Chip ID & Topology & Available & Available & Annealing time\\
            & name & qubits & couplers & (min, max) microseconds\\
            \hline
            \hline
            \texttt{Advantage\_system4.1} & Pegasus $P_{16}$ & 5627 & 40279 & (0.5, 2000)\\
            \hline
            \texttt{Advantage2\_prototype1.1} & Zephyr $Z_{4}$ & 563 & 4790 & (1, 2000)\\
            \hline
            \texttt{Advantage2\_prototype2.3} & Zephyr $Z_{6, 4}$ & 1248 & 10829 & (0.5, 2000)\\
            \hline
        \end{tabular}
    \end{center}
    \caption{D-Wave Quantum Annealing processor summary. Due to manufacturing defects, the qubit and coupler counts are less than the logical graphs.}
    \label{tab:hardware_summary}
\end{table}

\subsection{Proof of Uniqueness of the Planted Solution}
\label{sec:proof}
The process outlined in Section~\ref{sec:improvement}, whereby smaller QUBOs (solved to optimality with CPLEX) are combined with the larger posiform planted QUBO (having a unique solution), preserves the uniqueness of the solution. This section gives a formal proof of this statement.

Although some of the random QUBOs being generated may have multiple solutions (i.e., multiple minima), the combined final QUBO has a unique one. To see that, consider the random QUBOs $R_1, \dots, R_k$ and the posiform QUBO $P$. Denote the variables of $P$ by $X = \{x_1, \dots, x_n\}$, and let $\sub_i(X)$ represent the subset of $X$ used as variables in each $R_i$, for $i = 1, \dots, k$. Let $X^*=\{x_1^*,\dots,x_n^*\}$ be the unique minimum of $P$. We claim that $X^*$ is also a unique minimum of the final QUBO $Q=\sum_i R_i+\alpha P$, where $\alpha$ is the posiform coefficients scale factor. 

To prove that claim, let $\hat{X}=\{\hat{x}_1,\dots,\hat{x}_n\}$ be a binary assignment of the variables such that $\hat{X}\neq X^*$. Since, by construction, $\sub_i(X^*)$ is a minimum of $R_i$ for each $i=1,\dots,k$, it follows that
\begin{equation}
R_i(\sub_i(X^*)) \leq R_i(\sub_i(\hat{X})),\quad\mbox{for } i=1,\dots,k.\label{eq:Ri}
\end{equation}
Furthermore, $X^*$ is a unique minimum of $P$ and $\hat{X}\neq X^*$. Then
\begin{equation}
P(X^*)<P(\hat{X}). \label{eq:P}
\end{equation}
Combining eq.~\eqref{eq:Ri} and eq.~\eqref{eq:P}, we get
$$Q(X^*)=\sum_i R_i(X^*)+\alpha P(X^*)<\sum_i R_i(\hat{X})+\alpha P(\hat{X})=Q(\hat{X}).$$
Hence, $Q(X^*)<Q(\hat{X})$, and since this holds for any $\hat{X}\neq X^*$, $X^*$ is a unique minimum of $Q$.

\subsection{Measuring the Compute Time to Optimally Solve a QUBO: Time-to-Solution}
\label{section:methods_TTS}
The compute time required to sample an optimal solution with high confidence, in particular for heuristic probabilistic solvers, is given by the \textit{Time-to-solution (TTS)} measure \cite{R_nnow_2014}, which is defined as
\begin{equation}
    TTS = T_{\text{anneal}} \frac{\log (0.01)}{\log (1-p)},
\end{equation}
where $T_{\text{anneal}}$ is the average QPU time used per anneal-readout cycle and $p$ is the proportion of samples that correspond to an optimal solution. $T_{\text{anneal}}$ is measured by dividing the total QPU time (total QPU time is defined as the \emph{QPU access time} from the D-Wave system, which includes programming time, readout time, and anneal time) by the number of anneals measured in the given dataset. This TTS measure quantifies the expected compute time required to sample an optimal solution at least once, with 99\% probability. Note that the only compute time used to measure the TTS is the QPU access time; this does not include local embedding and data processing CPU time.

\begin{figure}[t!]
    \centering
    \includegraphics[width=0.49\textwidth]{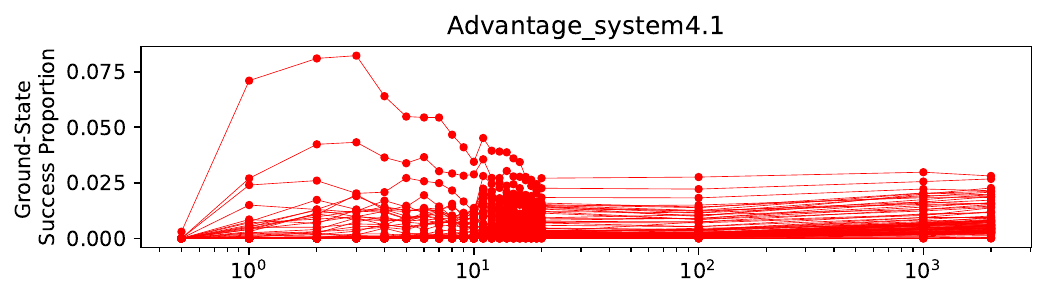}
    \includegraphics[width=0.49\textwidth]{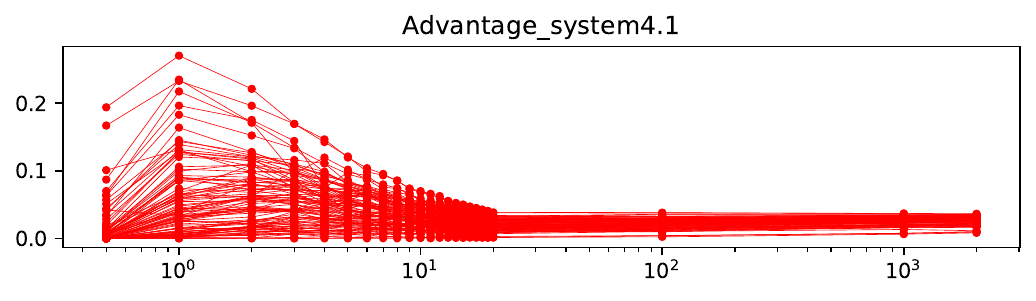}
    \includegraphics[width=0.49\textwidth]{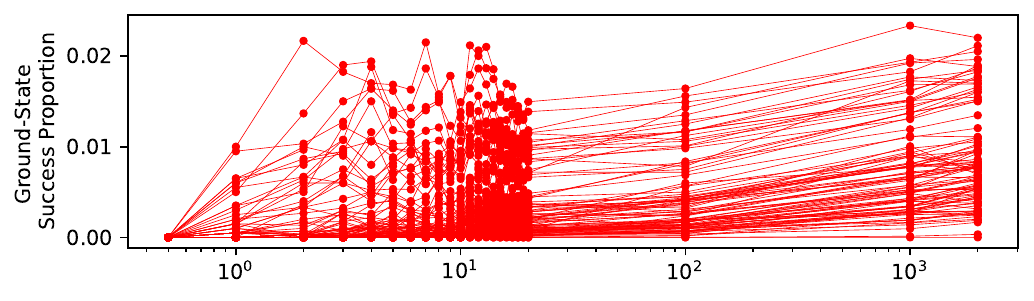}
    \includegraphics[width=0.49\textwidth]{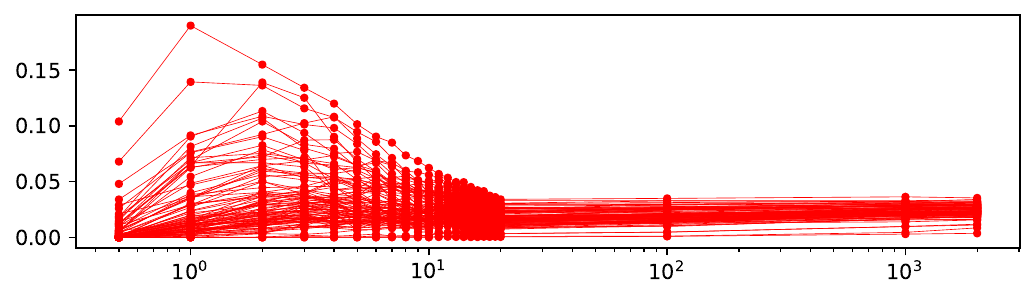}
    \includegraphics[width=0.49\textwidth]{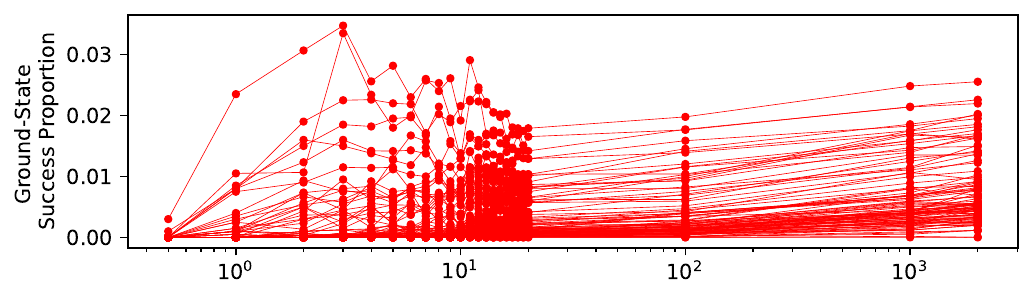}
    \includegraphics[width=0.49\textwidth]{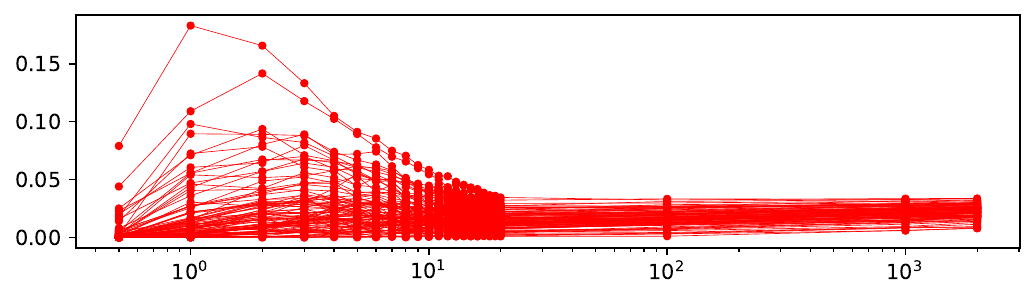}
    \includegraphics[width=0.49\textwidth]{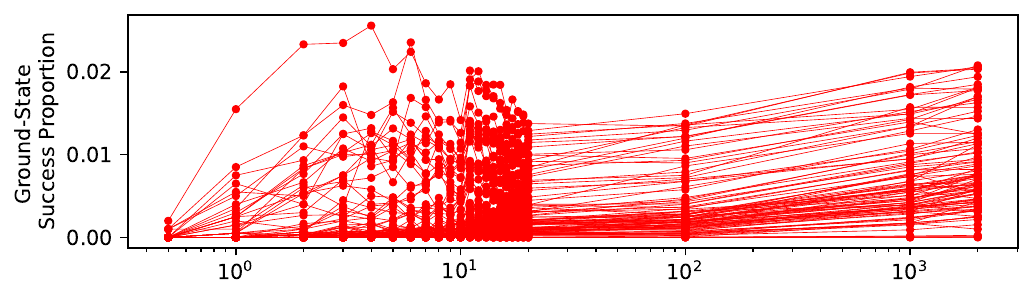}
    \includegraphics[width=0.49\textwidth]{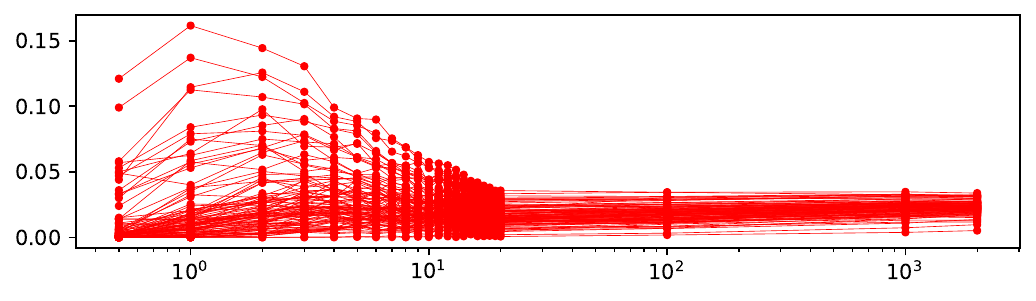}
    \includegraphics[width=0.49\textwidth]{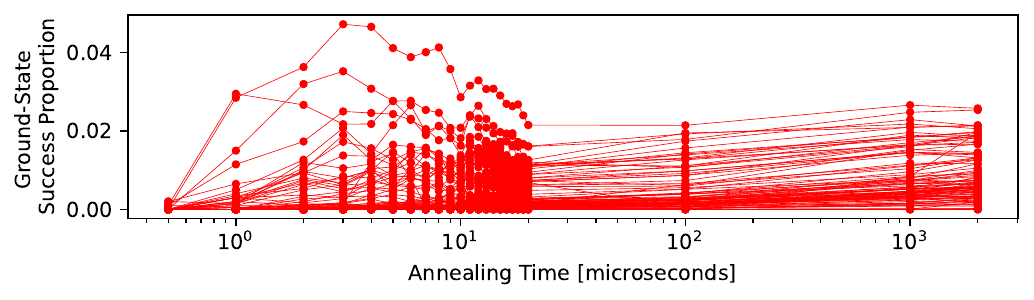}
    \includegraphics[width=0.49\textwidth]{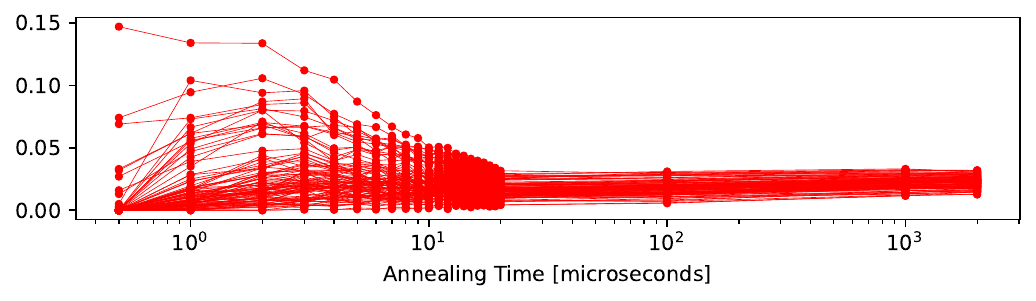}
    \caption{\texttt{Advantage\_system4.1} processor results showing the optimal solution sampling rate out of the number of anneals used, for $100$ different random QUBO instances, as a function of log-scale annealing time. Columns show QUBOs with coefficients from \emph{lin$_{2}$} (left) and \emph{lin$_{20}$} (right). Rows show results for QUBOs constructed with $128$ disjoint random QUBOs of size either $43$ or $44$ variables (first row), $64$ disjoint random QUBOs of size either $87$ or $88$ variables (second row), $32$ disjoint random QUBOs of size either $175$ or $176$ variables (third row), $16$ disjoint random QUBOs of size either $351$ or $352$ variables (fourth row), and $8$ disjoint random QUBOs of size either $703$ or $704$ variables (fifth row).}
    \label{fig:sampling_success_rate_Pegasus4.1}
\end{figure}

\subsection{Simulated Annealing}
\label{section:methods_simulated_annealing}
The generated QUBO problems are also solved using simulated annealing \cite{kirkpatrick1983optimization} to compare the performance of quantum annealers against a classical general-purpose solver. Simulated annealing performance is evaluated as a function of the number of Metropolis-Hastings update sweeps, ranging from 1 to 10,000. The comparison uses the Python 3 package \texttt{dwave-neal}\footnote{\url{https://github.com/dwavesystems/dwave-neal}}, a C++ implementation with a Python wrapper. The default geometric annealing schedule is used, generating $10,000$ samples for each parameter setting and problem instance. In this implementation, a sweep of variable updates is performed in a fixed order for each step of $\beta$, where $\beta$ corresponds to the number of Metropolis-Hastings updates in the simulated annealing schedule. The simulated annealing CPU time is measured as process time.

\subsection{Gurobi Settings}
\label{section:methods_Gurobi}

The Gurobi optimization software \cite{gurobi} is used to solve the same QUBO instances that are solved using simulated annealing and quantum annealing (up to problem sizes where it is still feasible). Gurobi provides an optimality guarantee of found solutions, and can deterministically find an optimal variable assignment of the optimization problem. We use Gurobi to solve the QUBO problems as binary Quadratic Programs. Gurobi version 11.0.3 is used for all simulations, and the compute platform is a Red Hat Linux node with Intel(R) Xeon(R) CPU E5-2695 v4 2.10GHz. Gurobi settings are all default except the following; a single thread is used, the compute time limit is $4,000,000$ seconds, and the MIP gap is $1\mathrm{e}{-8}$. The compute time reported as the classical CPU time time. 

\begin{figure}[th!]
    \centering
    \includegraphics[width=0.49\textwidth]{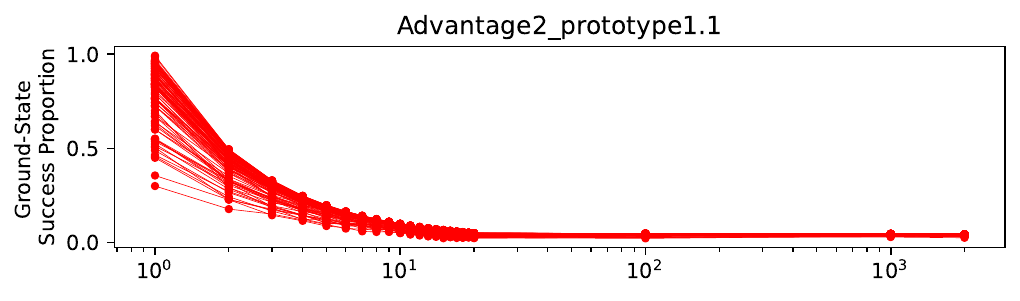}
    \includegraphics[width=0.49\textwidth]{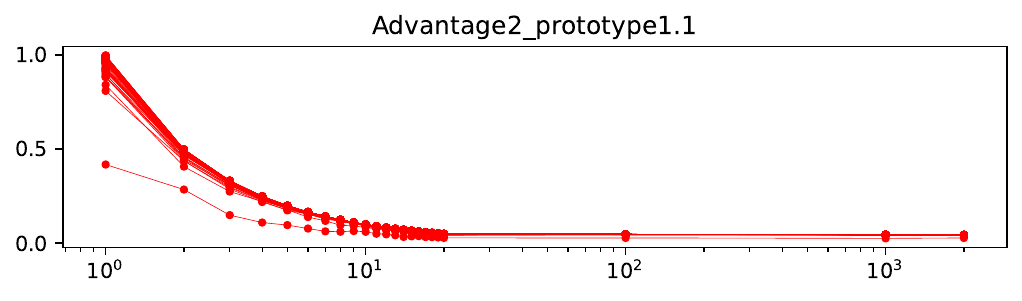}
    \includegraphics[width=0.49\textwidth]{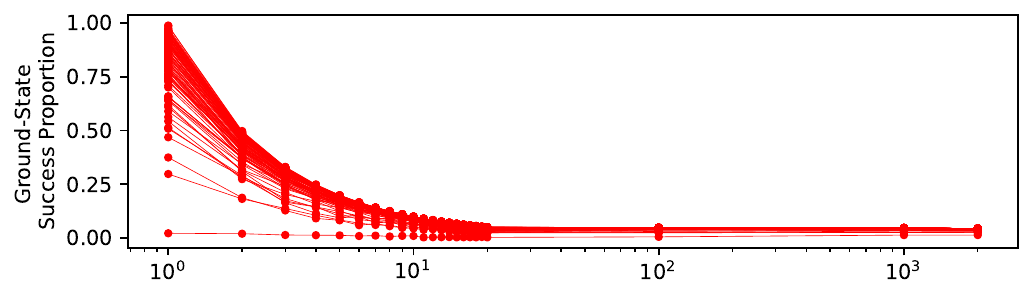}
    \includegraphics[width=0.49\textwidth]{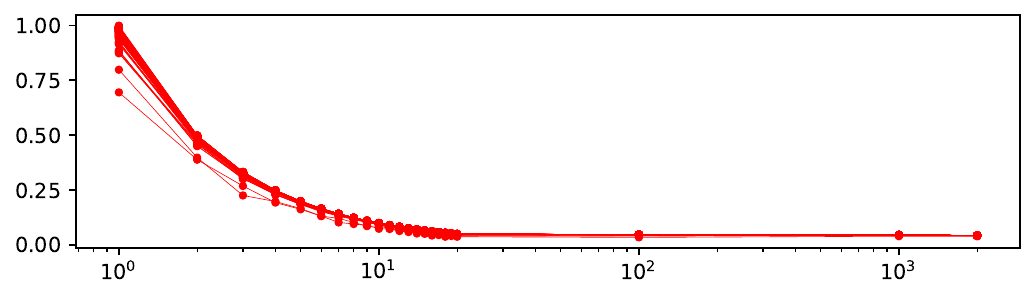}
    \includegraphics[width=0.49\textwidth]{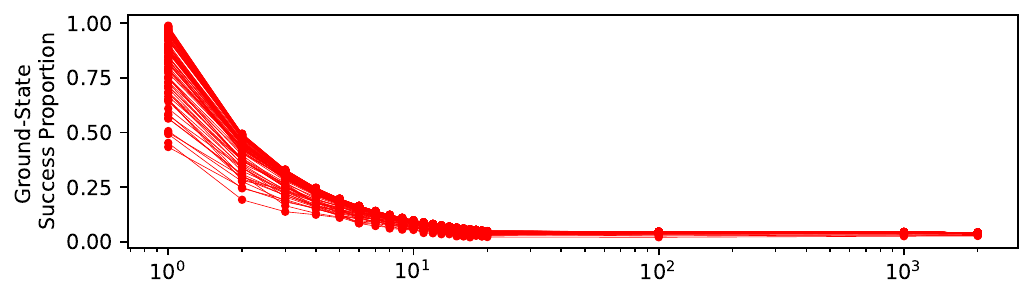}
    \includegraphics[width=0.49\textwidth]{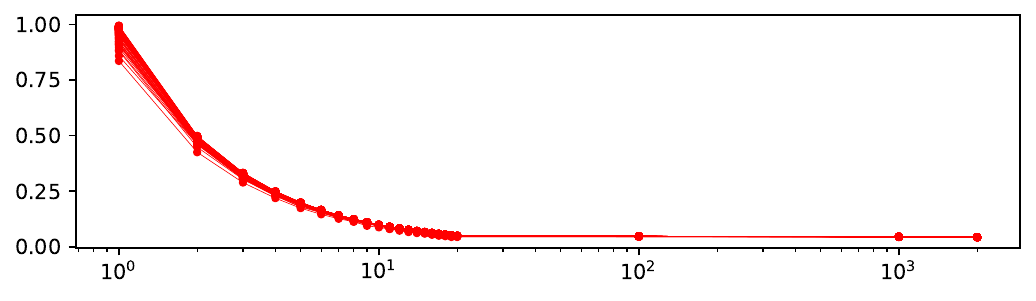}
    \includegraphics[width=0.49\textwidth]{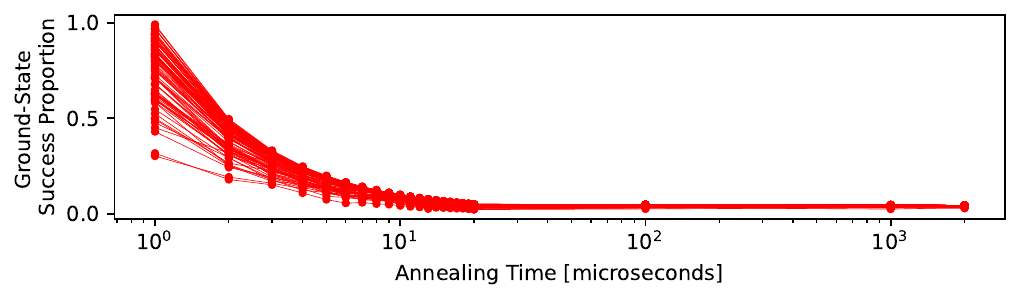}
    \includegraphics[width=0.49\textwidth]{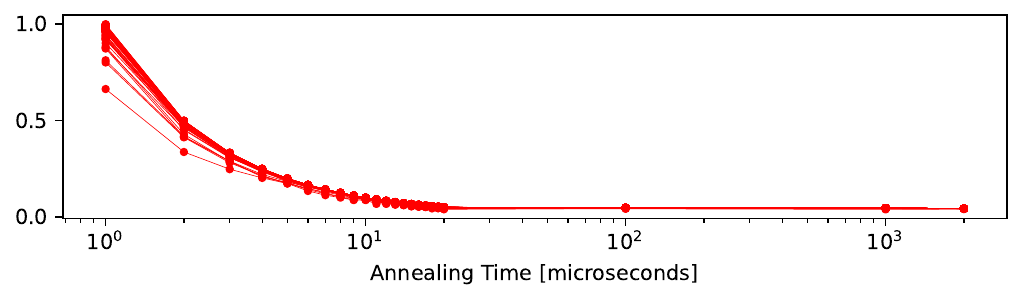}
    \caption{\texttt{Advantage2\_prototype1.1} processor results showing the optimal solution sampling rate out of the number of anneals used, for $100$ different random QUBO instances, as a function of log-scale annealing time. Columns show QUBOs with coefficients from \emph{lin$_{2}$} (left) and \emph{lin$_{20}$} (right). Posiform QUBO scaling coefficient is $0.1$. Rows show results for QUBOs constructed with maximum random glued QUBO sizes of, in order from top to bottom, $36$, $71$, $141$, and $282$ variables. }
    \label{fig:sampling_success_rate_Zephyr1.1_0.1}
\end{figure}

\section{Results}
\label{section:results}
This section presents simulation results measuring the accuracy (in terms of ground state probability) and runtime (in the TTS metric) of the instances generated with the algorithm of Section~\ref{sec:improvement}. In particular, we start by giving details of the simulation setting in Section~\ref{section:results_setting}. Afterwards, we examine the success rates for optimal solution sampling on D-Wave (Section~\ref{section:results_optimal_solution_sampling_DWave}) and simulated annealing (Section~\ref{section:results_optimal_solution_sampling_SA}), as well as TTS results (Section~\ref{section:results_TTS}). We conclude the section with Gurobi results (Section~\ref{section:results_Gurobi}).

\subsection{Simulation Setting}
\label{section:results_setting}
The annealing times used to sample the problem instances is varied over $\{1, 2, 3, 4, \ldots 19, 20, 100, 1000, 2000\}$ microseconds. In the case of \texttt{Advantage\_system4.1} and \texttt{Advantage2\_prototype2.3}, annealing time of $500$ nanoseconds is also used. The measured sample distributions are computed by executing $100$ anneal-readout cycles for each job, which is then repeated multiple times for each parameter configuration so as to build a large sample distribution for each QUBO problem instance. Some characteristics of the three D-Wave systems are given in Table~\ref{tab:hardware_summary}. The D-Wave QPUs used in this study typically operate at or near $16$ milliKelvin. 

The \texttt{Advantage2\_prototype1.1} with posiform scaling coefficient $0.01$ hardware compatible QUBO models (for both \emph{lin$_{20}$} and \emph{lin$_{2}$} cases) are sampled on the D-Wave hardware using $10000$ samples for each of the random $100$ QUBO models and annealing times. All other QUBO models are sampled with $1000$ samples per parameter combination. The \texttt{Advantage2\_prototype1.1} QUBOs with posiform scaling $0.01$ are solved with a larger number of samples because the sampling success rate was found to be especially low ($0$ for many problem instances), and thus we increased the sample count with the goal of being able to quantify the TTS. 

The largest random QUBOs that are exactly solved using CPLEX is limited by the compute time required to solve the models. We limit the compute time to be a maximum of 2 days on a single compute node, using a single thread. For any larger Zephyr graphs, CPLEX reached this compute time limit, as did larger Pegasus graph instances. Therefore, larger random QUBOs could be solved, but would require either using a faster optimization software, or using much more compute time. 

\begin{figure}[th!]
    \centering
    \includegraphics[width=0.49\textwidth]{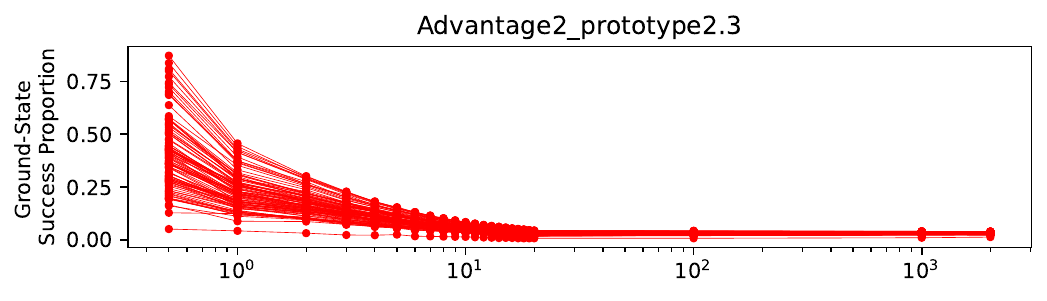}
    \includegraphics[width=0.49\textwidth]{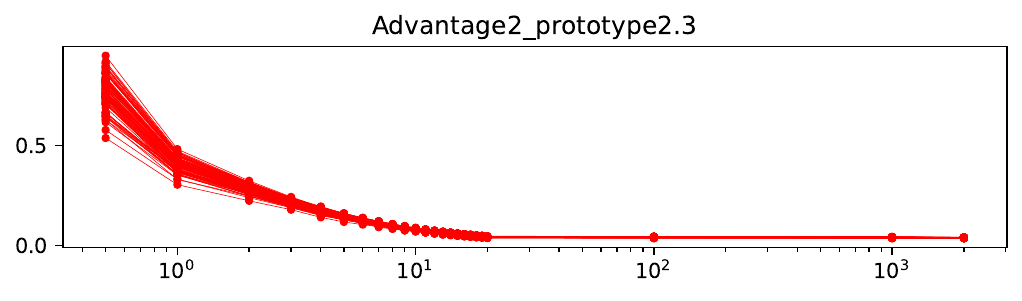}
    \includegraphics[width=0.49\textwidth]{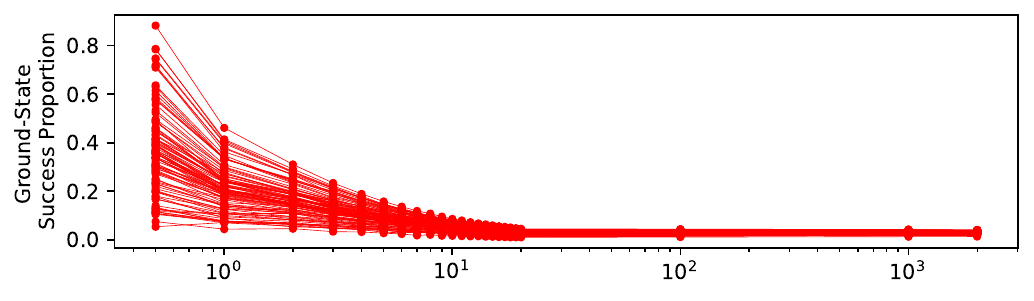}
    \includegraphics[width=0.49\textwidth]{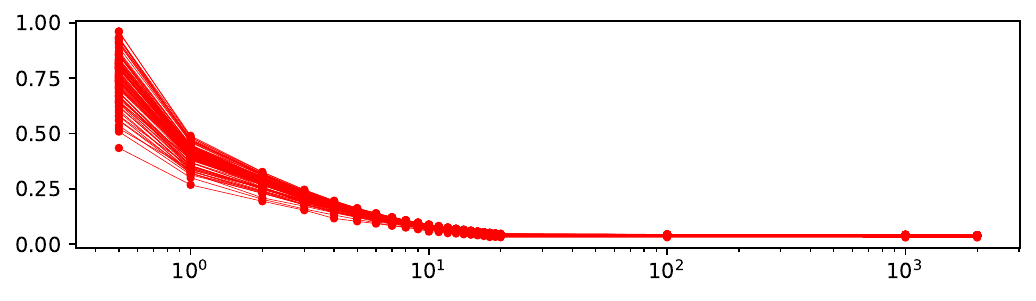}
    \includegraphics[width=0.49\textwidth]{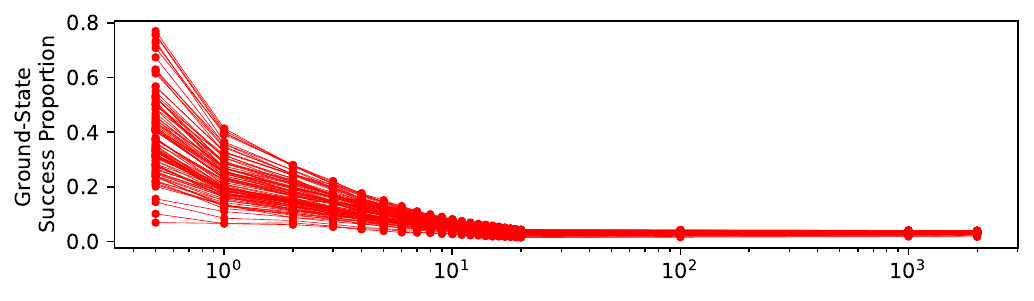}
    \includegraphics[width=0.49\textwidth]{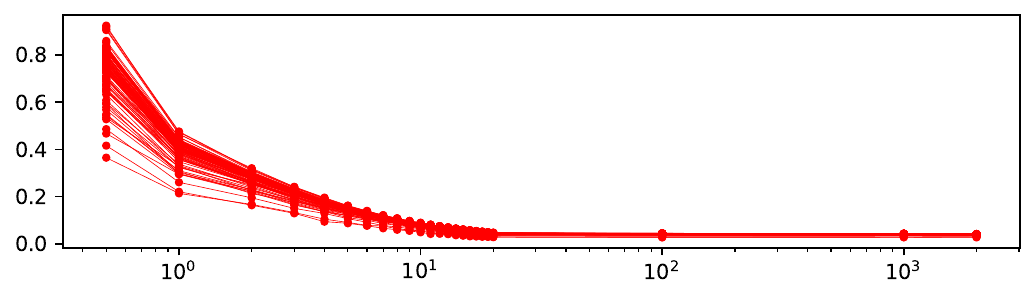}
    \includegraphics[width=0.49\textwidth]{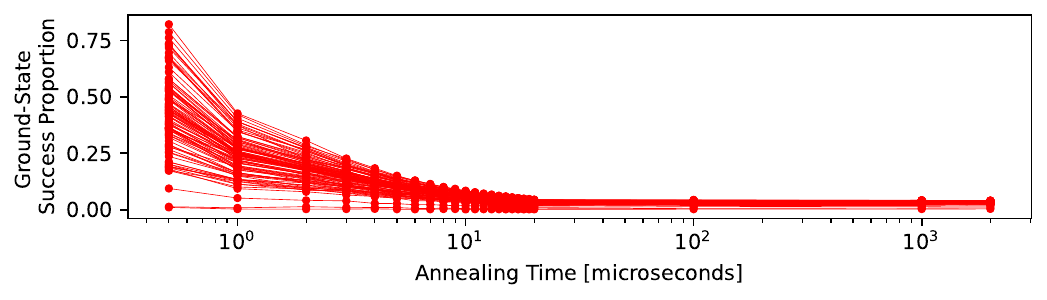}
    \includegraphics[width=0.49\textwidth]{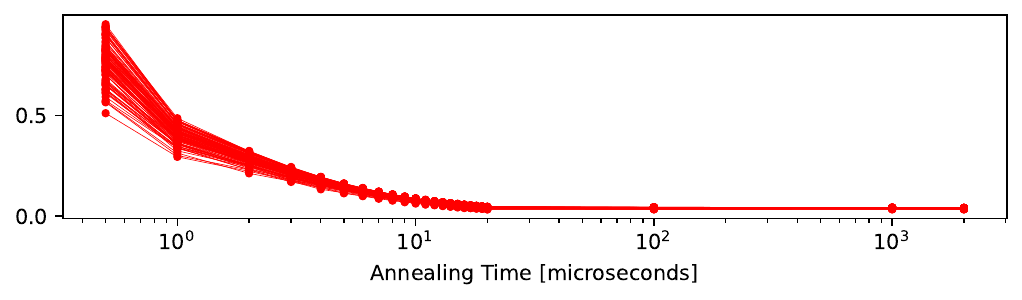}
    \caption{\texttt{Advantage\_prototype2.3} results showing the optimal solution sampling rate out of the number of anneals used, for $100$ different random QUBO instances, as a function of log-scale annealing time. Columns show QUBOs with coefficients from \emph{lin$_{2}$} (left) and \emph{lin$_{20}$} (right). Posiform QUBO scaling coefficient is $0.1$. Rows show results for QUBOs constructed with maximum random glued QUBO sizes of, in order from top to bottom, $39$, $77$, $153$, and $305$ variables.  }
    \label{fig:sampling_success_rate_Zephyr2.3_0.1}
\end{figure}

\subsection{Success Rates of Optimal Solution Sampling on D-Wave Quantum Annealers}
\label{section:results_optimal_solution_sampling_DWave}

In this section we investigate the success rate with which the optimal (that is, the planted) solution is being sampled on different D-Wave devices. The devices being considered are \texttt{Advantage\_system4.1}, \texttt{Advantage2\_prototype1.1}, and \texttt{Advantage\_prototype2.3}. To this end, we generate $100$ random posiform planted QUBOs which always cover the entire hardware graph of the respective architecture being used, and glue on a varying number of smaller random-coefficient QUBOs having appropriately chosen sizes.

We start with \texttt{Advantage\_system4.1}. When gluing on smaller QUBOs to the posiform planted QUBO, we employ a posiform QUBO scaling coefficient of $0.1$. Figure~\ref{fig:sampling_success_rate_Pegasus4.1} reports the success rate for \texttt{Advantage\_system4.1} in finding the unique planted optimal solution. All proportions being reported are with respect to the generated ensemble of $100$ posiform planted QUBOs. Figure~\ref{fig:sampling_success_rate_Pegasus4.1} presents success rates as a function of annealing time. We observe that the optimal solution sampling success rate is generally low, with a maximum of approximately $0.25$. Although the shapes of the plots in the two columns for {lin$_{2}$} and {lin$_{20}$} look similar, the scales on the left are much lower, indicating smaller success probabilities. We observe that smaller annealing times lead to a higher success rate on the \texttt{Advantage\_system4.1}, and that especially for coefficients chosen from \emph{lin$_{20}$} (right column), success rates flatten off closer to $0$ with longer annealing times.

We repeat the same experiment on the \texttt{Advantage2\_prototype1.1} device using a posiform QUBO scaling coefficient of $0.1$ when combining the posiform planted QUBO with smaller QUBOs to disguise the solution. Similarly to the previous experiment, Figure~\ref{fig:sampling_success_rate_Zephyr1.1_0.1} again shows the the ground-state sampling success rates. We observe in Figure~\ref{fig:sampling_success_rate_Zephyr1.1_0.1} that in contrast to the previous experiment on \texttt{Advantage\_system4.1}, the success probabilities for \texttt{Advantage2\_prototype1.1} are much higher for low annealing times, with many generated QUBOs being solvable with probability $1.0$. Moreover, there is a clear decrease to zero in probability as the annealing times increase. This pattern in consistent across the two coefficient choices (left and right columns), as well as the number and sizes of the smaller QUBOs being glued onto the posiform planted QUBO.

\begin{figure}[th!]
    \centering
    \includegraphics[width=0.49\textwidth]{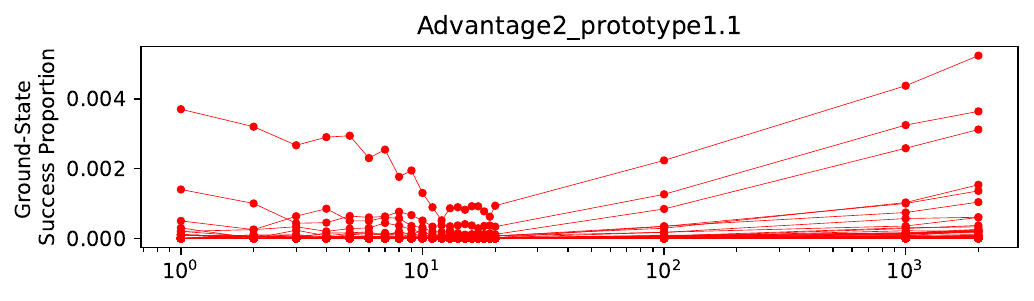}
    \includegraphics[width=0.49\textwidth]{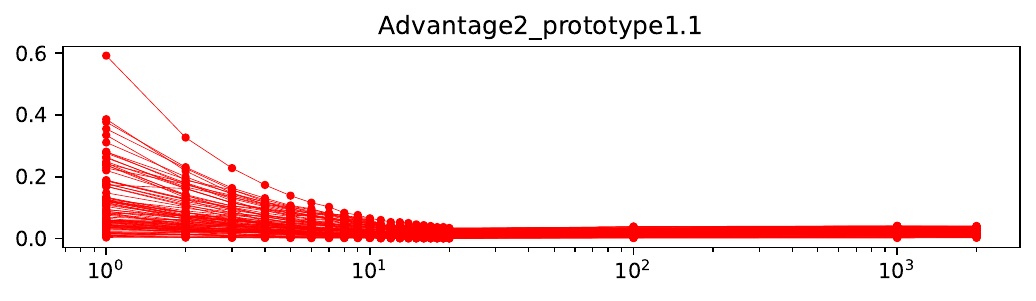}
    \includegraphics[width=0.49\textwidth]{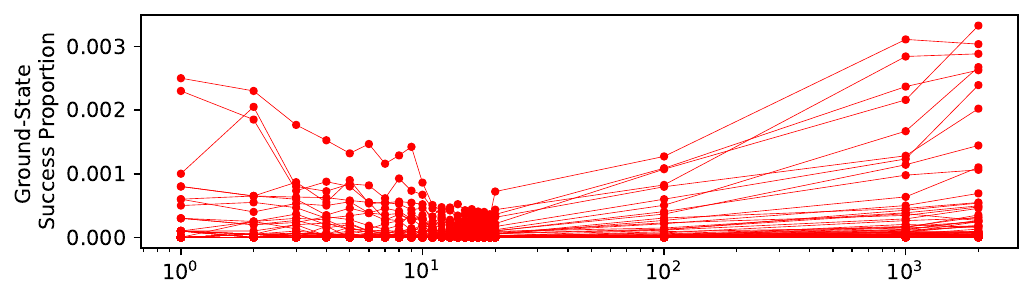}
    \includegraphics[width=0.49\textwidth]{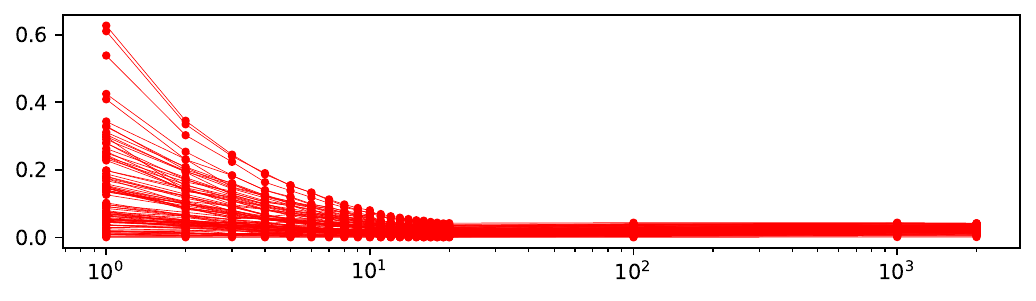}
    \includegraphics[width=0.49\textwidth]{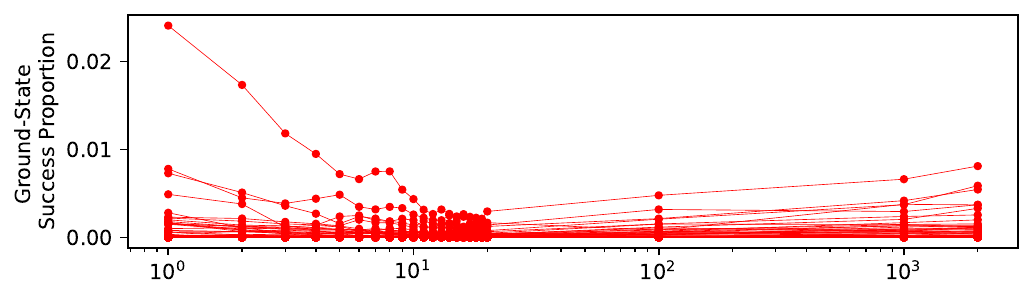}
    \includegraphics[width=0.49\textwidth]{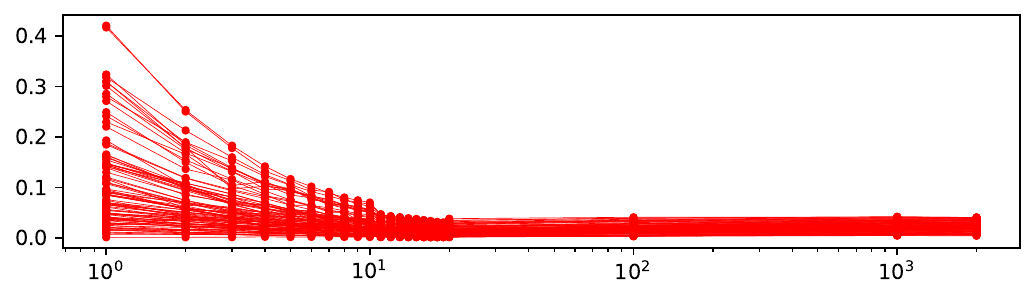}
    \includegraphics[width=0.49\textwidth]{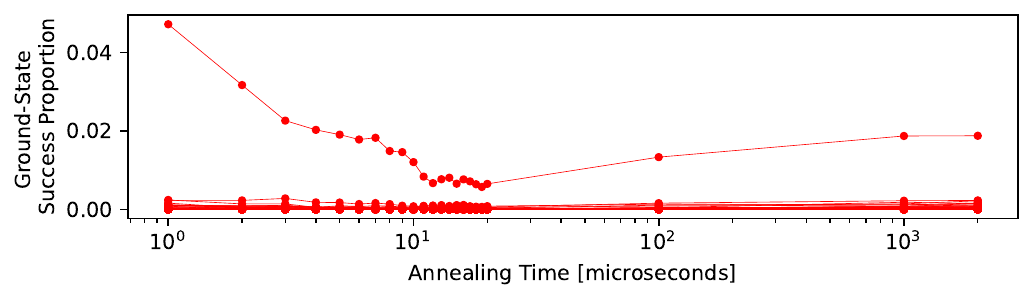}
    \includegraphics[width=0.49\textwidth]{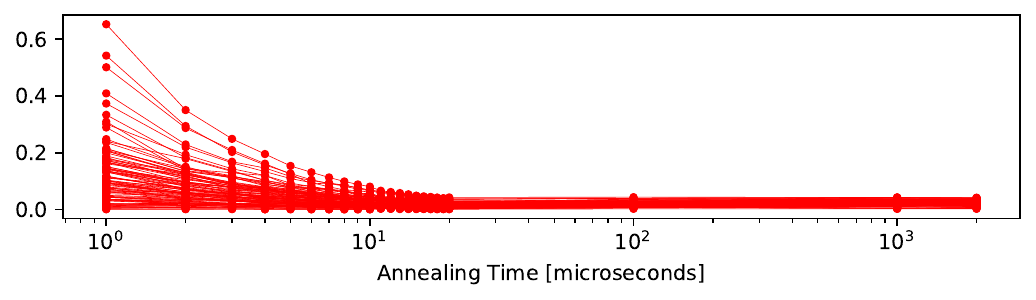}
    \caption{\texttt{Advantage2\_prototype1.1} results, where each subfigure shows the optimal solution sampling rate out of the number of anneals used, for $100$ different random QUBO instances, as a function of log-scale annealing time. Posiform QUBO scaling coefficient is $0.01$. \emph{lin$_{2}$} QUBO instances are shown in the left-hand side column and \emph{lin$_{20}$} QUBO instances are shown in the right-hand side column. Rows show results for QUBOs constructed with maximum random glued QUBO sizes of, in order from top to bottom, $36$, $71$, $141$, and $282$ variables.  }
    \label{fig:sampling_success_rate_Zephyr1.1_0.01}
\end{figure}

\begin{figure}[th!]
    \centering
    \includegraphics[width=0.49\textwidth]{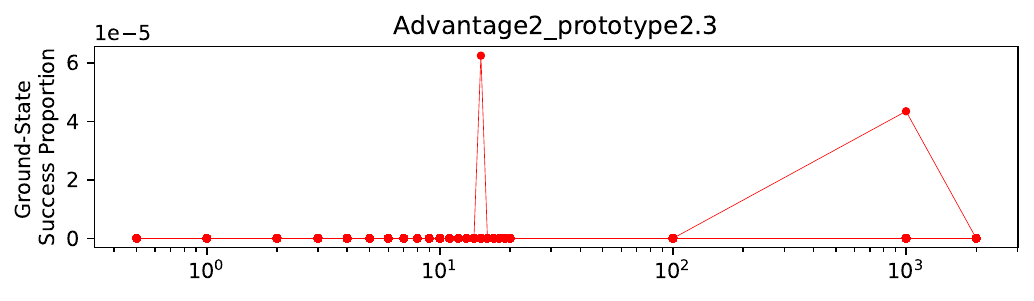}
    \includegraphics[width=0.49\textwidth]{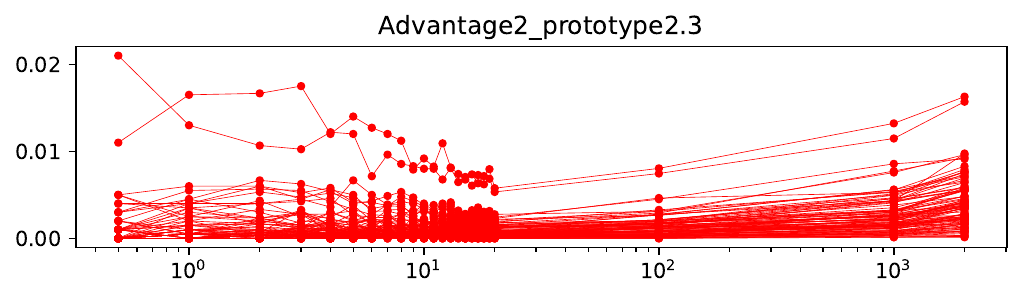}
    \includegraphics[width=0.49\textwidth]{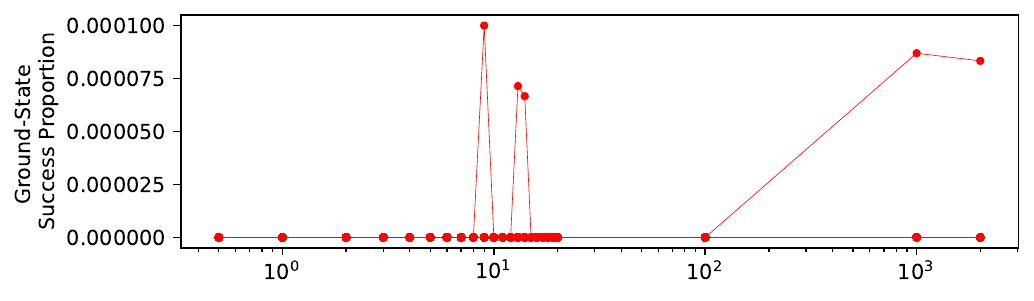}
    \includegraphics[width=0.49\textwidth]{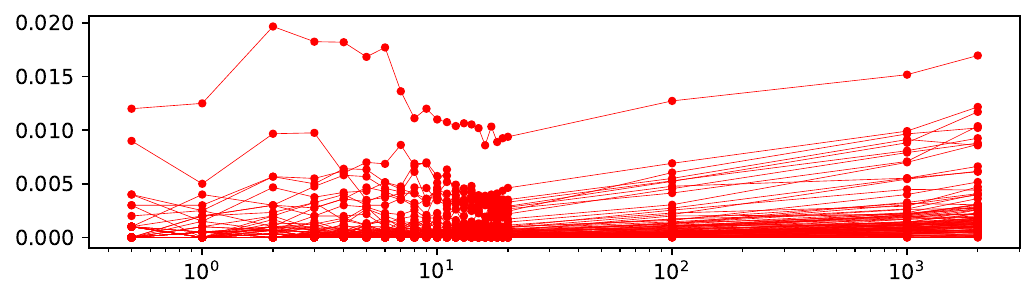}
    \includegraphics[width=0.49\textwidth]{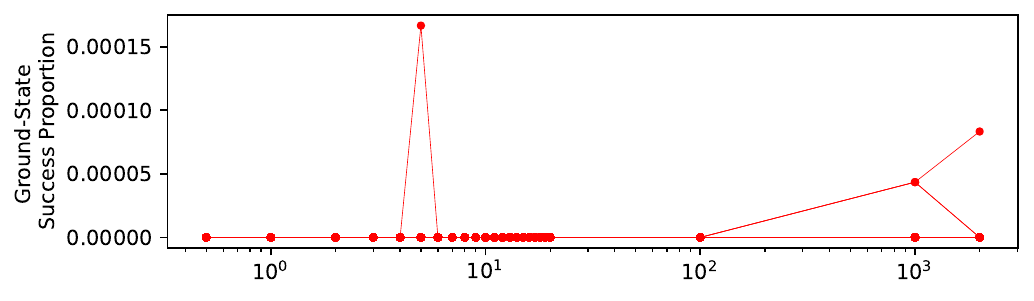}
    \includegraphics[width=0.49\textwidth]{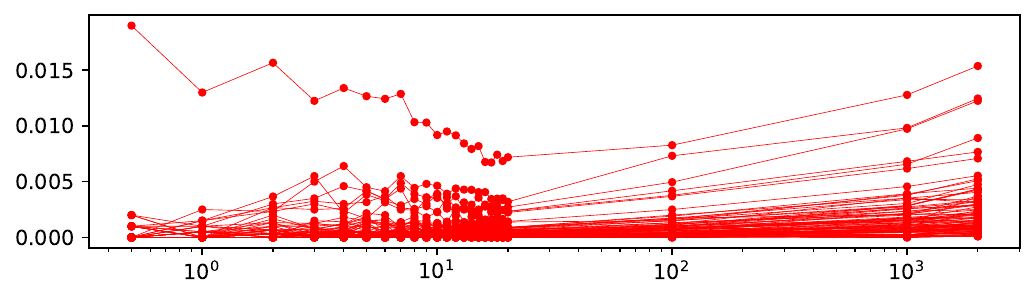}
    \includegraphics[width=0.49\textwidth]{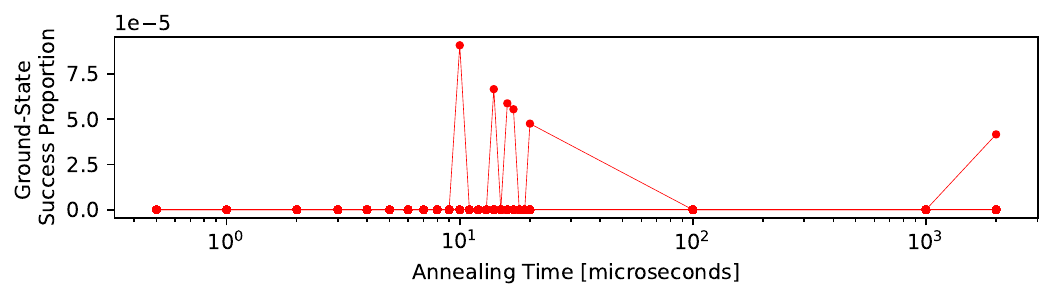}
    \includegraphics[width=0.49\textwidth]{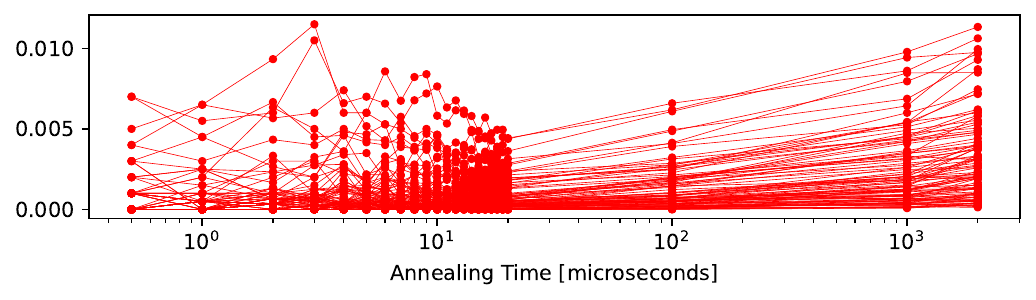}
    \caption{\texttt{Advantage\_prototype2.3} sampling results, where each subfigure shows the optimal solution sampling rate out of the number of anneals used, for $100$ different random QUBO instances, as a function of log-scale annealing time. Posiform QUBO scaling coefficient is $0.01$. \emph{lin$_{2}$} QUBO instances are shown in the left-hand side column and \emph{lin$_{20}$} QUBO instances are shown in the right-hand side column. Rows show results for QUBOs constructed with maximum random glued QUBO sizes of, in order from top to bottom, $39$, $77$, $153$, and $305$ variables.  }
    \label{fig:sampling_success_rate_Zephyr2.3_0.01}
\end{figure}

This experiment is repeated for~\texttt{Advantage\_prototype2.3} using a posiform QUBO scaling coefficient is $0.1$ in Figure~\ref{fig:sampling_success_rate_Zephyr2.3_0.1}. The results in Figure~\ref{fig:sampling_success_rate_Zephyr2.3_0.1} for the \texttt{Advantage\_prototype2.3} device confirm the ones seen for \linebreak\texttt{Advantage2\_prototype1.1} in that the achieved success probabilities are close to $1$ for low annealing times and decrease to closer to $0$ as the annealing times increase. As before, all subplots show a qualitative similar behavior for the different coefficient choices (columns) or the number and sizes of the smaller QUBOs being glued onto the posiform planted QUBO (rows).

Figure~\ref{fig:sampling_success_rate_Zephyr1.1_0.01} shows the quantum annealing ground-state sampling success rate for QUBOs tailored to the \linebreak\texttt{Advantage2\_prototype1.1} hardware graph, but now using a posiform scaling coefficient of $0.01$. This shows a clear decrease of ground-state sampling success as compared to Figure~\ref{fig:sampling_success_rate_Zephyr1.1_0.1} with a larger posiform scaling coefficient. We observe that by making the posiform scaling coefficient small and making the random QUBO coefficient discrete in $\{+1, -1\}$, we can make these solution planted QUBOs much harder for the D-Wave hardware to solve optimally. 

Figure~\ref{fig:sampling_success_rate_Zephyr2.3_0.01} shows quantum annealing ground-state sampling success rates for the QUBOs defined on the \linebreak\texttt{Advantage2\_prototype2.3} hardware graph, again with a posiform scaling coefficient of $0.01$ (as opposed to Figure~\ref{fig:sampling_success_rate_Zephyr2.3_0.1}). Here we see a significant decrease in optimal solution sampling rate, where for the \emph{lin$_{2}$} QUBO instances (left column) the unique optimal is rarely sampled. Thus we again observe that the posiform scaling coefficient attenuates the computational hardness of these planted QUBOs. The overall consistent finding is that smaller posiform QUBO coefficients with respect to the fused random QUBOs make the problems harder for the D-Wave quantum annealers to solve.

Figures~\ref{fig:sampling_success_rate_Pegasus4.1}, \ref{fig:sampling_success_rate_Zephyr1.1_0.1}, \ref{fig:sampling_success_rate_Zephyr2.3_0.1} all show a consistent trend of higher ground-state sampling rates at smaller annealing times, and lower ground-state sampling rates at longer annealing times. This is a very interesting property of quantum annealing and these particular optimization problems -- for many other optimization problems solved with quantum annealing in other studies, this trend is exactly reversed; usually better optimal solution sampling occurs at longer annealing times~\cite{Albash_2018, pelofske2023comparinggenerationsdwavequantum, tasseff2022emerging, willsch2022benchmarking, QA_QAOA_pelofske, pelofske2023shortdepth, king2017quantumannealingamidlocal}. 

Figures~\ref{fig:sampling_success_rate_Pegasus4.1}, \ref{fig:sampling_success_rate_Zephyr1.1_0.1}, \ref{fig:sampling_success_rate_Zephyr2.3_0.1}, \ref{fig:sampling_success_rate_Zephyr1.1_0.01}, \ref{fig:sampling_success_rate_Zephyr2.3_0.01} all show a very consistent trend that the sampling success rate, although it does depend on annealing time, does not change noticeably as a function of the varying random QUBO sizes that were fused to the posiform planted QUBO. This can be seen by looking at the differences between the rows of all of these figures -- each row denotes a different random QUBO size threshold. This is a counter-intuitive finding since larger random QUBOs generally require more compute time to solve to optimality, and thereby we would expect to see lower sampling success rates.

\begin{figure}[th!]
    \centering
    \includegraphics[width=0.49\textwidth]{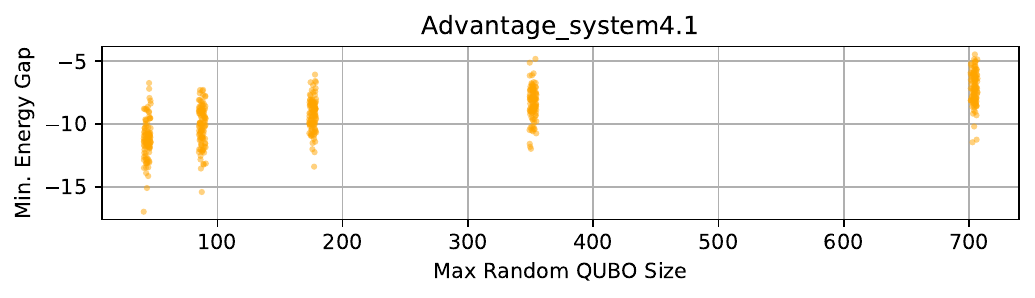}
    \includegraphics[width=0.49\textwidth]{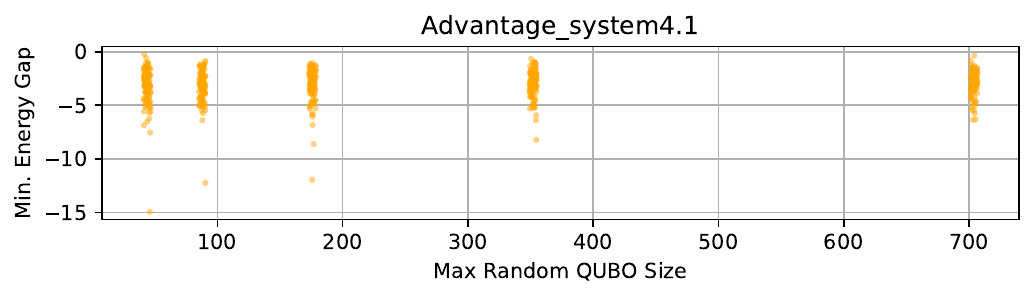}
    \caption{Difference between the energy of the unique planted solution and the minimum energy sample found on the D-Wave hardware (y-axis) as a function of the glued random QUBO size (x-axis), for QUBO coefficient choice \emph{lin$_{2}$} (left) and \emph{lin$_{20}$} (right). All of these QUBOs had a posiform scaling factor of $0.01$, and were defined on the \texttt{Advantage\_system4.1} hardware graph -- the notable property of these QUBOs is that the D-Wave hardware never sampled the unique planted solution. The minimum energy gap for the best solution found using the D-Wave hardware is indicated on the y-axis; values close to $0$ denote samples very close to the optimal solution energy. Plot uses jittering on the x-axis for better visualization. }
    \label{fig:Advantage_system4.1_0.01_energy_gap_sampling}
\end{figure}

\begin{figure}[th!]
    \centering
    \includegraphics[width=0.49\textwidth]{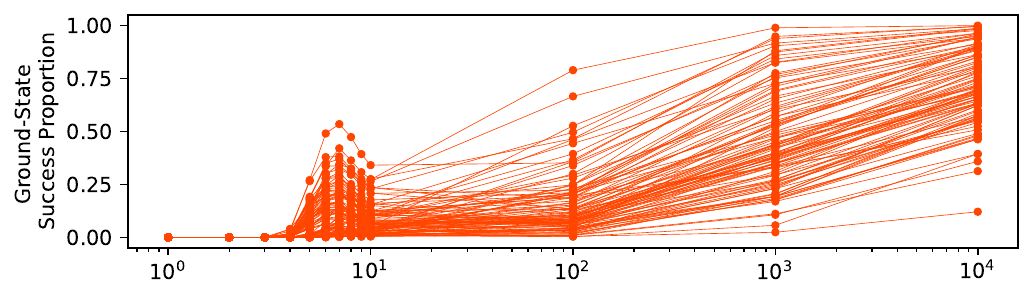}
    \includegraphics[width=0.49\textwidth]{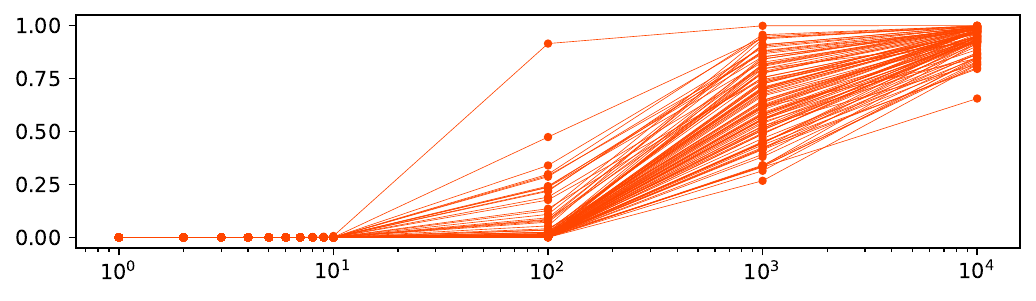}
    \includegraphics[width=0.49\textwidth]{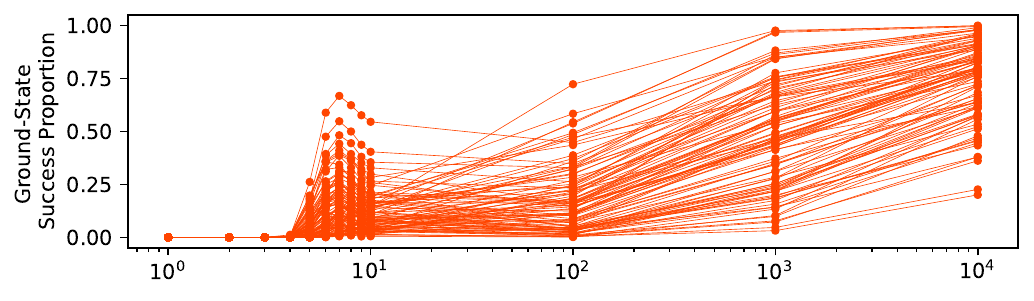}
    \includegraphics[width=0.49\textwidth]{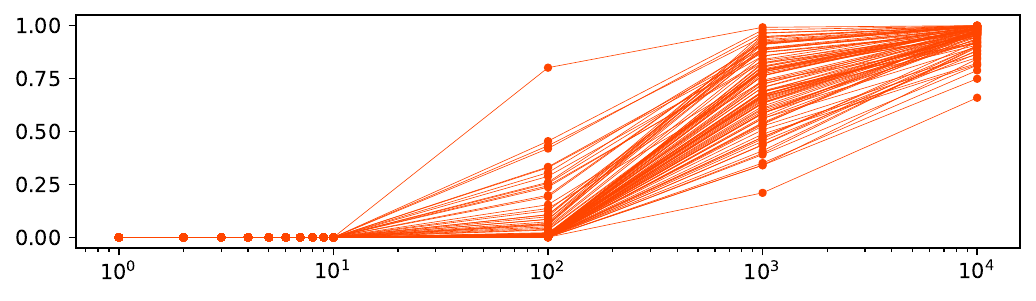}
    \includegraphics[width=0.49\textwidth]{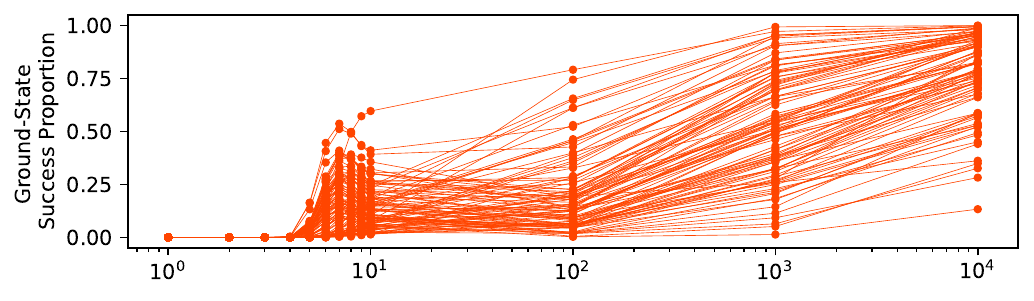}
    \includegraphics[width=0.49\textwidth]{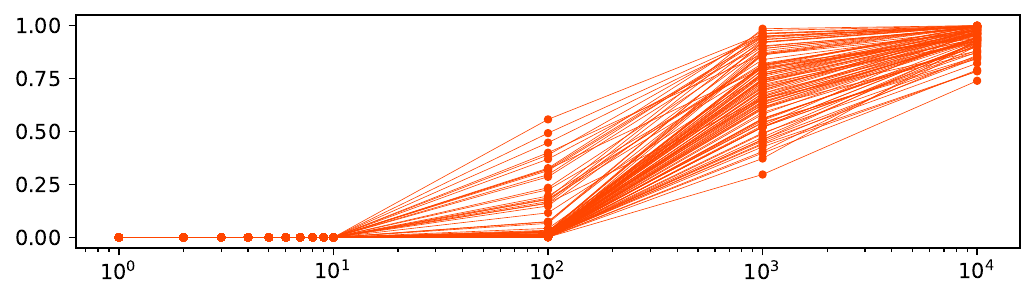}
    \includegraphics[width=0.49\textwidth]{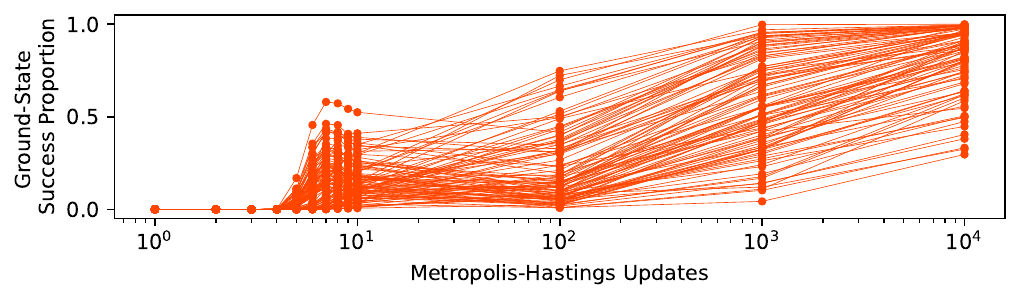}
    \includegraphics[width=0.49\textwidth]{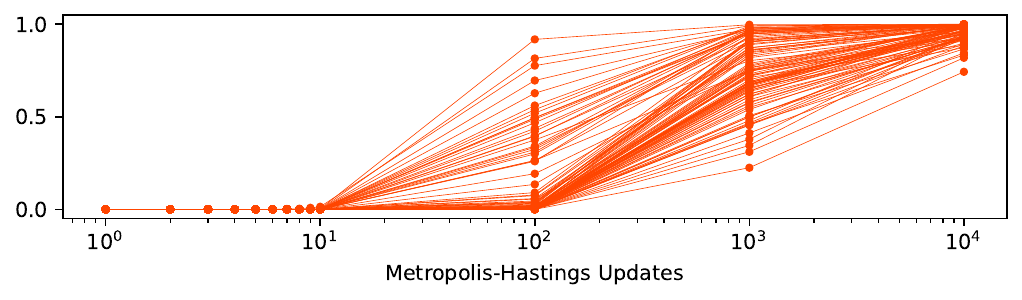}
    \caption{Simulated annealing applied to \texttt{Advantage2\_prototype2.3} hardware compatible QUBOs, where each subfigure shows the optimal solution sampling rate for the same $100$ different random QUBO instances already used in Section~\ref{section:results_optimal_solution_sampling_DWave}, as a function of log-scale number of complete sweeps of Metropolis-Hastings updates. Columns show QUBOs with coefficients from \emph{lin$_{2}$} (left) and \emph{lin$_{20}$} (right). Posiform scaling coefficient of $0.01$. The same QUBOs are solved in Figure~\ref{fig:sampling_success_rate_Zephyr2.3_0.01}.}
    \label{fig:SA_success_proportion_Zephyr2.3_0.01}
\end{figure}

For all problem instances generated on \texttt{Advantage\_system4.1} using posiform scale factor of $0.01$, the D-Wave hardware never sampled any of the optimal solutions out of the $1000$ samples obtained for each individual instance and QUBO setting configuration. For this reason, there are no ground-state sampling probability figures for this particular QUBO configuration, as there is for posiform coefficient scaling of $0.1$ on the \texttt{Advantage\_system4.1} instances (Figure \ref{fig:sampling_success_rate_Pegasus4.1}). However, there is a natural question of how close the D-Wave the hardware was to sampling the planted solution. Figure \ref{fig:Advantage_system4.1_0.01_energy_gap_sampling} answers this question by plotting the energy difference between the planted solution energy and the lowest energy sample found on D-Wave hardware. The more negative this quantity is, the further away the quantum annealing hardware sample was from the planted solution, and the closer to $0$ this quantity is, the closer the hardware sampling was to the optimal solution.

\subsection{Success Rates of Optimal Solution Sampling with Simulated Annealing}
\label{section:results_optimal_solution_sampling_SA}
In order to have a base for comparison we repeat the experiments of the previous section and solve the same QUBOs with the classical simulated annealing algorithm \cite{kirkpatrick1983optimization}. In particular, we solve the same set of QUBOs as in Section~\ref{section:results_optimal_solution_sampling_DWave}.

\begin{figure}[th!]
    \centering
    \includegraphics[width=0.49\textwidth]{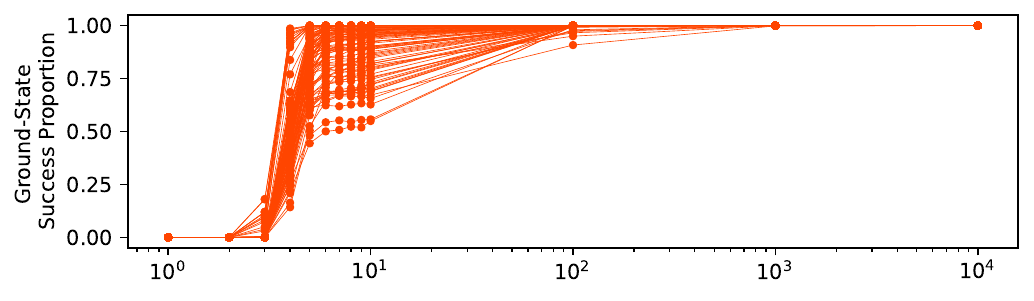}
    \includegraphics[width=0.49\textwidth]{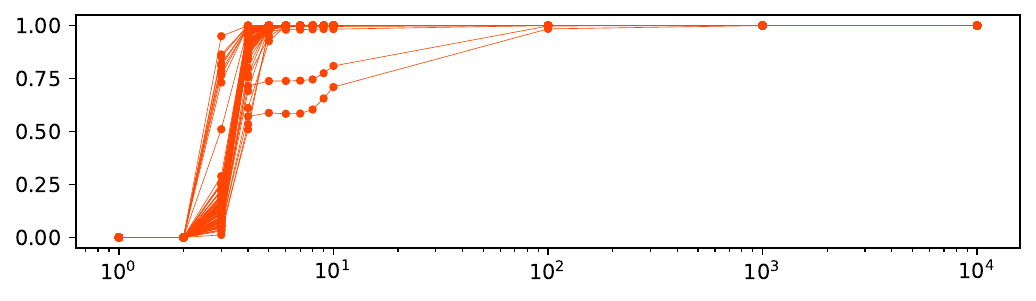}
    \includegraphics[width=0.49\textwidth]{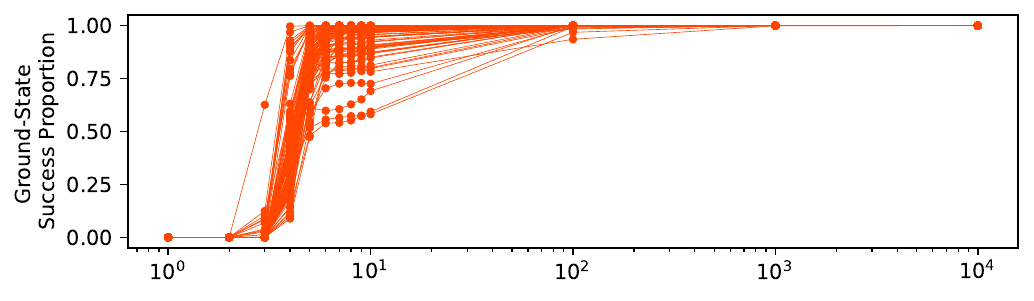}
    \includegraphics[width=0.49\textwidth]{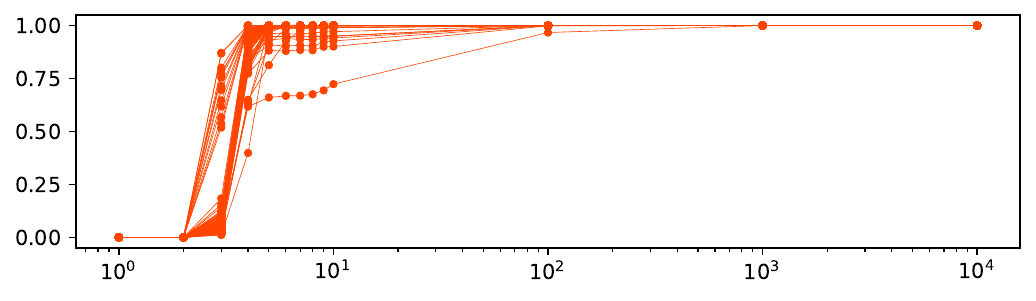}
    \includegraphics[width=0.49\textwidth]{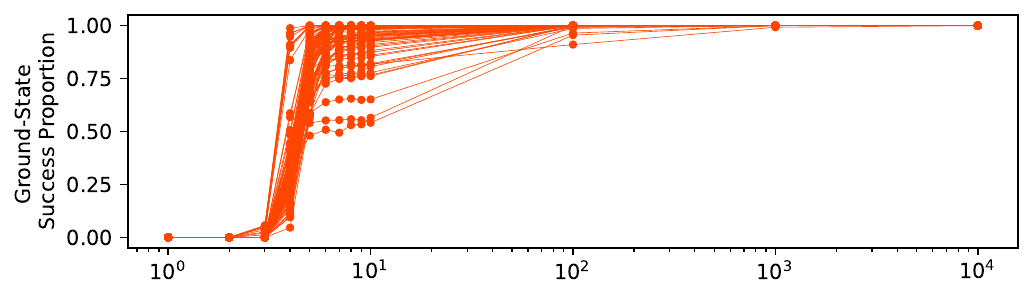}
    \includegraphics[width=0.49\textwidth]{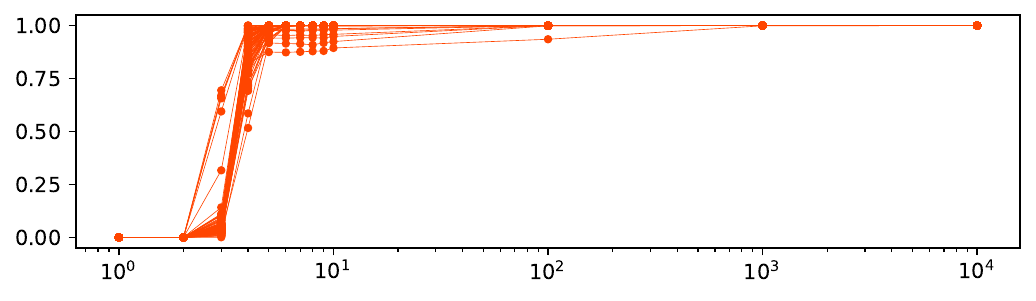}
    \includegraphics[width=0.49\textwidth]{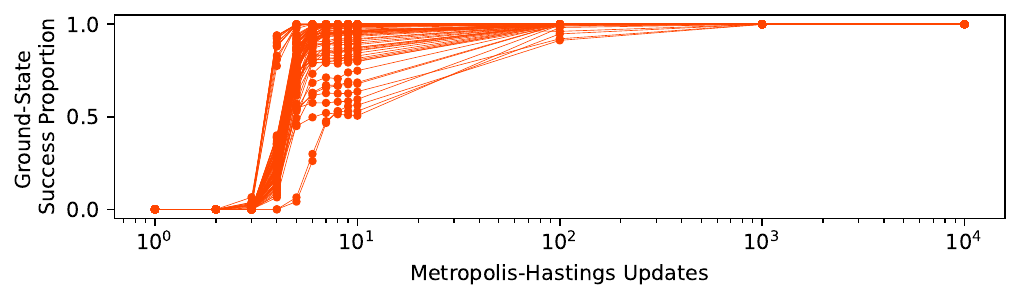}
    \includegraphics[width=0.49\textwidth]{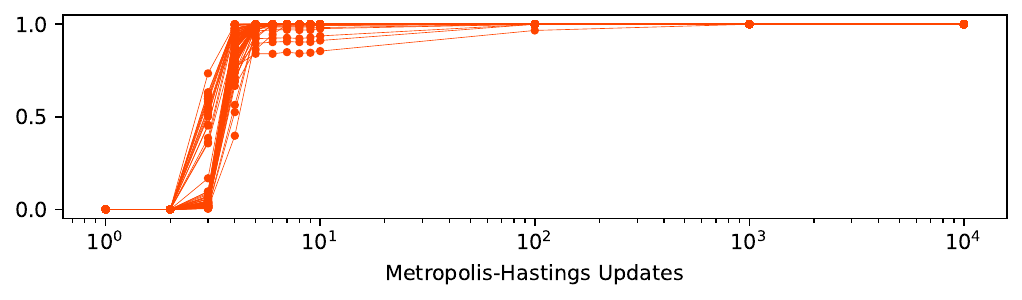}
    \caption{Simulated annealing sampling results. \texttt{Advantage2\_prototype2.3} hardware compatible QUBO results, where each subfigure shows the optimal solution sampling rate for the $100$ different random QUBO instances, as a function of log-scale number of complete sweeps of Metropolis-Hastings updates. Posiform QUBO scaling coefficient is $0.1$. \emph{lin$_{2}$} QUBO instances are shown in the left-hand side column and \emph{lin$_{20}$} QUBO instances are shown in the right-hand side column. The same QUBOs are solved in Figure~\ref{fig:sampling_success_rate_Zephyr2.3_0.1}. }
    \label{fig:SA_success_proportion_Zephyr2.3_0.1}
\end{figure}

Figure~\ref{fig:SA_success_proportion_Zephyr2.3_0.01} shows simulation results when solving the same set of QUBOs generated for Figure~\ref{fig:sampling_success_rate_Zephyr2.3_0.01}, that is QUBOs tailored to the \texttt{Advantage2\_prototype2.3} hardware graph. Two observations are noteworthy. First, the displayed plots are qualitatively similar, in that they all show that the success probability increases with the number of Metropolis-Hastings updates, which is expected. In particular, the plots are similar for the QUBOs generated for QUBOs with coefficients from \emph{lin$_{2}$} (left column) and \emph{lin$_{20}$} (right column). Second, we observe that for a low number of Metropolis-Hastings updates, the success probability is (close to) zero, showing that the generated QUBOs are not trivial, and that moreover, more Metropolis-Hastings updates are needed for a non-zero success probability when the coefficients come \emph{lin$_{20}$} (right column).

Figure~\ref{fig:SA_success_proportion_Zephyr2.3_0.1} shows qualitatively similar results to Figure~\ref{fig:SA_success_proportion_Zephyr2.3_0.01}, where a small number of spin updates yields low ground-state success rate and a larger number of spin updates yeilds ground-state success rates closer to $1$. However, Figure~\ref{fig:SA_success_proportion_Zephyr2.3_0.1} with posiform scaling of $0.1$ has a much higher ground-state success probability compared to Figure~\ref{fig:SA_success_proportion_Zephyr2.3_0.01} with posiform scaling of $0.01$. This shows that the posiform scale factor can attenuate the hardness of these QUBO problems for simulated annealing. This same quality of the larger posiform scaling coefficients being easier was seen in the quantum annealing sampling results.

Appendix \ref{section:appendix_SA_plots} shows three additional simulated annealing sampling figures (Figures \ref{fig:SA_success_proportion_Pegasus4.1_0.1}, \ref{fig:SA_success_proportion_Zephyr1.1_0.01}, \ref{fig:SA_success_proportion_Zephyr1.1_0.1}). All of the simulated annealing sampling plots show that the optimal solution is able to be reliably sampled for all QUBO configurations if a sufficiently large number of variable spin updates is utilized.

An interesting property of the ground-state sampling curves from simulated annealing for the posiform scaling $0.01$ and \emph{lin$_{2}$} coefficient distributions (see the left columns of Figures~\ref{fig:sampling_success_rate_Zephyr2.3_0.01} and \ref{fig:SA_success_proportion_Zephyr2.3_0.01}) is a slight down-turn of success probability just before $10$ Metropolis-Hastings spin updates, before the success probability continues to increase. This non-monotonic behavior seems to be a characteristic of these specific problem instances as it is not seen in the other posiform solution planted QUBO configurations, and the cause is not known.

\begin{figure}[th!]
    \centering
    \includegraphics[width=0.49\textwidth]{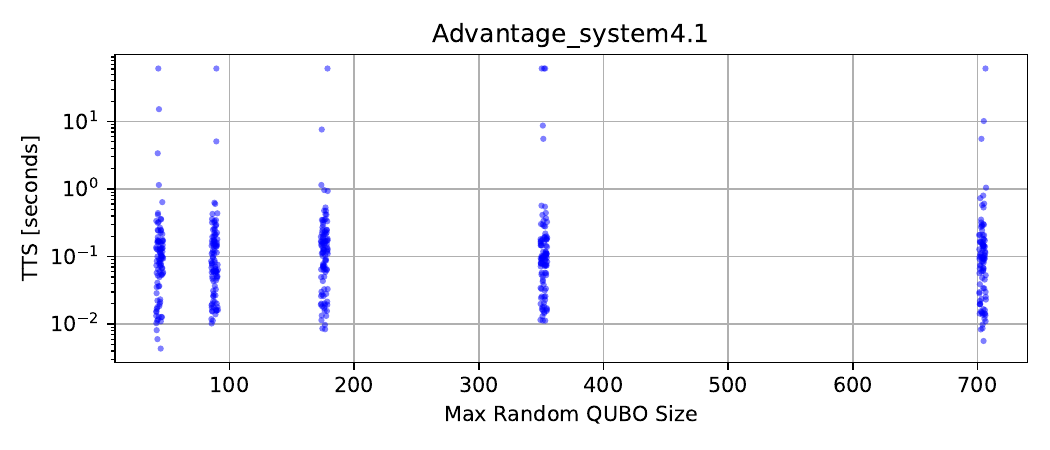}
    \includegraphics[width=0.49\textwidth]{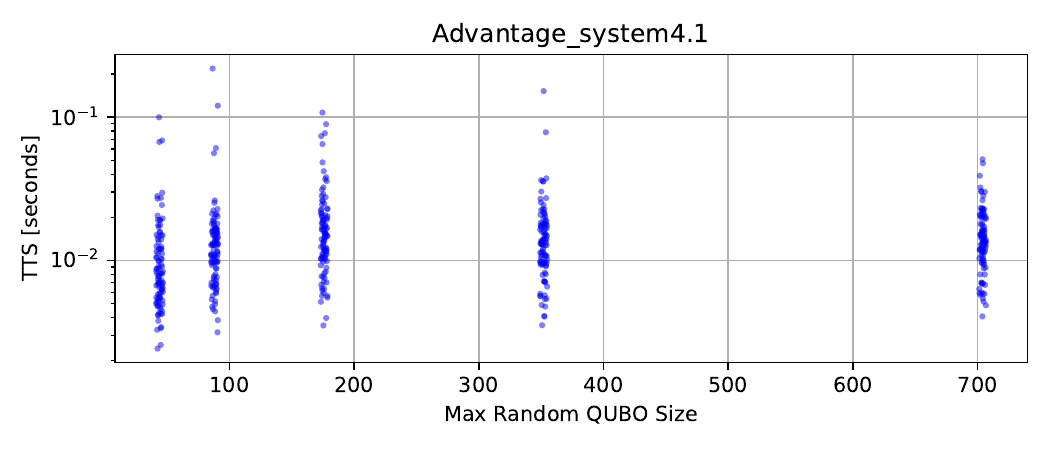}
    \includegraphics[width=0.49\textwidth]{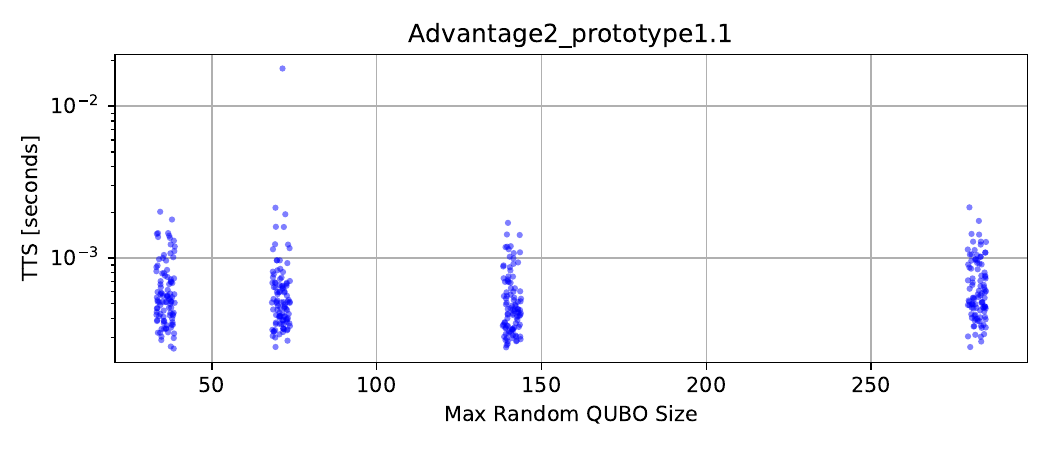}
    \includegraphics[width=0.49\textwidth]{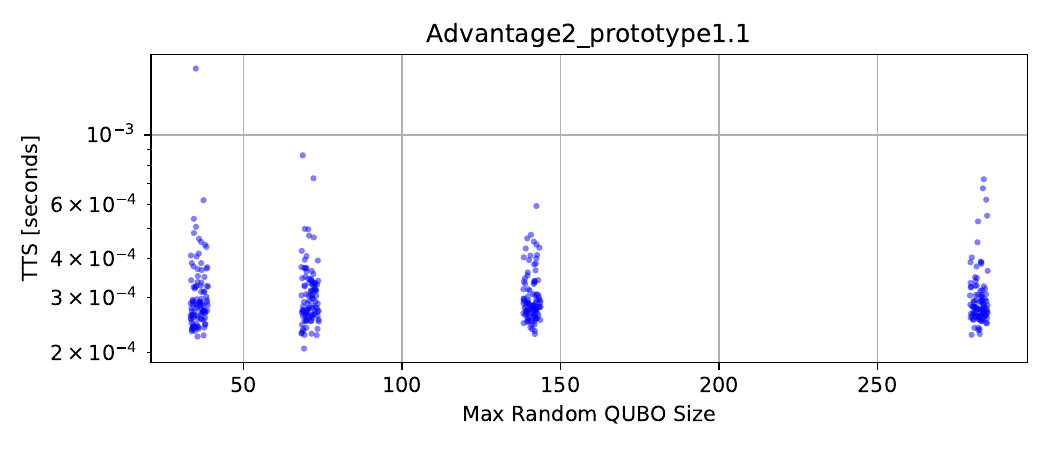}
    \includegraphics[width=0.49\textwidth]{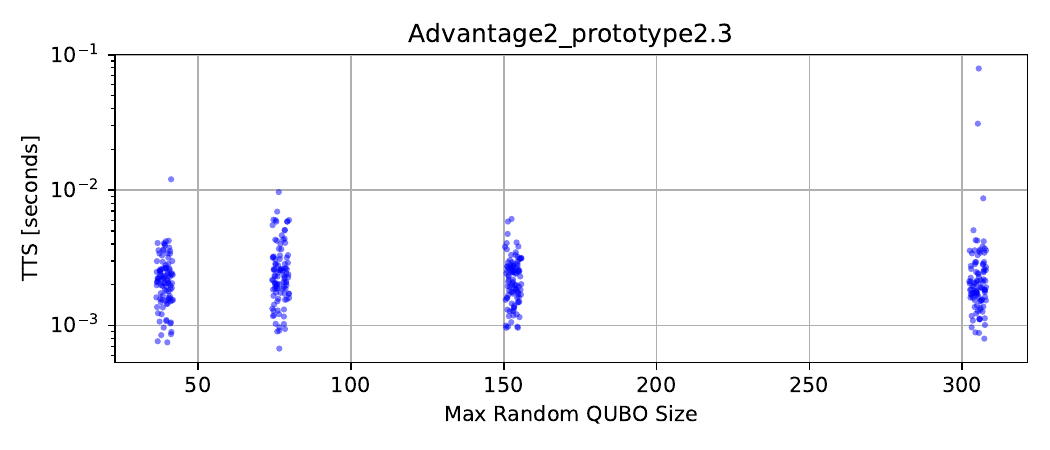}
    \includegraphics[width=0.49\textwidth]{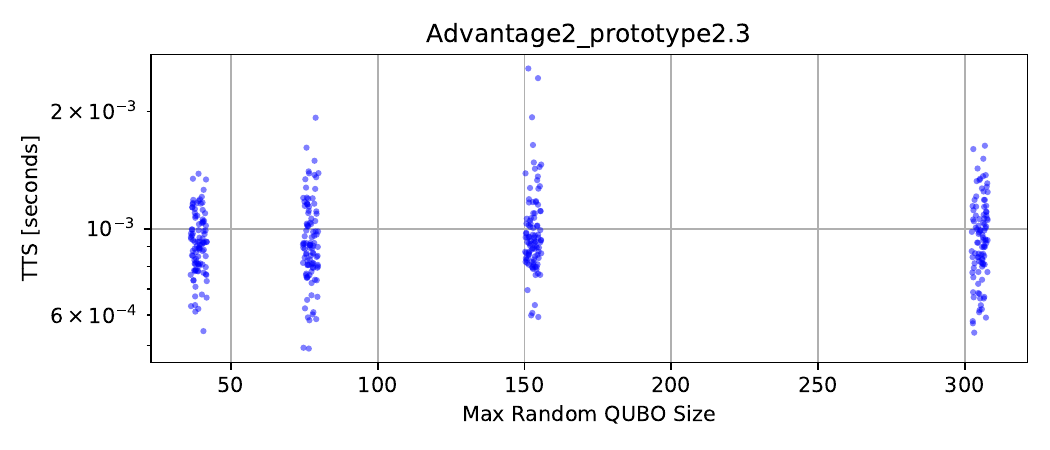}
    \caption{TTS in seconds as a function of the maximum random QUBO sizes that were glued to the whole-chip posiform QUBO. Columns show QUBOs with coefficients from \emph{lin$_{2}$} (left) and \emph{lin$_{20}$} (right). Posiform coefficient scale factor $0.1$. Rows show the three different D-Wave devices under consideration, those are \texttt{Advantage\_system4.1} (top), \texttt{Advantage2\_prototype1.1} (middle), \texttt{Advantage\_prototype2.3} (bottom). Plot uses jittering on the x-axis for better visualization. Log scale on the y-axis.}
    \label{fig:TTS_0.1}
\end{figure}

\subsection{Time to Solution on D-Wave QPUs}
\label{section:results_TTS}
In this section we record the TTS metric for solving the previously considered ensembles of QUBOs. The TTS is computed as in Section~\ref{section:methods_TTS}. Here TTS is computed for each individual hardware tailored QUBO instance using the entirety of the parameter sweep and all samples obtained for each QUBO, the results for which are described in Section~\ref{section:results_optimal_solution_sampling_DWave}. In particular, this means that the sample success rate is measured from the entirety of the samples across all annealing times used, and therefore the QPU time per anneal is the average QPU time across different annealing times. This approach to measure TTS was taken because the sampling success rate can vary quite significantly as a function of the annealing time. In particular, for some annealing times the measured optimal solution sampling success rate is $0$.

\begin{figure}[th!]
    \centering
    \includegraphics[width=0.49\textwidth]{figures/TTS_function_of_random_QUBO_sizes_Advantage2_prototype1.1_0.1_v1.pdf}
    \includegraphics[width=0.49\textwidth]{figures/TTS_function_of_random_QUBO_sizes_Advantage2_prototype1.1_0.1_v2.pdf}
    \includegraphics[width=0.49\textwidth]{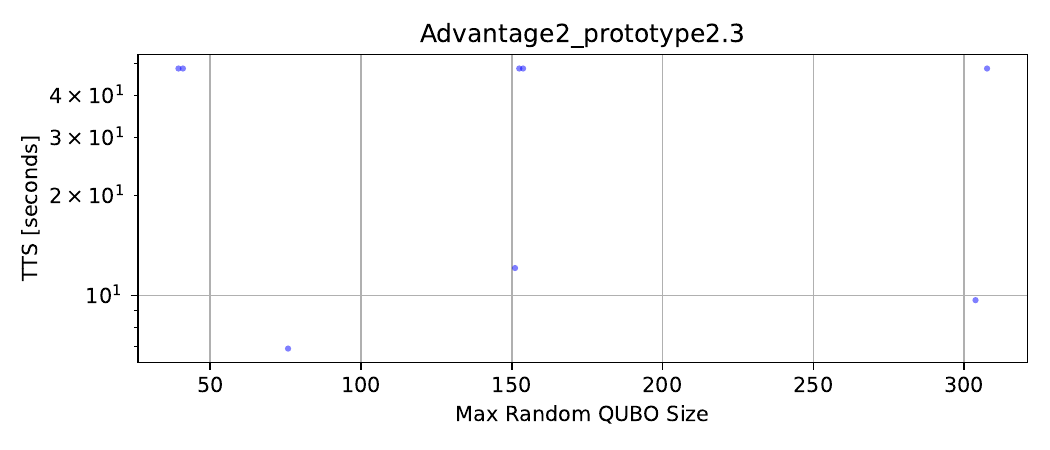}
    \includegraphics[width=0.49\textwidth]{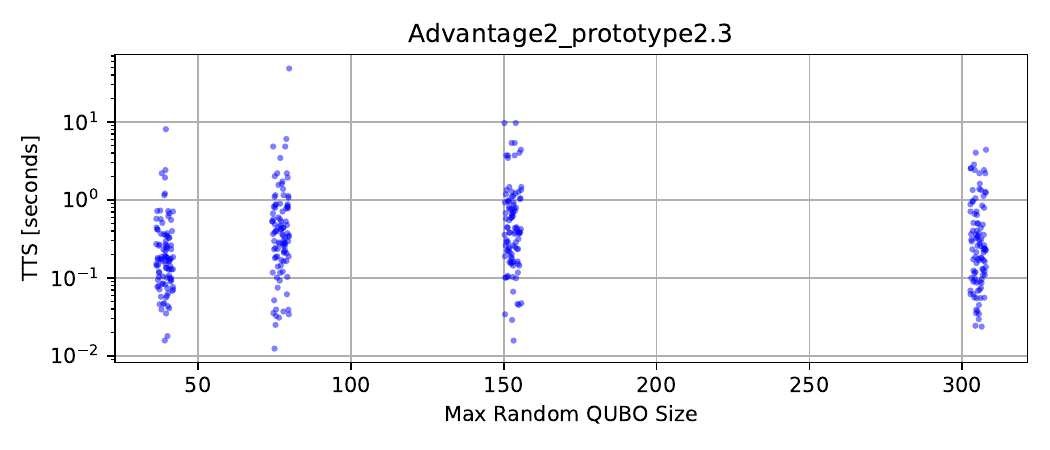}
    \caption{TTS in seconds as a function of the maximum random QUBO sizes that were glued to the whole-chip posiform QUBO. Columns show QUBOs with coefficients from \emph{lin$_{2}$} (left) and \emph{lin$_{20}$} (right). Posiform coefficient scale factor $0.01$. Rows show the two different D-Wave devices under consideration, those are \texttt{Advantage2\_prototype1.1} (top) and \texttt{Advantage\_prototype2.3} (bottom). Plot uses jittering on the x-axis for better visualization. Log scale on the y-axis.}
    \label{fig:TTS_0.01}
\end{figure}

\begin{table}[th!]
\scriptsize
\begin{center}
\begin{tabular}{ |p{1.9cm}||p{4.7cm}|p{4.9cm}|p{4.7cm}| } 
 \hline
 QUBO Setting Configuration & Advantage\_system4.1 & Advantage2\_prototype1.1 & Advantage2\_prototype2.3 \\ 
 \hline
 \hline
 lin$_{2}$, 0.1, 50 & 99\%: (0.004348, 0.93124, 60.972384) & 100\%: (0.000253, 0.000648, 0.002017) & 100\%: (0.000751, 0.00232, 0.012021)  \\ 
 \hline
 lin$_{2}$, 0.1, 100 & 100\%: (0.010235, 0.781875, 60.965955) & 100\%: (0.000259, 0.000771, 0.017736) & 100\%: (0.000676, 0.002706, 0.009672) \\ 
 \hline
 lin$_{2}$, 0.1, 200 & 100\%: (0.008407, 0.850835, 60.964776) & 100\%: (0.000257, 0.000553, 0.001703) & 100\%: (0.000963, 0.002303, 0.006125) \\ 
 \hline
 lin$_{2}$, 0.1, 400 & 100\%: (0.01127,6 2.085293, 60.974281) & 100\%: (0.000259, 0.000677, 0.002156) & 100\%: (0.0008, 0.003271, 0.079056) \\ 
 \hline
 lin$_{2}$, 0.1, 800 & 98\%: (0.005632, 0.918252, 60.96889) & &  \\ 
 \hline
 lin$_{2}$, 0.01, 50 & 0 & 80\%: (0.05632, 62.179882, 290.988536) & 2\%: (48.36274, 48.363182, 48.363624) \\ 
 \hline
 lin$_{2}$, 0.01, 100 & 0 & 90\%: (0.094933, 47.553529, 290.707565) & 1\%: (6.908099, 6.908099, 6.908099) \\ 
 \hline
 lin$_{2}$, 0.01, 200 & 0 & 96\%: (0.022682, 28.709637, 290.874972) & 3\%: (12.089156, 36.27184, 48.364066) \\ 
 \hline
 lin$_{2}$, 0.01, 400 & 0 & 100\%: (0.00854, 18.433104, 290.870422) & 2\%: (9.671388, 29.015738, 48.360088) \\ 
 \hline
 lin$_{20}$, 0.1, 50 & 100\%: (0.001896, 0.009023, 0.081443) & 100\%: (0.000225, 0.000319, 0.001641) & 100\%: (0.000546, 0.000928, 0.001386) \\ 
 \hline
 lin$_{20}$, 0.1, 100 & 100\%: (0.002404, 0.01277, 0.232345) & 100\%: (0.000206, 0.000314, 0.000862) & 100\%: (0.000492, 0.000951, 0.001931) \\ 
 \hline
 lin$_{20}$, 0.1, 200 & 100\%: (0.002644, 0.014972, 0.112691) & 100\%: (0.000229, 0.000305, 0.000592) & 100\%: (0.000593, 0.001009, 0.002584) \\ 
 \hline
 lin$_{20}$, 0.1, 400 & 100\%: (0.002577, 0.011956, 0.136661) & 100\%: (0.000228, 0.000306, 0.000722) & 100\%: (0.000541, 0.000964, 0.001636) \\ 
 \hline
 lin$_{20}$, 0.1, 800 & 100\%: (0.002972, 0.01056, 0.040907) &  &  \\ 
 \hline
 lin$_{20}$, 0.01, 50 & 0 & 100\%: (0.001037, 0.008054, 0.047374) & 100\%: (0.015793, 0.373501, 8.059249) \\ 
 \hline
 lin$_{20}$, 0.01, 100 & 0 & 100\%: (0.000851, 0.008995, 0.226093) & 99\%: (0.012445, 1.167636, 48.360088) \\ 
 \hline
 lin$_{20}$, 0.01, 200 & 0 & 100\%: (0.001318, 0.007832, 0.032513) & 100\%: (0.015753, 0.968868, 9.671654) \\ 
 \hline
 lin$_{20}$, 0.01, 400 & 0 & 100\%: (0.000827, 0.006612, 0.060201) & 100\%: (0.023779, 0.571787, 4.395777) \\ 
 \hline
\end{tabular}
\end{center}
\caption{Summary metrics of the D-Wave hardware sampling results. Each cell contains a percentage indicating the percent of the 100 random problem instances which were solved optimally at least once, and three TTS measurements (in seconds) for this set of hardware experiments: the minimum TTS, the mean TTS, and the maximum TTS. Necessarily, TTS is only defined when the optimal solution is found at least once, and therefore the TTS measures are only defined for the problem instances which were solved at least once. Empty cells denote data that was not collected (namely, when the maximum size $800$ variable row entry for \texttt{Advantage2\_prototype2.3}). }
\label{table:TTS_quantities}
\end{table}

Figure~\ref{fig:TTS_0.1} shows TTS measurements for the full ensemble of posiform planted QUBOs with posiform scale of $0.1$, sampled on the three D-Wave QPUs. Figure \ref{fig:TTS_0.01} shows the same for the posiform scalng of $0.01$ (not including \texttt{Advantage\_system4.1} because the optimal solution sampling rate was always zero). Each subplot shows a scatterplot of the TTS measurements as a function of the maximum random QUBO sizes that were glued to the whole-chip posiform QUBO. As the TTS cannot be measured if the optimal solution is never found, there are missing data points in these TTS measures.

Figures~\ref{fig:TTS_0.1} and \ref{fig:TTS_0.01} show that the measured TTS largely stays in the same range irrespective of the size (and number) of the QUBOs glued onto the whole-chip posiform QUBO. 

\begin{figure}[th!]
    \centering
    \includegraphics[width=0.49\textwidth]{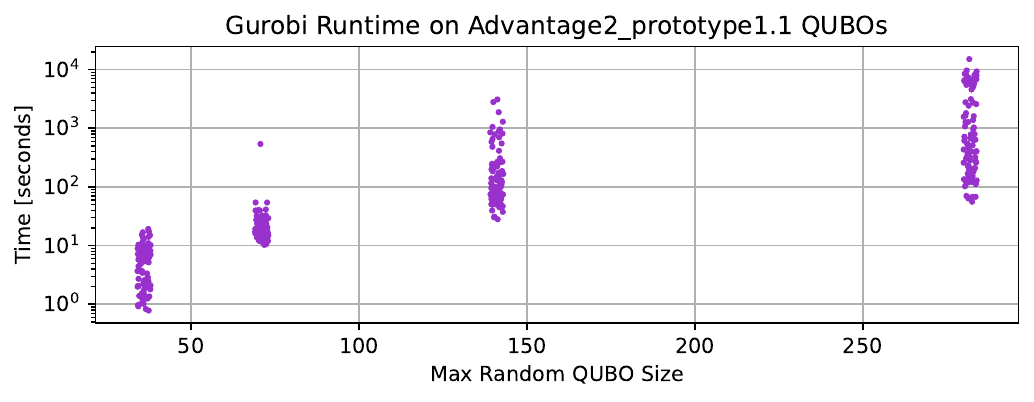}
    \includegraphics[width=0.49\textwidth]{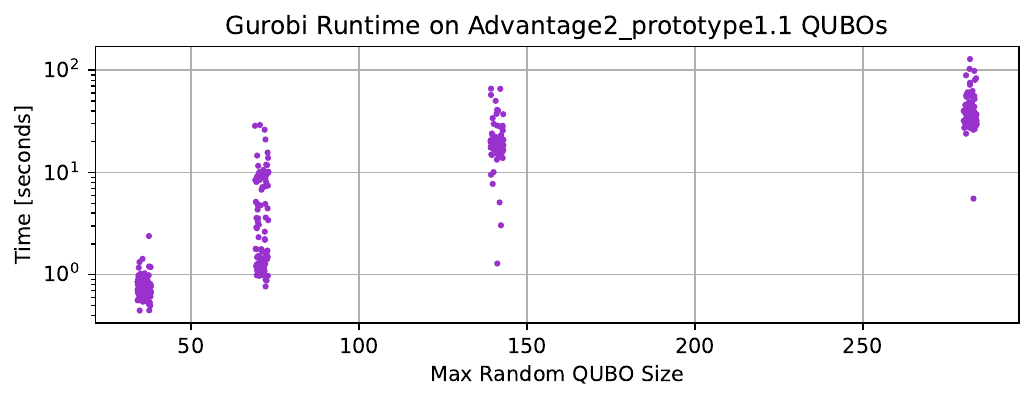}
    \includegraphics[width=0.49\textwidth]{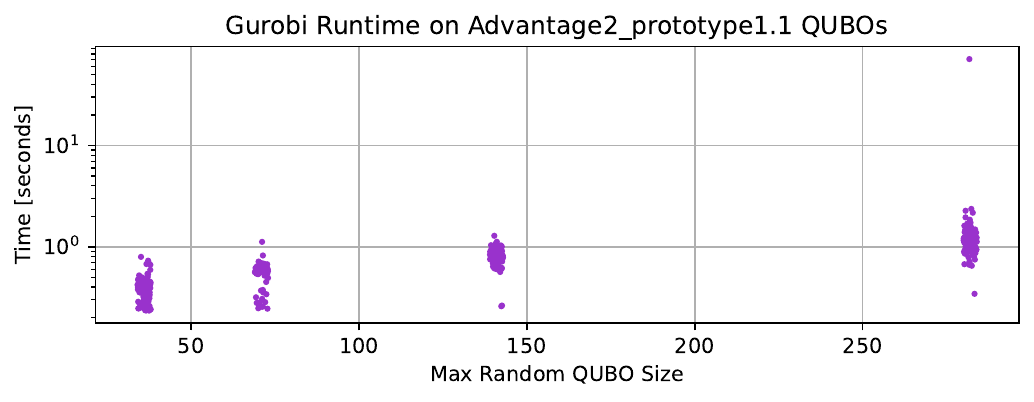}
    \includegraphics[width=0.49\textwidth]{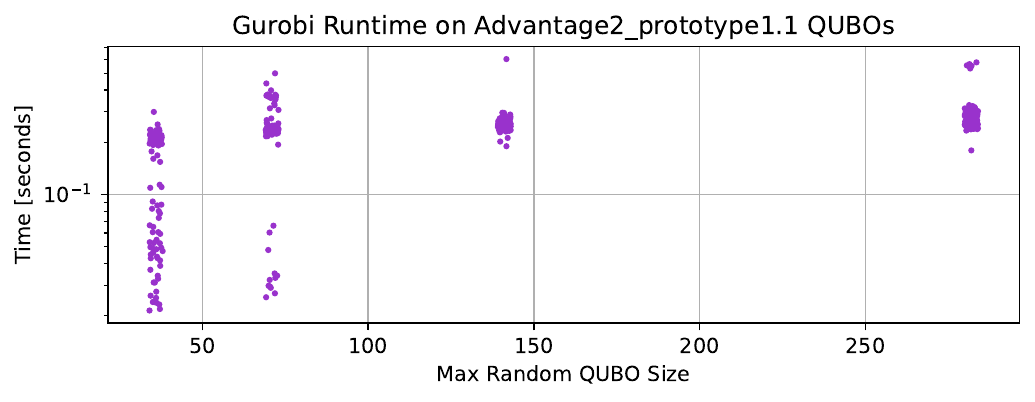}
    \caption{Gurobi runtimes in seconds (log-scale y-axis) as a function of the the random QUBO sizes (x-axis) for the \texttt{Advantage2\_prototype1.1} defined QUBOs. QUBO configurations are as follows: $0.01$ posiform scaling \emph{lin$_{2}$} (top left), $0.01$ posiform scaling \emph{lin$_{20}$} (top right), $0.1$ posiform scaling \emph{lin$_{2}$} (bottom left), $0.1$ posiform scaling \emph{lin$_{20}$} (bottom right). Here, we observe a clear dependence on total CPU runtimes with respect to the random QUBO size. x-axis coordinates have a small amount of noise jitter added for improved visualization. }
    \label{fig:Gurobi_Advantage2_prototype1.1}
\end{figure}

Table~\ref{table:TTS_quantities} gives exact TTS quantities, and range of TTS quantities, for each QUBO parameter type.

\subsection{Gurobi Runtime}
\label{section:results_Gurobi}

\begin{figure}[th!]
    \centering
    \includegraphics[width=0.49\textwidth]{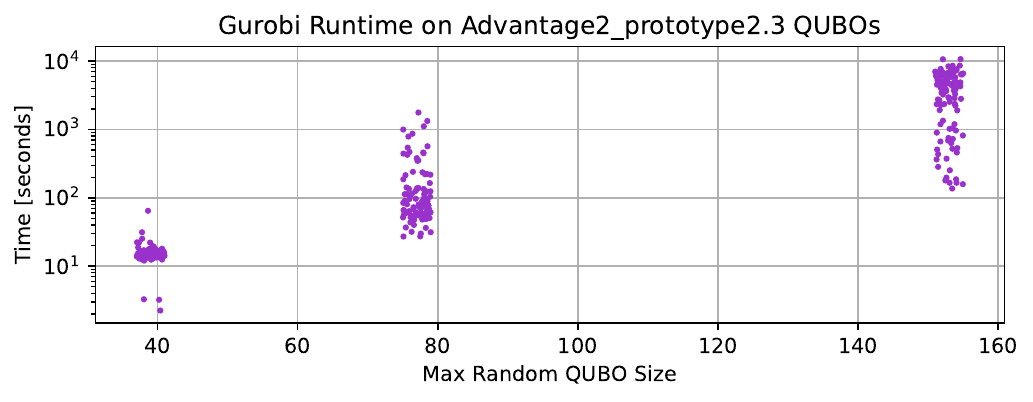}
    \includegraphics[width=0.49\textwidth]{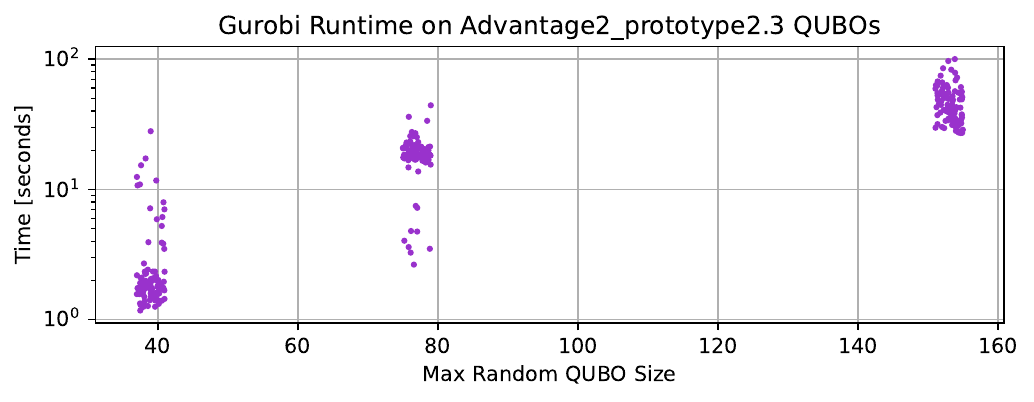}
    \includegraphics[width=0.49\textwidth]{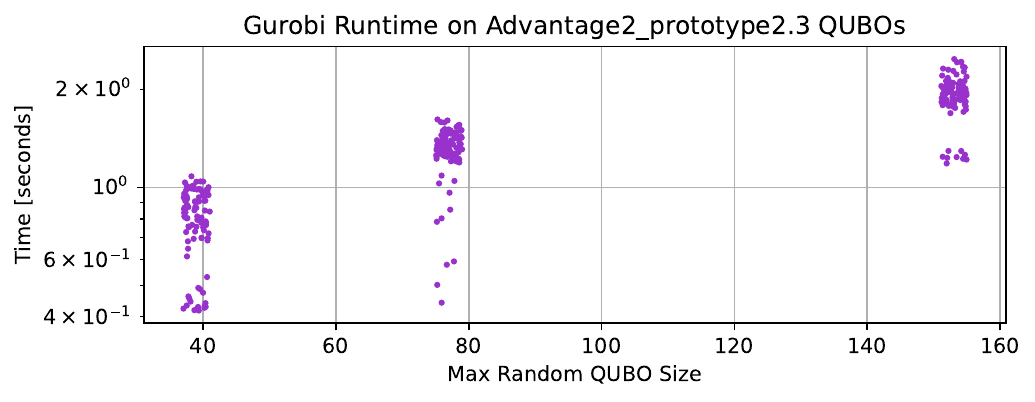}
    \includegraphics[width=0.49\textwidth]{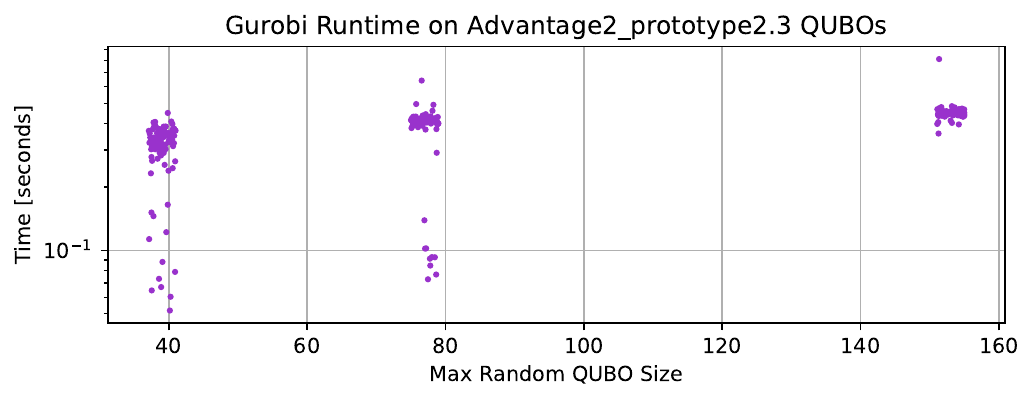}
    \caption{Gurobi runtimes in seconds (log-scale on the y-axis) as a function of the the random QUBO sizes (x-axis) for the \texttt{Advantage2\_prototype2.3} defined QUBOs. QUBO configurations are as follows: $0.01$ posiform scaling \emph{lin$_{2}$} (top left), $0.01$ posiform scaling \emph{lin$_{20}$} (top right), $0.1$ posiform scaling \emph{lin$_{2}$} (bottom left), $0.1$ posiform scaling \emph{lin$_{20}$} (bottom right). The x-axis coordinates have a small amount of noise jitter added for improved visualization. }
    \label{fig:Gurobi_Advantage2_prototype2.3}
\end{figure}

\begin{figure}[th!]
    \centering
    \includegraphics[width=0.49\textwidth]{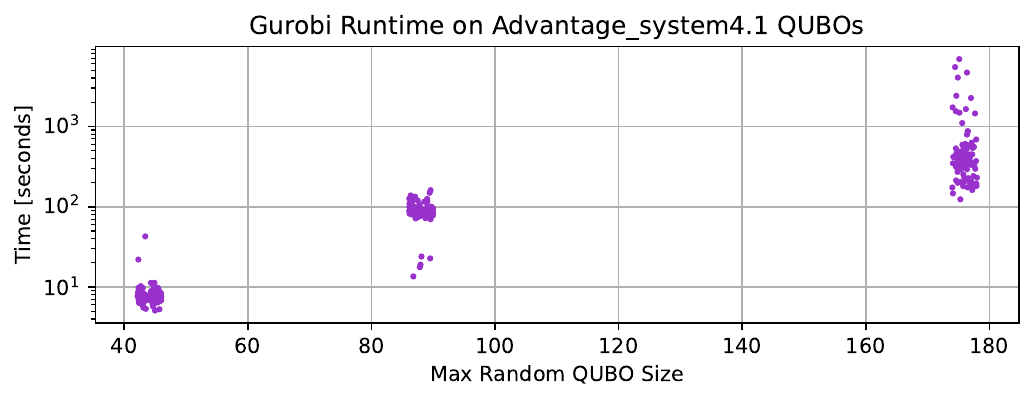}
    \includegraphics[width=0.49\textwidth]{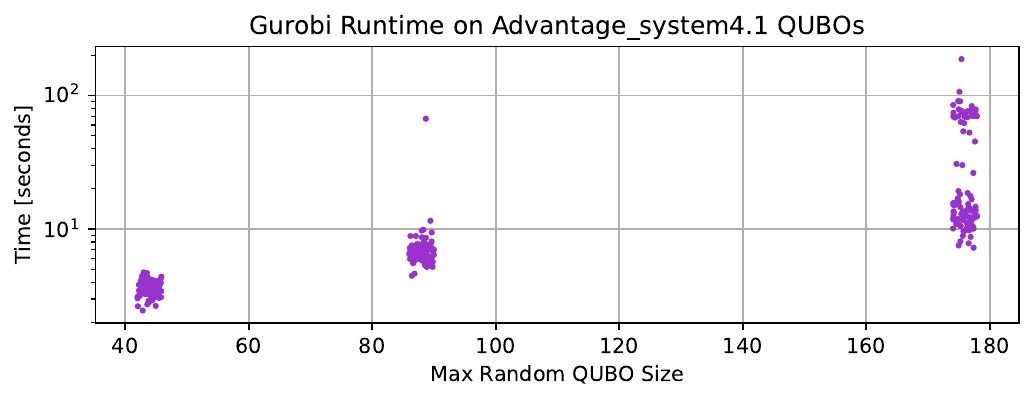}
    \includegraphics[width=0.49\textwidth]{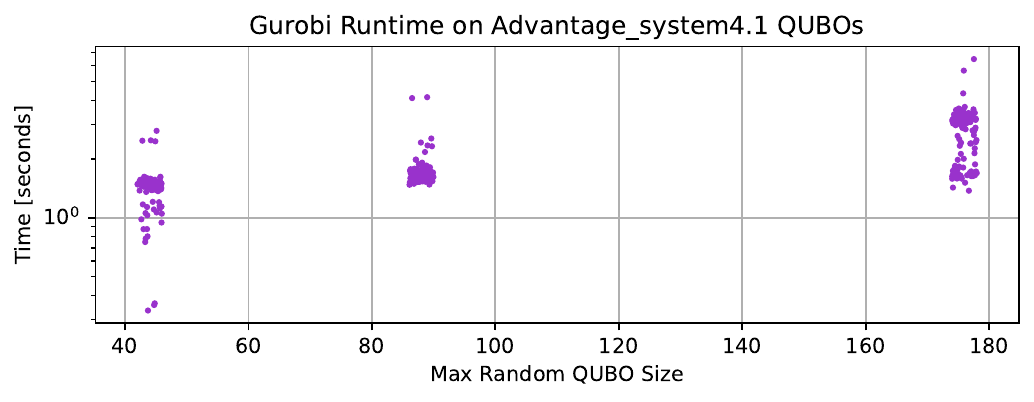}
    \includegraphics[width=0.49\textwidth]{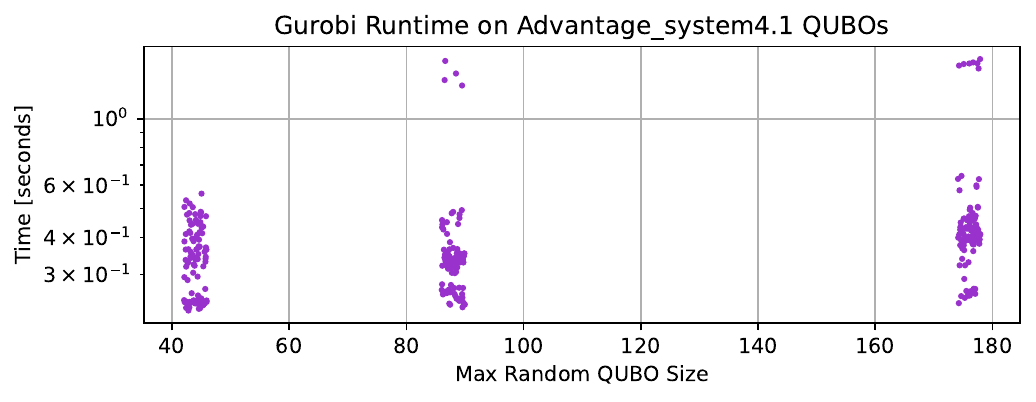}
    \caption{Gurobi runtimes in seconds (log-scale on the y-axis) as a function of the the random QUBO sizes (x-axis) for the \texttt{Advantage\_system4.1} defined QUBOs. QUBO configurations are as follows: $0.01$ posiform scaling \emph{lin$_{2}$} (top left), $0.01$ posiform scaling \emph{lin$_{20}$} (top right), $0.1$ posiform scaling \emph{lin$_{2}$} (bottom left), $0.1$ posiform scaling \emph{lin$_{20}$} (bottom right). The x-axis coordinates have a small amount of noise jitter added for improved visualization. }
    \label{fig:Gurobi_Advantage_system4.1}
\end{figure}

Figure~\ref{fig:Gurobi_Advantage2_prototype1.1} shows the Gurobi classical compute time required to deterministically find the unique optimal solution of the posiform planted instances, defined on \texttt{Advantage2\_prototype1.1}. Figures~\ref{fig:Gurobi_Advantage2_prototype2.3} and \ref{fig:Gurobi_Advantage_system4.1} shows the same Gurobi runtime scaling for the QUBOs defined on the other two D-Wave QPUs. However, in these cases, the cumulative compute time was too large to compute in a reasonable amount of time for the larger random QUBO subproblem sizes, and therefore these two figures are limited to at most the $200$ variable size cutoff. 

Fourth observations are noteworthy. First, for Gurobi there is a clear dependence on the size of the random QUBO instances for the required runtime to find the optimal solution. This is consistent with previous studies that have used integer programming tools such as Gurobi to solve random spin glass instances defined on D-Wave hardware graphs \cite{tasseff2022emerging}. In general, these random spin glasses get significantly harder to solve optimally as the problem size increases. This is significant because it shows that despite the gluing of the relatively easy-to-solve posiform planted QUBOs, the random QUBOs do make the overall problem very computationally challenging for state of the art optimization software such as Gurobi. Notably, this characteristic of random QUBO size dependence is not shared by either quantum annealing or simulated annealing ground-state sampling results. 

Second, we see that the more discrete coefficient problem instances are harder to solve using Gurobi, and the finer coefficient precision problem instances are easier to solve using Gurobi. This property is also seen in the simulated annealing and quantum annealing results. 

Third, the QUBOs with posiform scaling coefficient $0.01$ and $\pm 1$ coefficients are clearly the hardest instances to solve using Gurobi. This scaling property is consistent with both the quantum annealing results and the simulated annealing results. 

Fourth, the exact runtimes used to solve each instance vary quite significantly for each random QUBO size. For some instances, the range of solution times spans several orders of magnitude.

\section{Discussion}
\label{section:discussion}
This study presents methodology to create solution planted QUBOs that are based on the posiform planting technique \cite{Hahn_2023}, but are computationally harder than the default posiform planted QUBOs. Posiform planting has several useful features, namely that the connectivity of the generated QUBO can be tailored to (essentially) arbitrary graphs, and the planted solution is guaranteed to be unique. These properties of posiform planting are inherited by the QUBO problem generation used in this study.

Our proposed method starts by generating a posiform planted QUBO which is tailored to some architecture of interest, for instance the hardware graph of a D-Wave annealer. To increase the hardness of posiform planted QUBOs, we propose to add smaller QUBOs with a known solution to them, thus changing the coefficients of the whole-chip posiform QUBO. We demonstrate that the success rate of sampling the planted optimal solution for this class of problem instances can be lowered, meaning that the problems are indeed harder, than the posiform planting QUBOs by themselves \cite{Hahn_2023}. This increase in computational hardness was seen by both quantum annealing and simulated annealing. Among the two choices of coefficients which we considered for the generated QUBOs, we find that choosing coefficients in $\{+1, -1\}$ and using a posiform QUBO scaling coefficient of $0.01$ produced the most computationally challenging optimization problem instances. A possible cause for this is because of a large number of local minima whose energy is very close to the true ground-state. 

All three of the D-Wave quantum annealing processors used were able to sample the optimal solutions of the easier of the generated QUBOs, but for many of the small posiform scale factor QUBOs the success rate was low, especially for the $5627$ qubit \texttt{Advantage\_system4.1}. This shows that these QUBO problems can be used as a concise benchmark for heuristic algorithms to solve combinatorial optimization problems, and in particular allows us to evaluate the capability of the \emph{entire} hardware graph of the quantum annealer to solve a combinatorial optimization problem with a significantly large combinatorial search space that has only a single optimal solution. This is a non-trivial benchmarking capability -- many existing reported D-Wave benchmarks do not allow for usage of all working components of the hardware graph \cite{Pelofske_noise_dynamics_quantum_annealers, Pelofske_2022_parallel, grant2022benchmarking, PhysRevApplied.15.014012, gilbert2024benchmarking}. These optimization problems are notably quite large in scale, and our results show that D-Wave quantum annealing hardware can correctly find the single, unique, planted solution for an optimization problem that has up to $5627$ binary decision variables.

Surprisingly, we found that the size of the smaller QUBOs glued onto the posiform planted QUBO (which is tailored to the whole D-Wave hardware) does not change the hardness, measured via the success rate of finding the ground state for either simulated annealing or quantum annealing. This result is unexpected, as it seems intuitive that larger random QUBOs would increase the difficulty of the overall problem compared to gluing together many smaller instances. On the practical side, this result implies that smaller random QUBO problems can be preferred to larger ones in order to improve the efficiency of the generating algorithm, as smaller problems require less time to solve on exact classical solvers compared to larger ones. Further investigation is needed to understand all factors influencing the computational hardness of these QUBO problems. However, notably the exact Gurobi runtime scaling did strongly depend on the size of the random QUBO problems. 

We also observed that the profiles of the ground-state sampling rates are different between the Pegasus chip D-Wave device (\texttt{Advantage\_system4.1}) and the two Zephyr chip D-Wave devices (\texttt{Advantage2\_prototype2.3} and \texttt{Advantage2\_prototype1.1}). The two Zephyr devices perform best at small annealing times and then the solution quality quickly drops off. The Pegasus hardware graph device shows a similar trend for the finer discretization random QUBOs, but for the coarser discretization random QUBOs the longer annealing times perform similarly to the median annealing times and very short annealing times do not produce good ground-state sampling rates.

\section*{Acknowledgments}
\label{sec:acknowledgments}

This work was supported by the U.S. Department of Energy through the Los Alamos National Laboratory. Los Alamos National Laboratory is operated by Triad National Security, LLC, for the National Nuclear Security Administration of U.S. Department of Energy (Contract No. 89233218CNA000001). This research used resources provided by the Los Alamos National Laboratory Institutional Computing Program, which is supported by the U.S. Department of Energy National Nuclear Security Administration under Contract No.~89233218CNA000001. Research presented in this article was supported by the NNSA's Advanced Simulation and Computing Beyond Moore's Law Program at Los Alamos National Laboratory. This research used resources provided by the Darwin testbed at Los Alamos National Laboratory (LANL) which is funded by the Computational Systems and Software Environments subprogram of LANL's Advanced Simulation and Computing program (NNSA/DOE). LANL report number LA-UR-24-26772.

\appendix

\section{Disjoint Partitioning of D-Wave Hardware Graphs}
\label{sec:appendix_partitioned_hardware_graphs}

\begin{figure}[th!]
    \centering
    \includegraphics[width=0.49\textwidth]{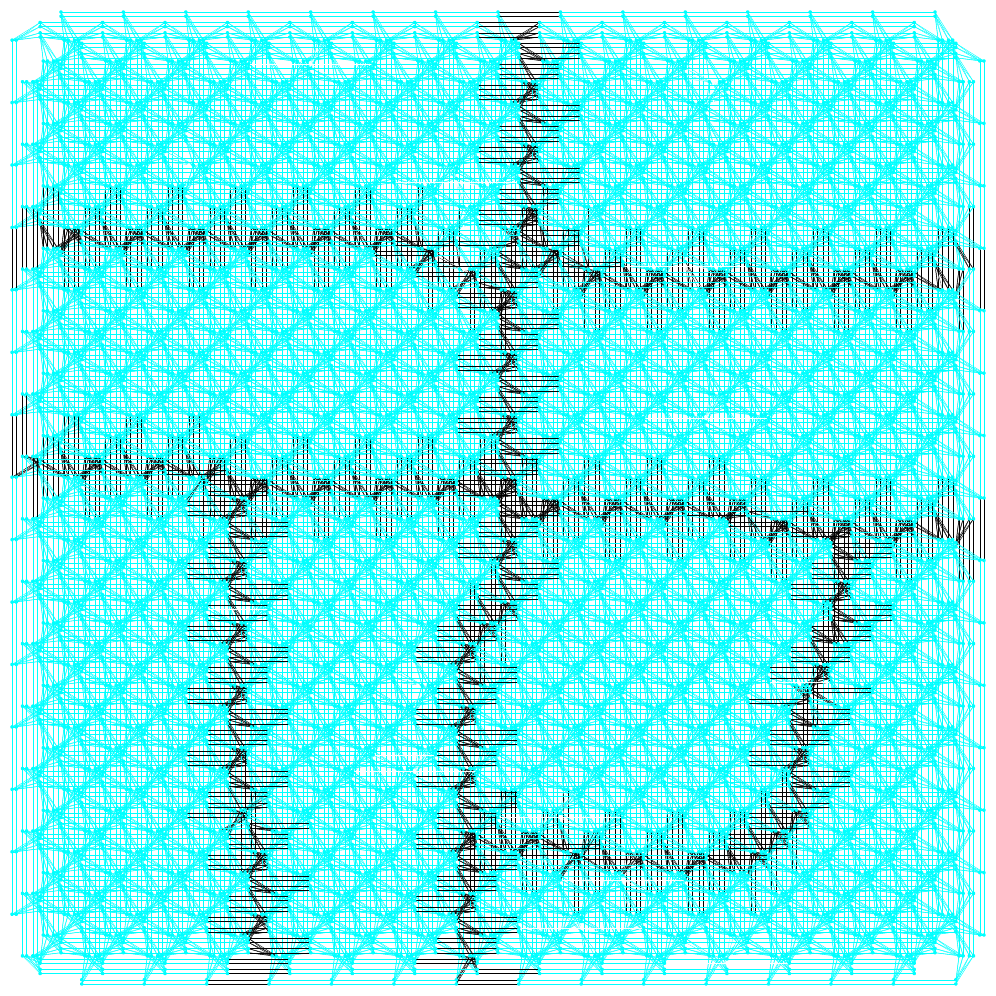}
    \includegraphics[width=0.49\textwidth]{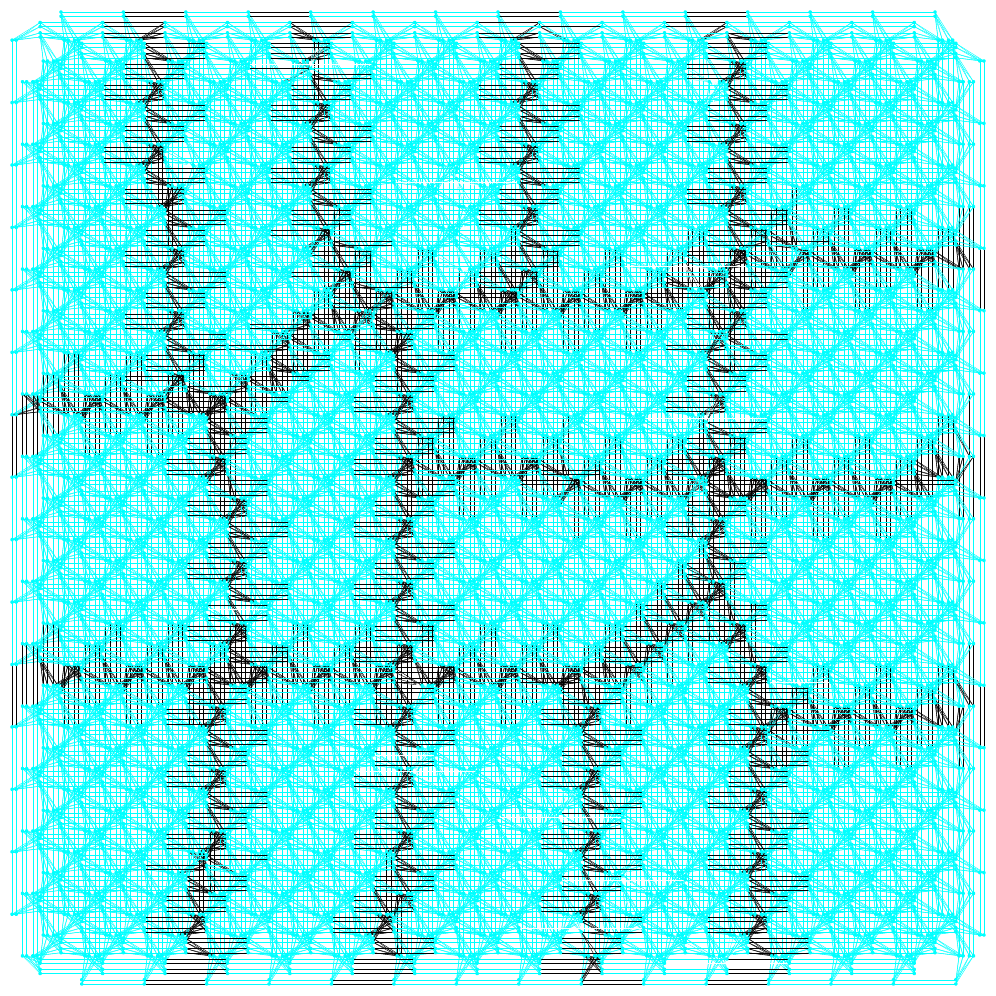}
    \includegraphics[width=0.32\textwidth]{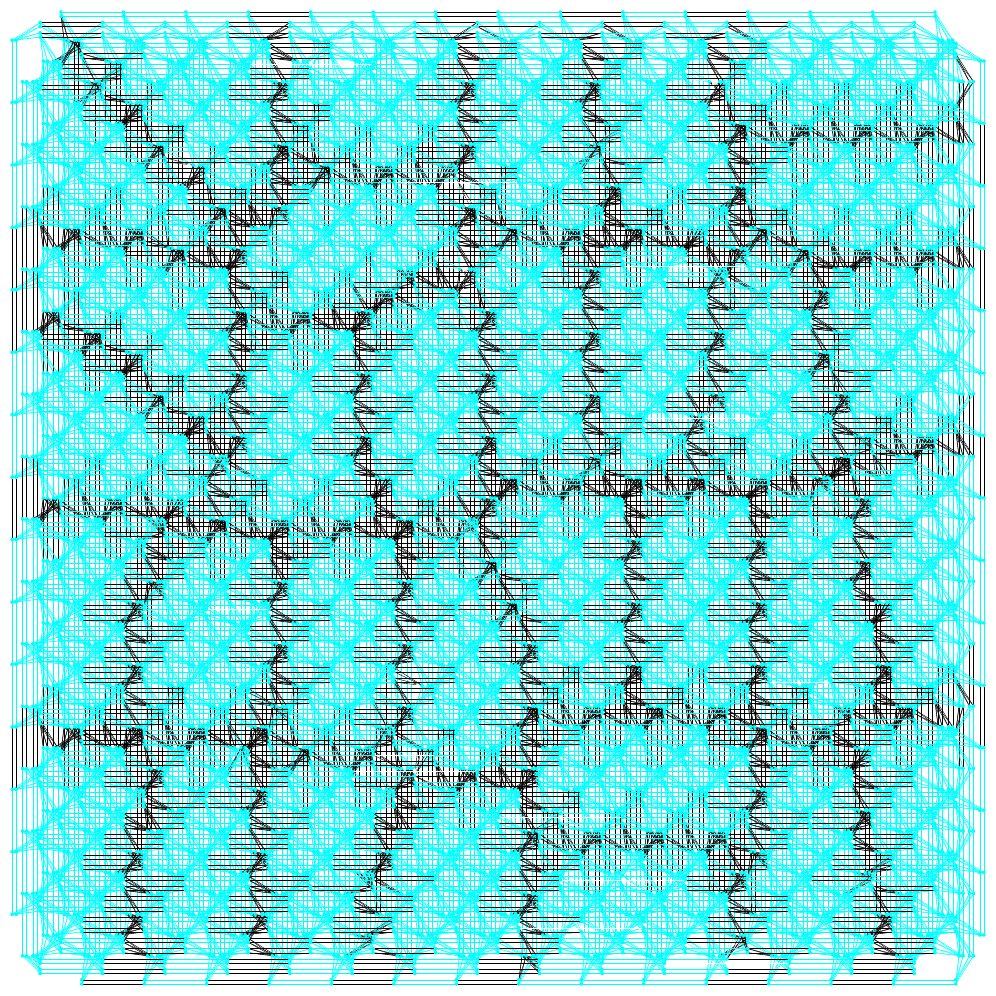}
    \includegraphics[width=0.32\textwidth]{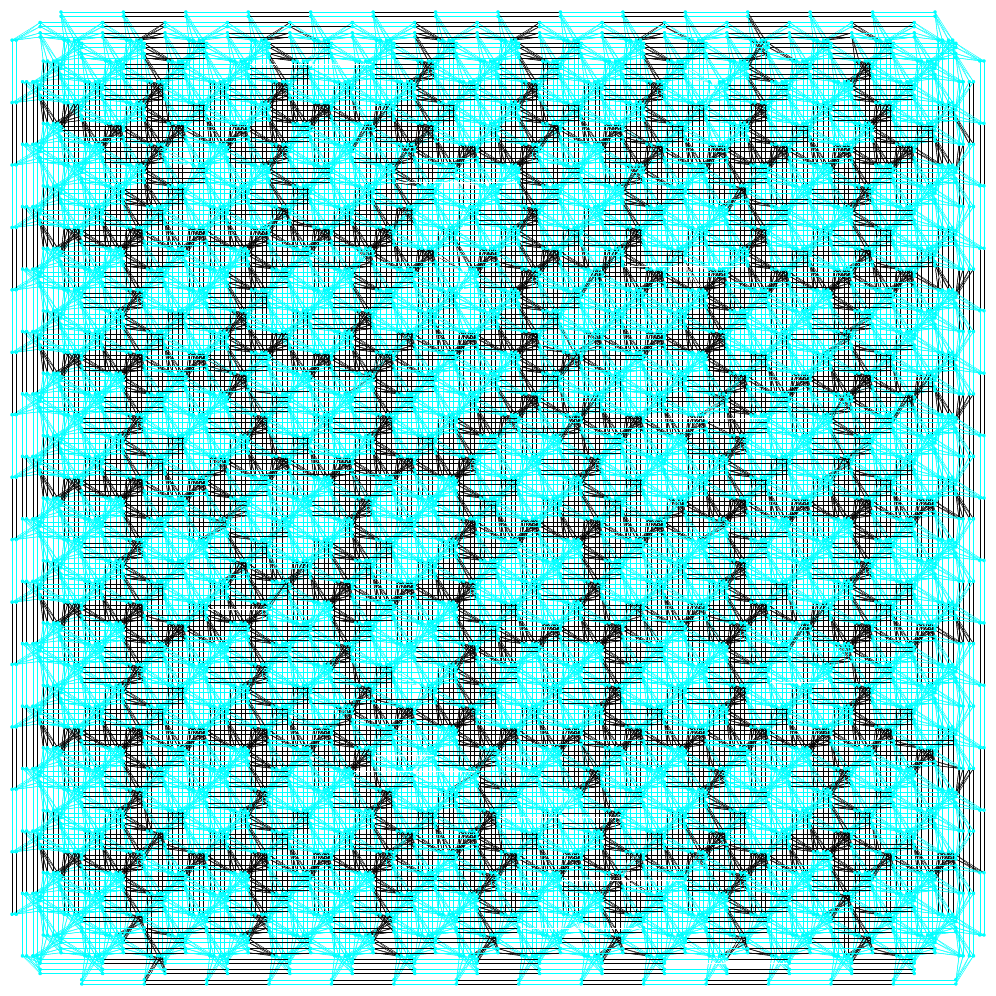}
    \includegraphics[width=0.32\textwidth]{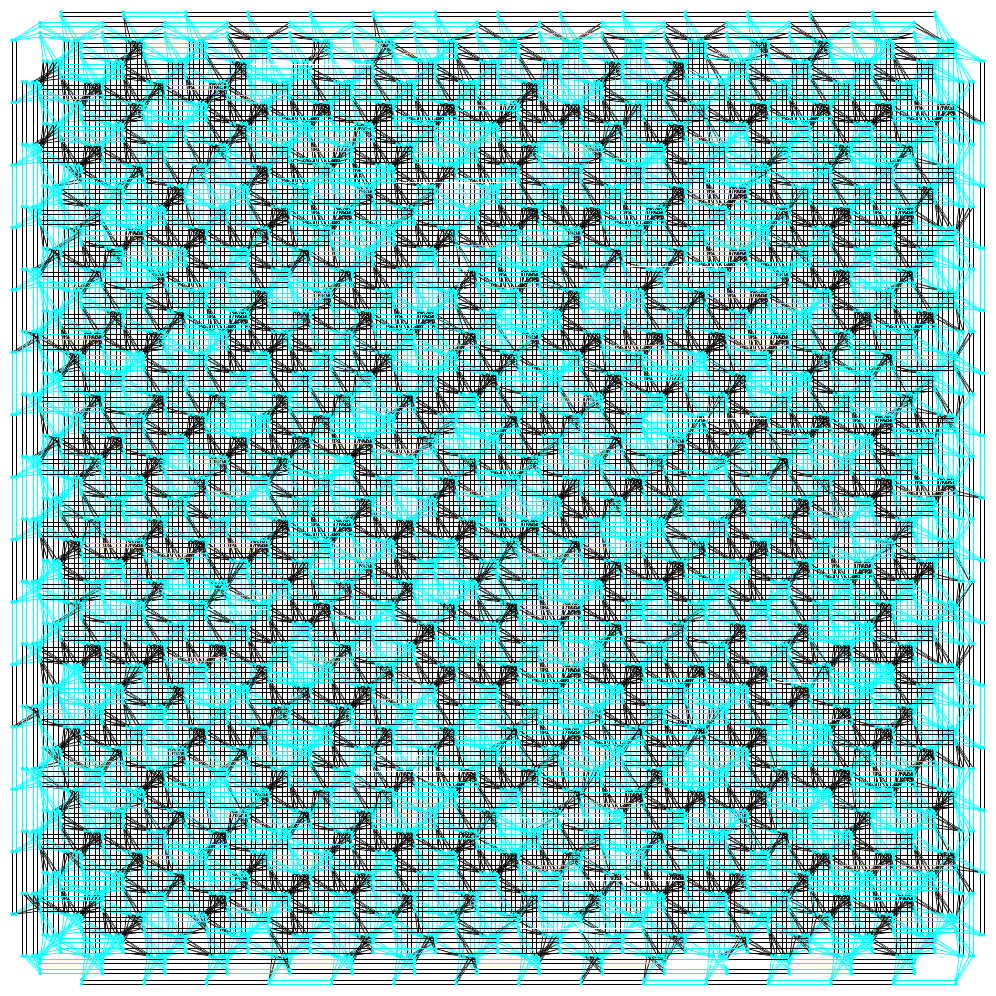}
    \caption{ Example hardware graph partitioning used for constructing the random QUBO instances for \texttt{Advantage\_system4.1}. Each of the disjoint subgraphs is colored cyan (both the nodes and the edges contained in the subgraph), and edges joining two disjoint subgraphs are black. In order from top-left to bottom-right; $8$ disjoint partitions containing either $703$ or $704$ variables, next $16$ disjoint partitions containing $352$ or $351$ variables, $32$ disjoint partitions containing $176$ or $175$ variables, $64$ disjoint partitions containing $88$ or $87$ variables, and lastly $128$ disjoint partitions each containing $43$ or $44$ variables. }
    \label{fig:hardware_graph_partitions_pegasus_4.1}
\end{figure}

\begin{figure}[th!]
    \centering
    \includegraphics[width=0.49\textwidth]{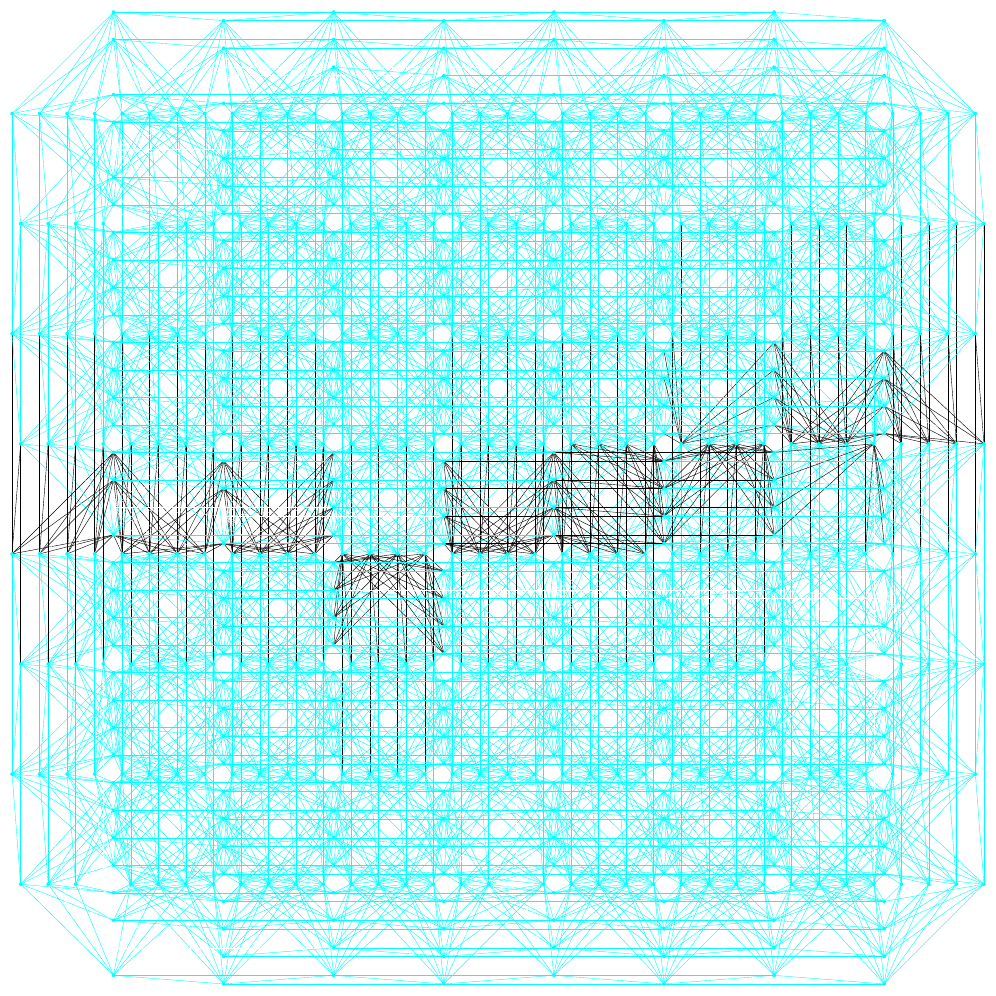}
    \includegraphics[width=0.49\textwidth]{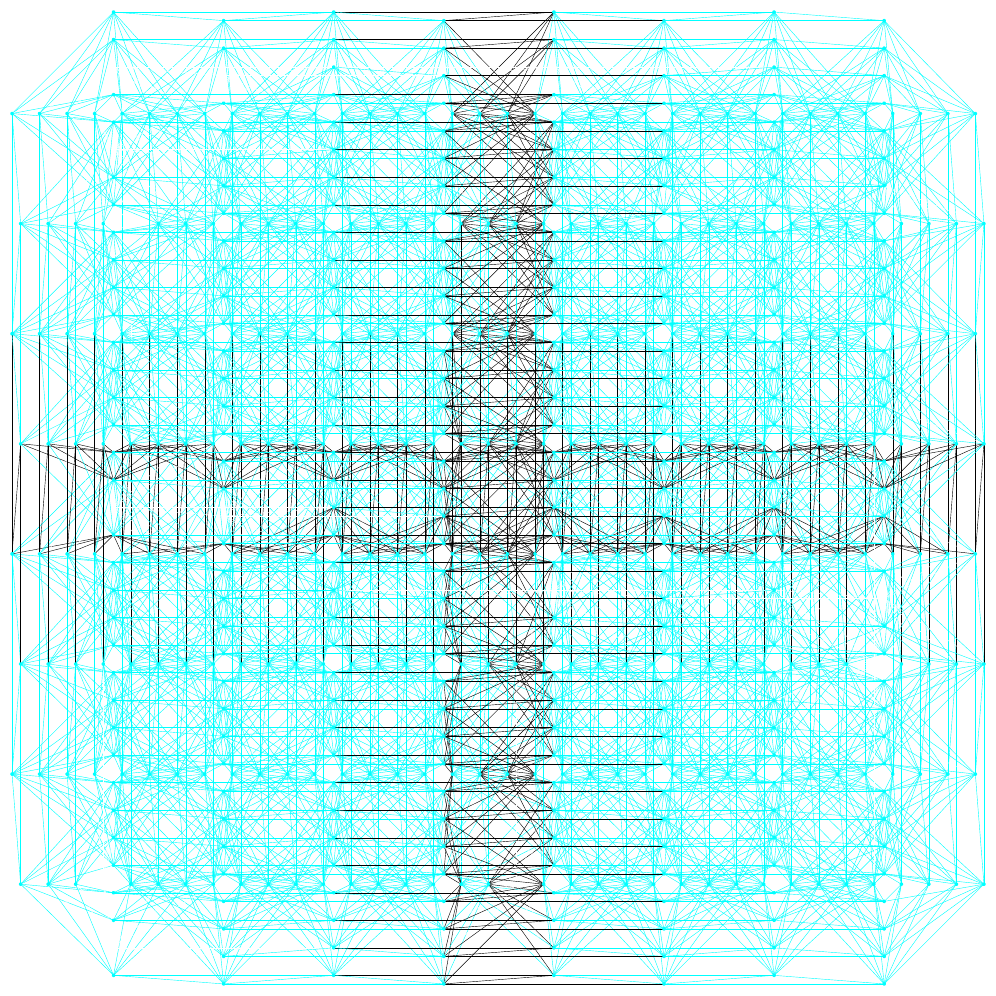}
    \includegraphics[width=0.49\textwidth]{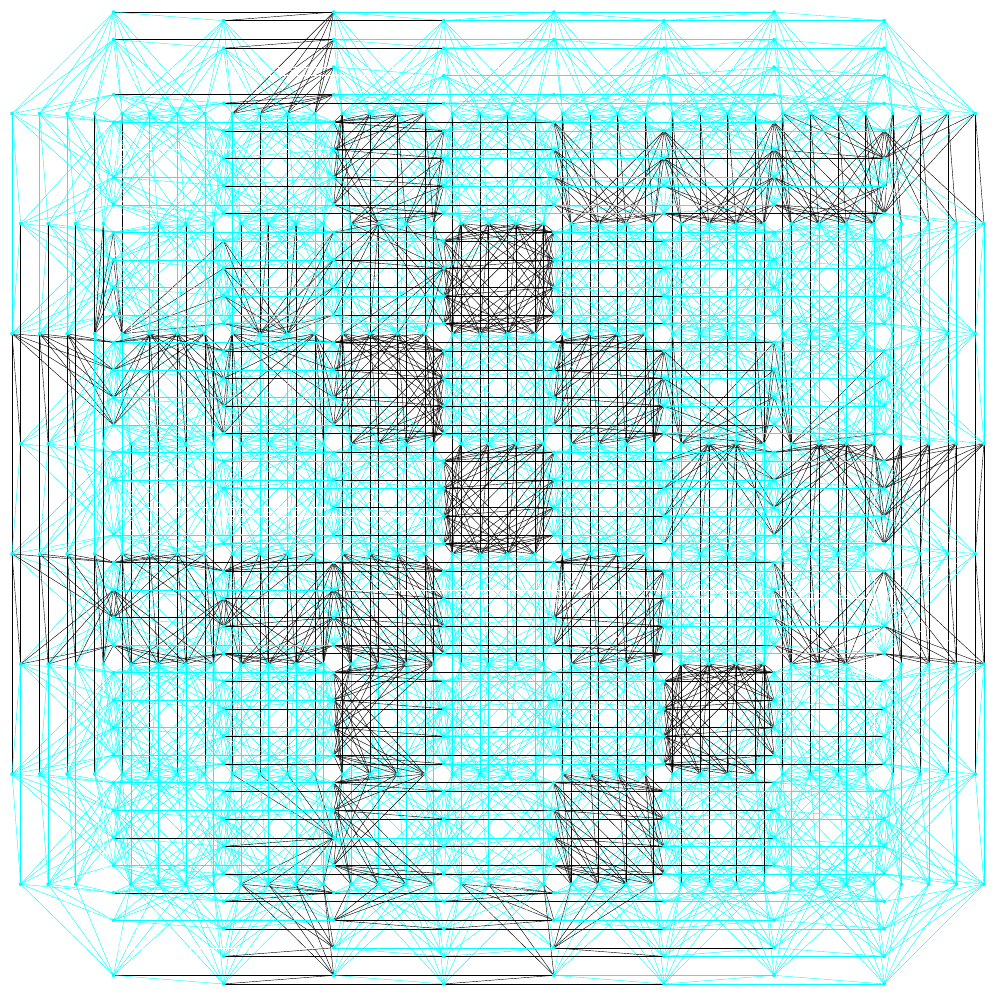}
    \includegraphics[width=0.49\textwidth]{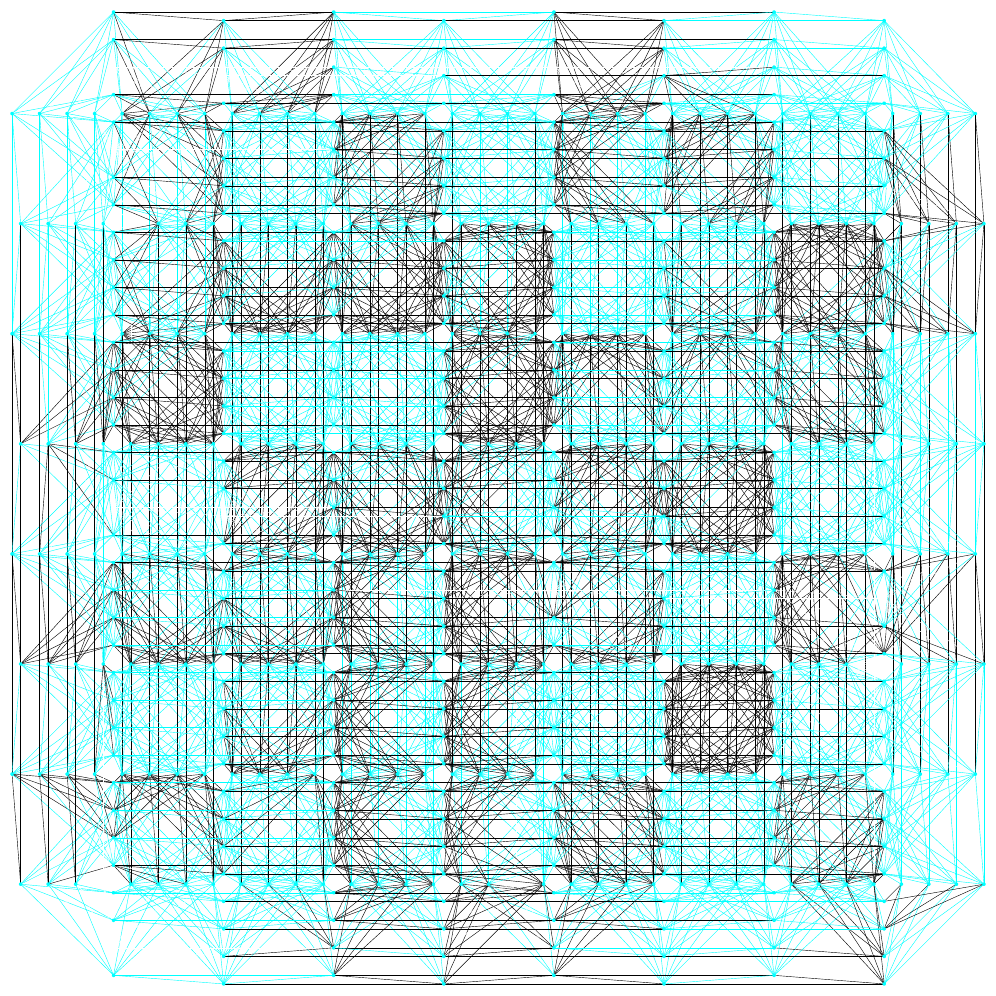}
    \caption{ Example hardware graph partitioning used for constructing the random QUBO instances for \texttt{Advantage2\_prototype1.1}. Each of the disjoint subgraphs is colored cyan (both the nodes and the edges contained in the subgraph), and edges joining two disjoint subgraphs are black. In order from top-left to bottom-right; $2$ disjoint partitions one containing $281$ variables and another containing $282$ variables, next $4$ disjoint partitions containing $140$ or $141$ variables, $8$ disjoint partitions containing $71$ or $70$ variables, and lastly $16$ disjoint partitions each containing $36$ or $35$ variables.  }
    \label{fig:hardware_graph_partitions_zephyr_prototype_1.1}
\end{figure}

Figures \ref{fig:hardware_graph_partitions_pegasus_4.1} and \ref{fig:hardware_graph_partitions_zephyr_prototype_1.1} show example disjoint partitioning of two of the D-Wave hardware graphs.

\section{Example QUBO Renderings}
\label{sec:appendix_example_QUBO_renderings}

\begin{figure}[th!]
    \centering
    \includegraphics[width=0.49\textwidth]{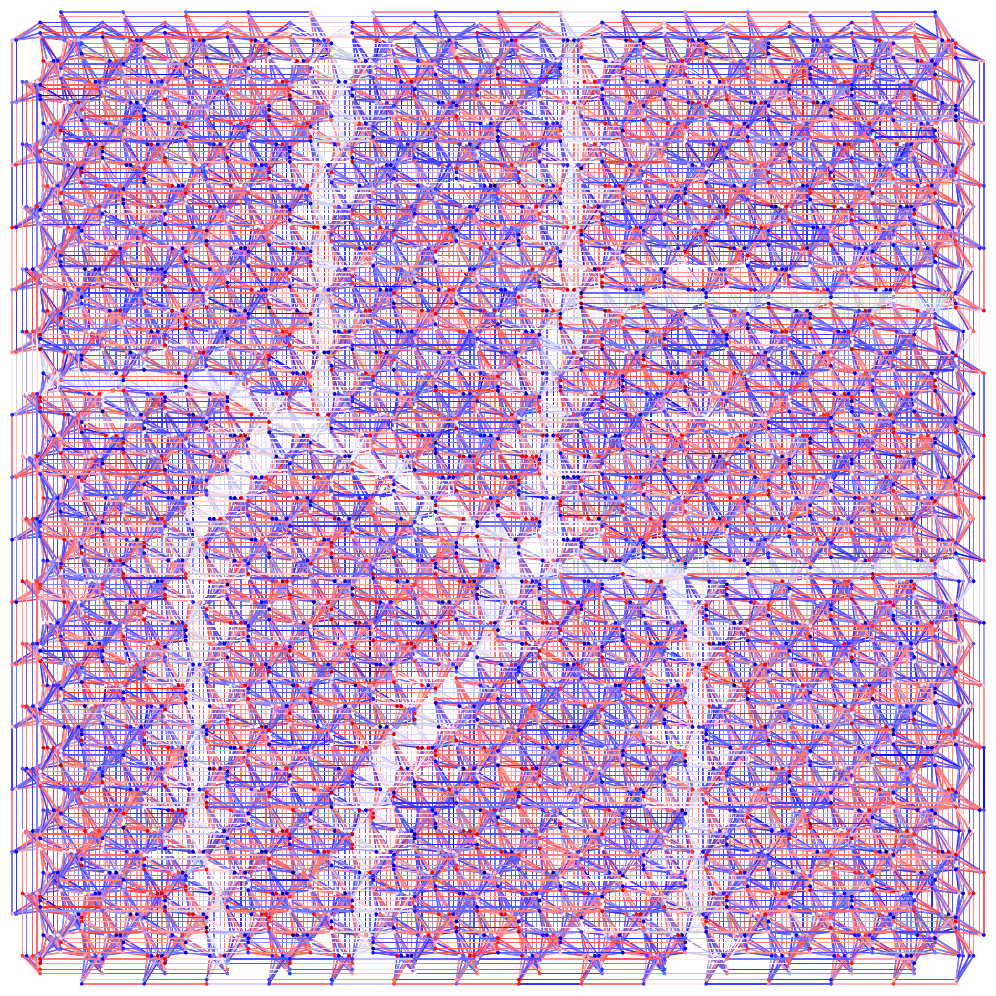}
    \includegraphics[width=0.49\textwidth]{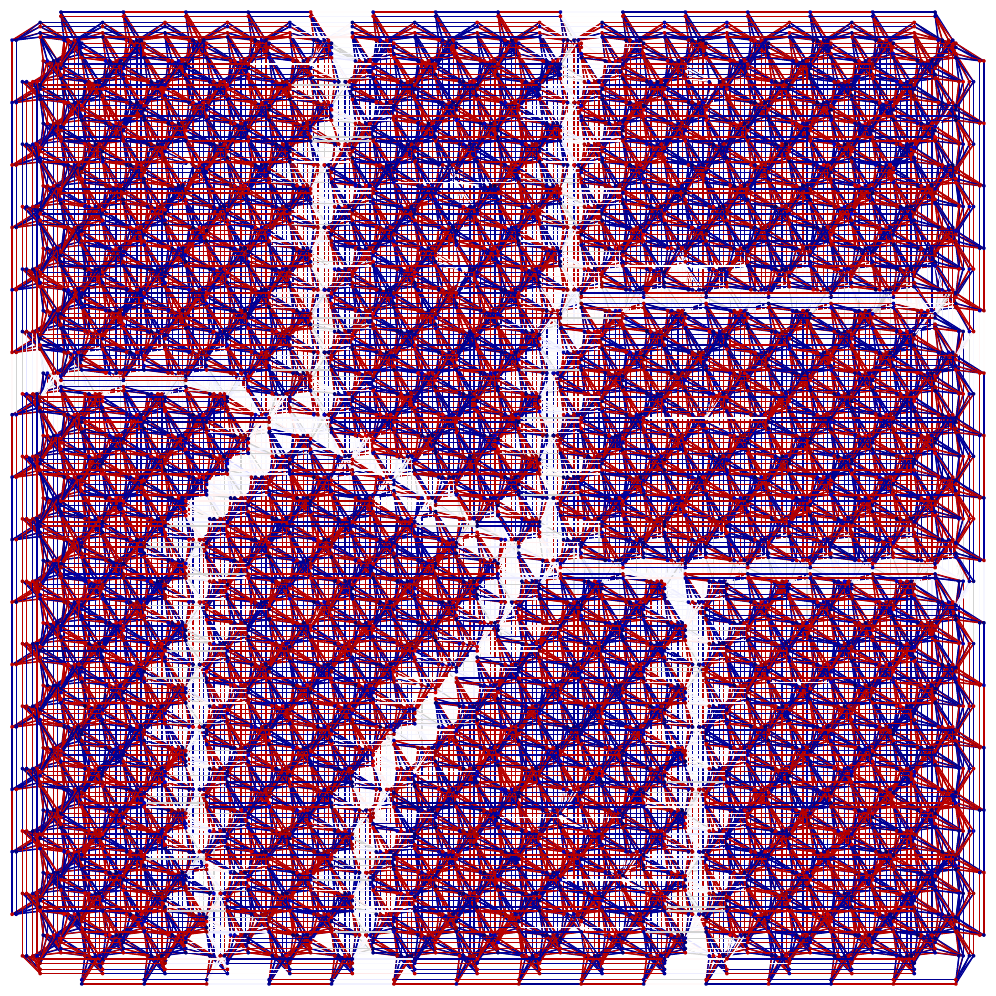}
    \includegraphics[width=0.49\textwidth]{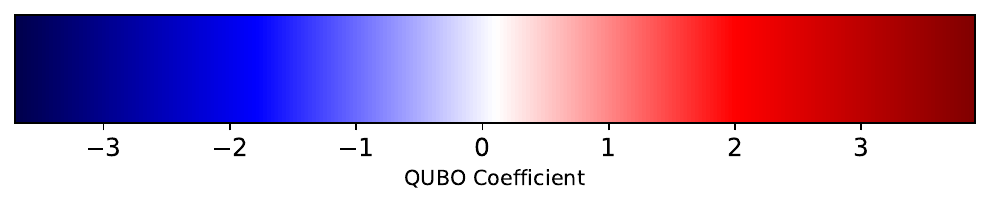}
    \includegraphics[width=0.49\textwidth]{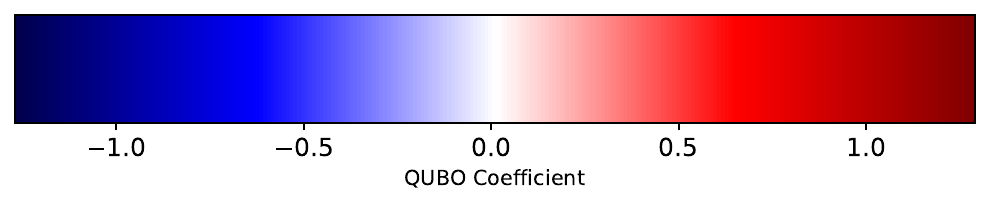}
    \caption{Whole-chip planted solution QUBO renderings for \texttt{Advantage\_system4.1} where the posiform QUBO scaling coefficient is $0.1$ (left) and $0.01$ (right) and \emph{lin$_{2}$} coefficient distribution. Blue denotes negative sign coefficients and red denotes positive sign coefficients.  }
    \label{fig:whole_chip_QUBO_v1}
\end{figure}

\begin{figure}[th!]
    \centering
    \includegraphics[width=0.49\textwidth]{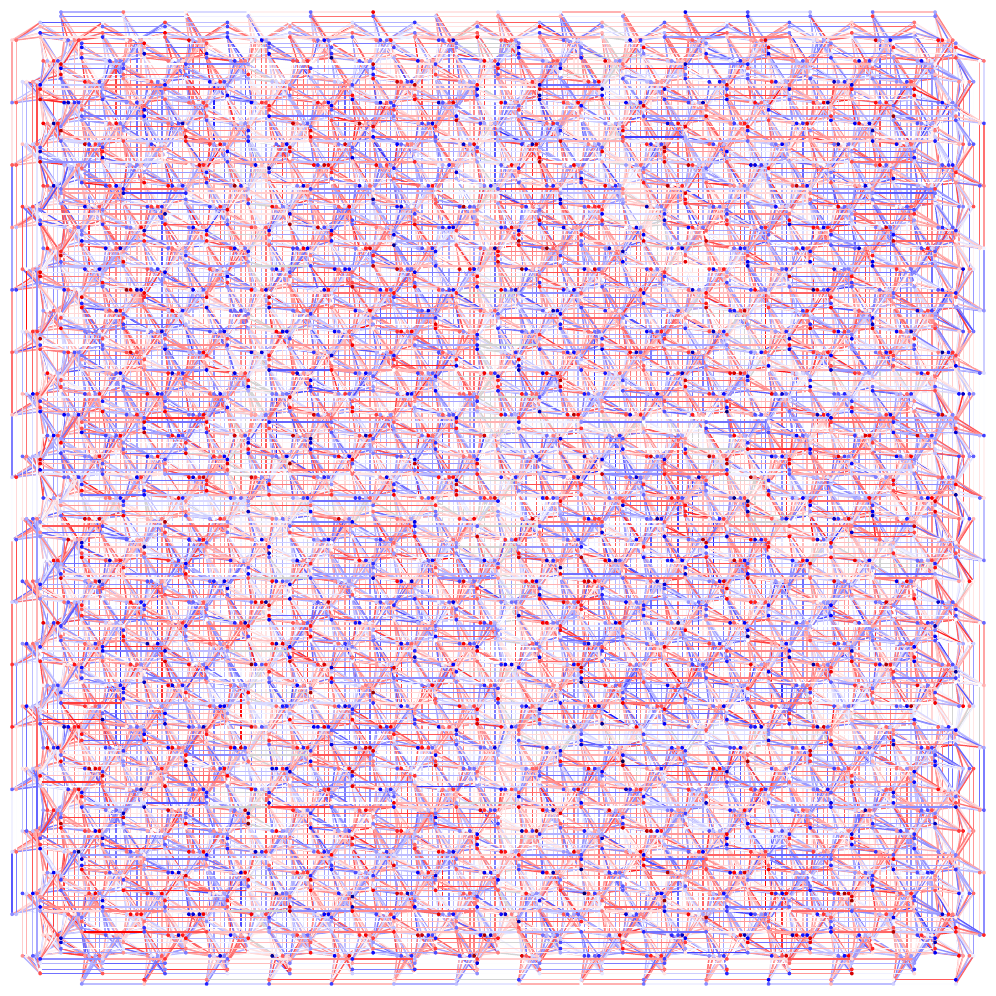}
    \includegraphics[width=0.49\textwidth]{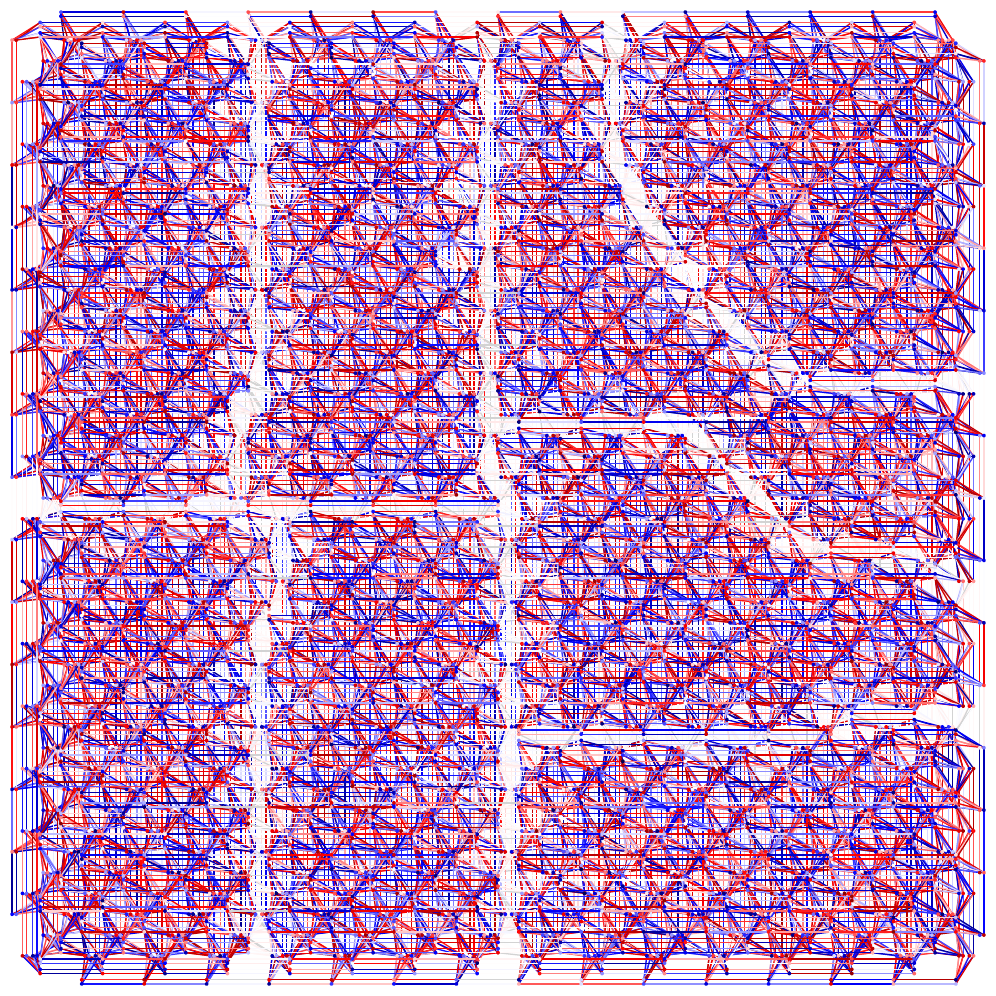}
    \includegraphics[width=0.49\textwidth]{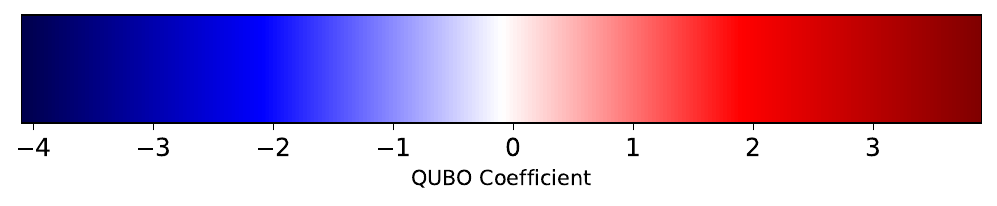}
    \includegraphics[width=0.49\textwidth]{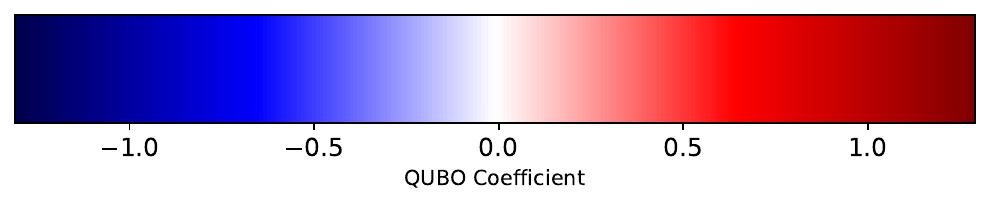}
    \caption{Whole-chip planted solution QUBO renderings for \texttt{Advantage\_system4.1} where the posiform QUBO scaling coefficient is $0.1$ (left) and $0.01$ (right) and \emph{lin$_{20}$} coefficient distribution. Blue denotes negative sign coefficients and red denotes positive sign coefficients.   }
    \label{fig:whole_chip_QUBO_v2}
\end{figure}

Figures \ref{fig:whole_chip_QUBO_v1} and \ref{fig:whole_chip_QUBO_v2} render specific examples of the posiform planted QUBO problems, having been fused to random QUBOs. 

The polynomial coefficients values are encoded by the colormaps, which are defined by the minimum and maximum coefficients that exist in each QUBO. 

\section{Additional Simulated Annealing Sampling Results}
\label{section:appendix_SA_plots}

\begin{figure}[th!]
    \centering
    \includegraphics[width=0.49\textwidth]{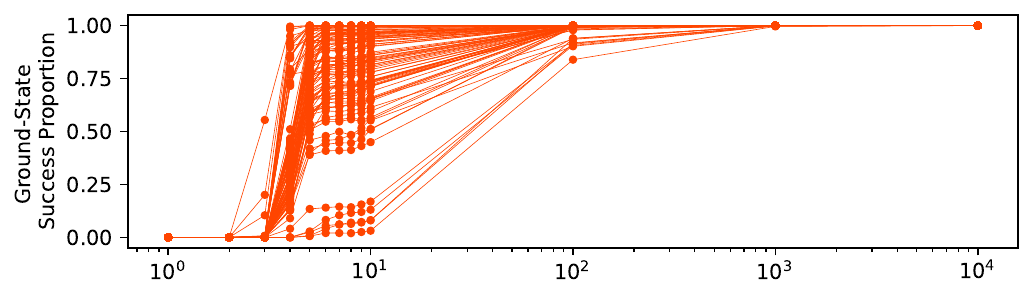}
    \includegraphics[width=0.49\textwidth]{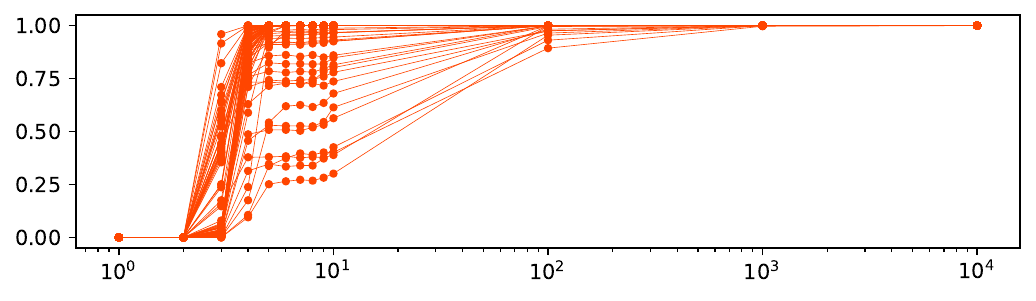}
    \includegraphics[width=0.49\textwidth]{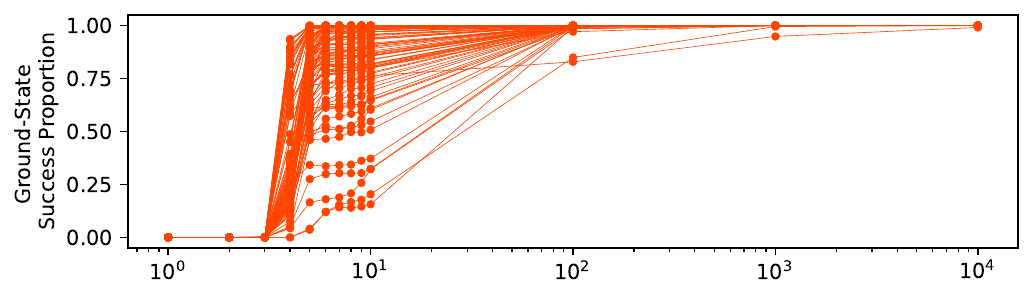}
    \includegraphics[width=0.49\textwidth]{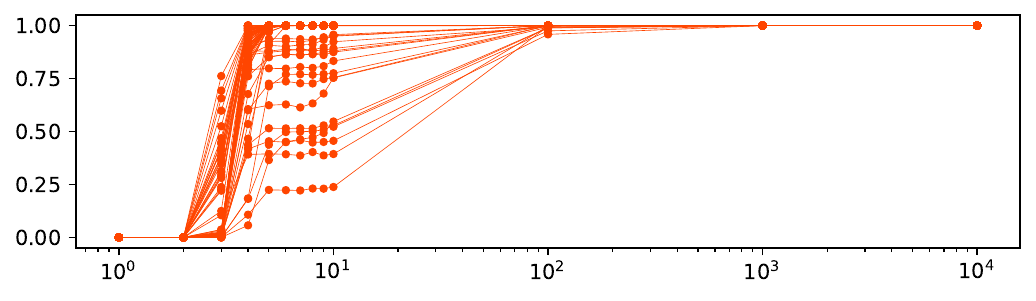}
    \includegraphics[width=0.49\textwidth]{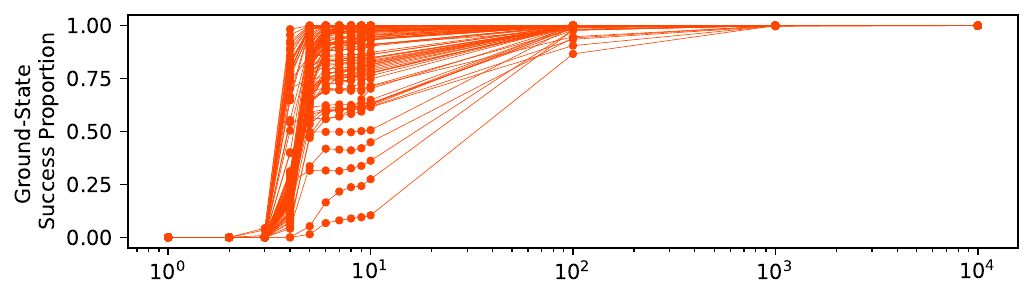}
    \includegraphics[width=0.49\textwidth]{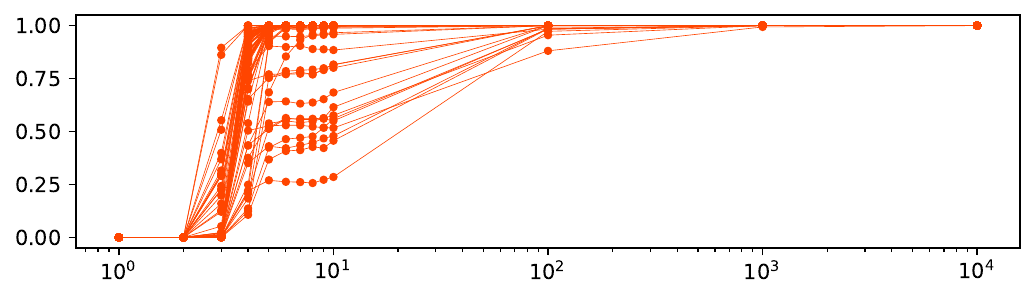}
    \includegraphics[width=0.49\textwidth]{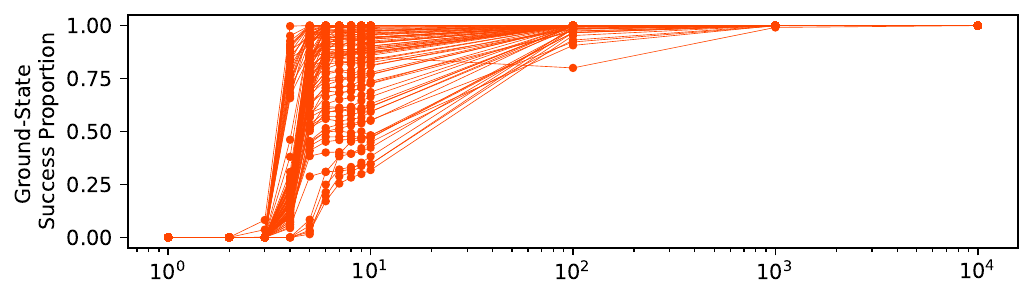}
    \includegraphics[width=0.49\textwidth]{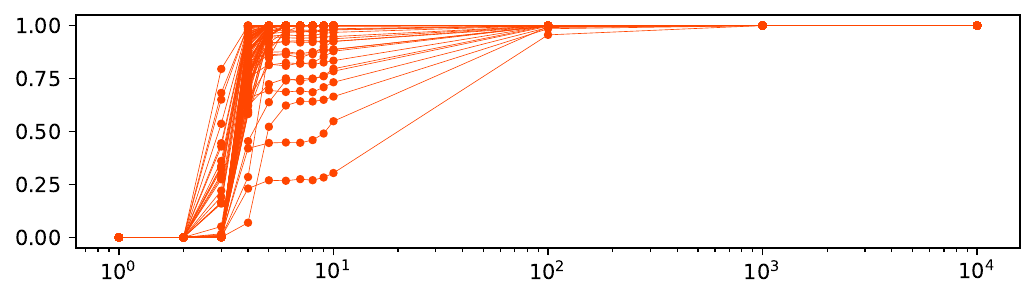}
    \includegraphics[width=0.49\textwidth]{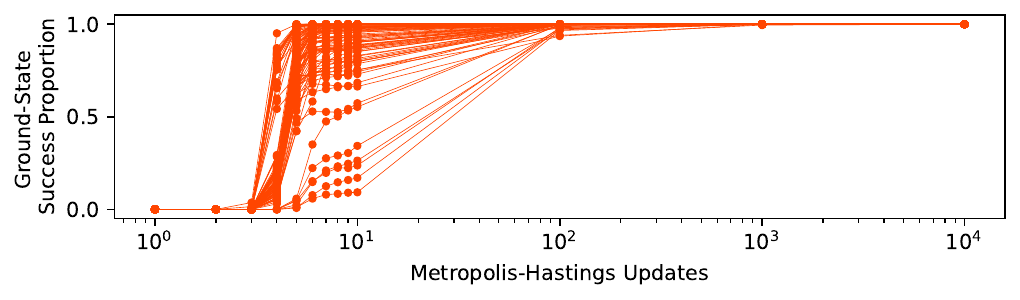}
    \includegraphics[width=0.49\textwidth]{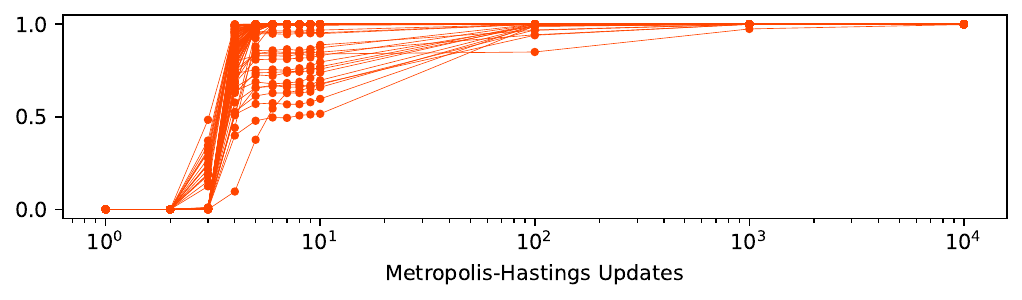}
    \caption{Simulated annealing applied to \texttt{Advantage\_system4.1} hardware compatible QUBOs, where each subfigure shows the optimal solution sampling rate for the $100$ different random QUBO instances, as a function of log-scale number of complete sweeps of Metropolis-Hastings updates. \emph{lin$_{2}$} Columns show QUBOs with coefficients from \emph{lin$_{2}$} (left) and \emph{lin$_{20}$} (right). Posiform scaling coefficient of $0.1$. Rows correspond to the QUBOs being solved in Figure~\ref{fig:sampling_success_rate_Pegasus4.1}.}
    \label{fig:SA_success_proportion_Pegasus4.1_0.1}
\end{figure}

\begin{figure}[th!]
    \centering
    \includegraphics[width=0.49\textwidth]{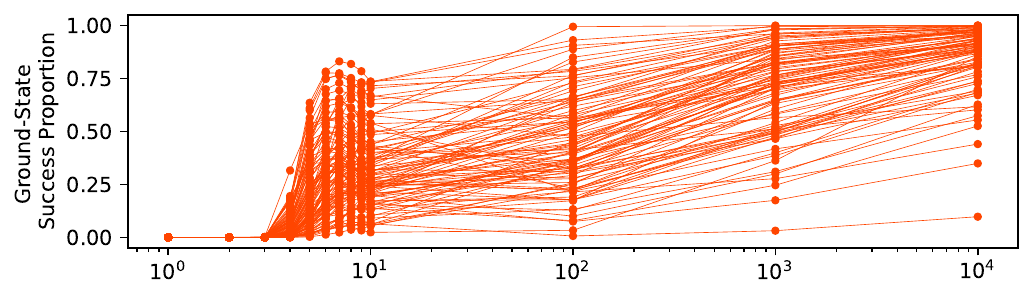}
    \includegraphics[width=0.49\textwidth]{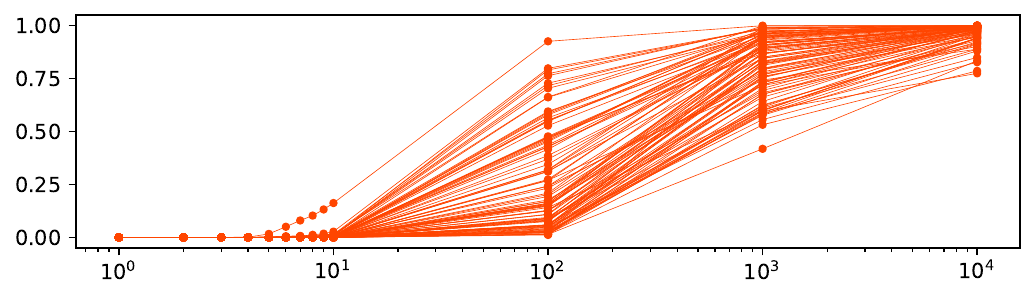}
    \includegraphics[width=0.49\textwidth]{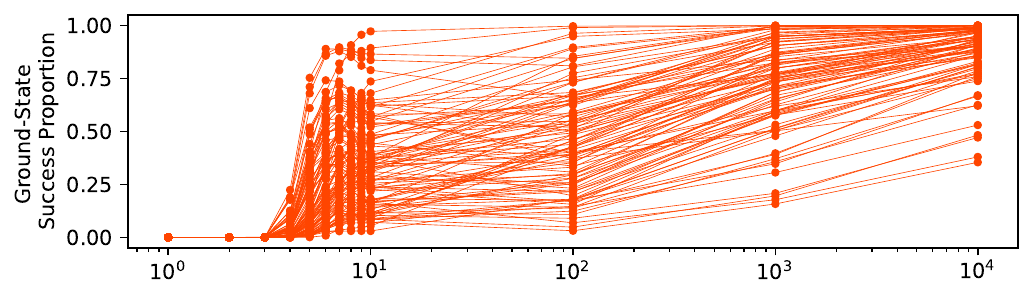}
    \includegraphics[width=0.49\textwidth]{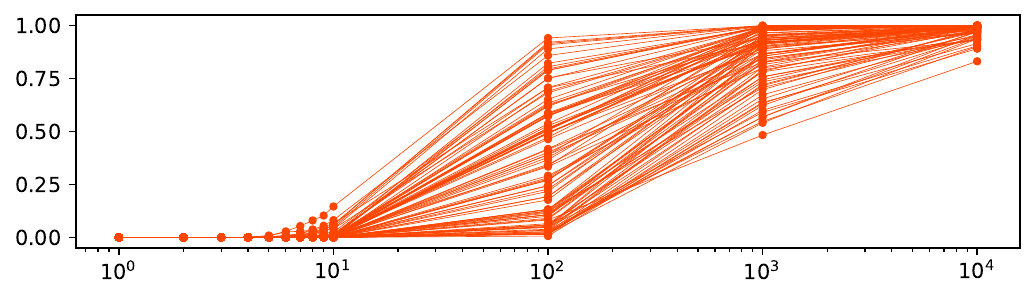}
    \includegraphics[width=0.49\textwidth]{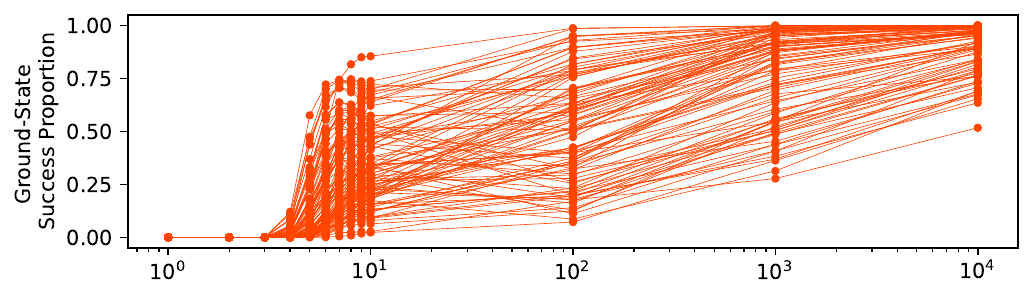}
    \includegraphics[width=0.49\textwidth]{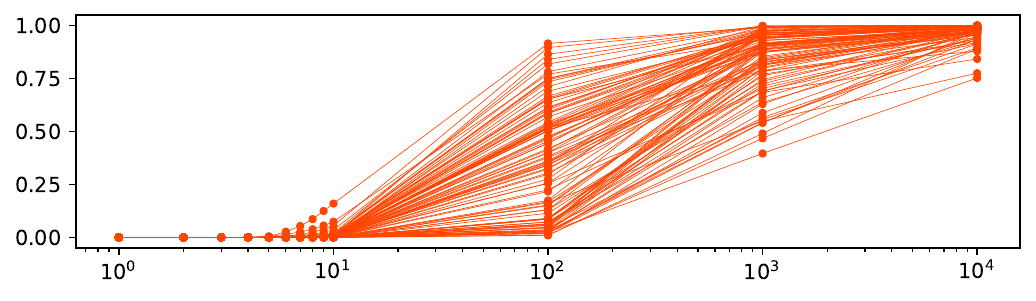}
    \includegraphics[width=0.49\textwidth]{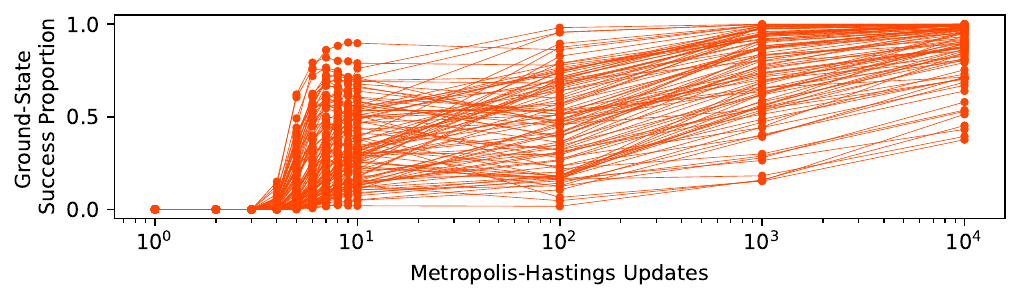}
    \includegraphics[width=0.49\textwidth]{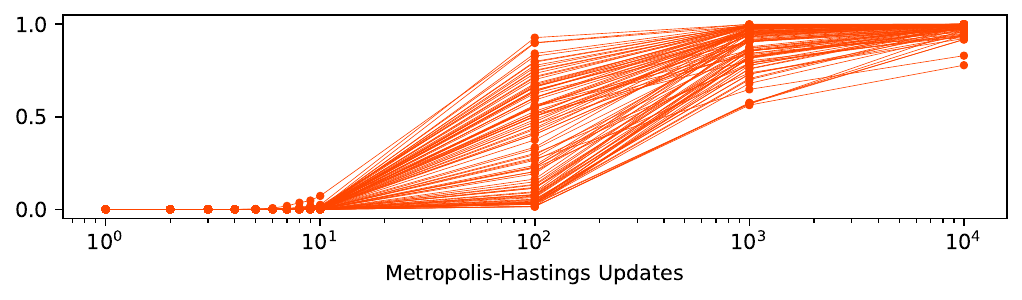}
    \caption{Simulated annealing sampling results. \texttt{Advantage2\_prototype1.1} hardware compatible QUBO results, where each subfigure shows the optimal solution sampling rate for $100$ different random QUBO instances, as a function of log-scale number of complete sweeps of Metropolis-Hastings updates. Posiform QUBO scaling coefficient is $0.01$. \emph{lin$_{2}$} QUBO instances are shown in the left-hand side column and \emph{lin$_{20}$} QUBO instances are shown in the right-hand side column. }
    \label{fig:SA_success_proportion_Zephyr1.1_0.01}
\end{figure}

\begin{figure}[th!]
    \centering
    \includegraphics[width=0.49\textwidth]{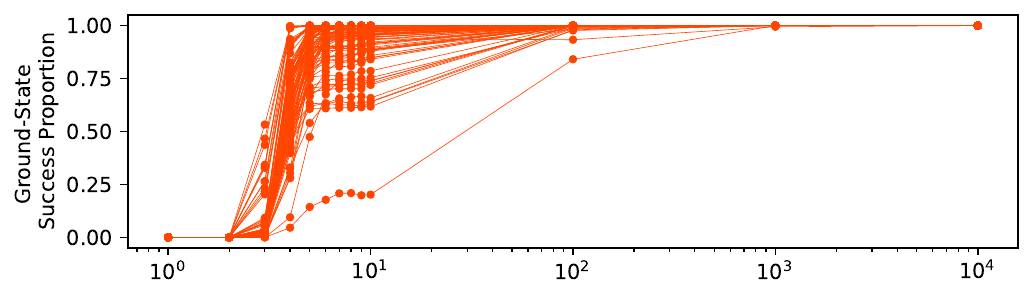}
    \includegraphics[width=0.49\textwidth]{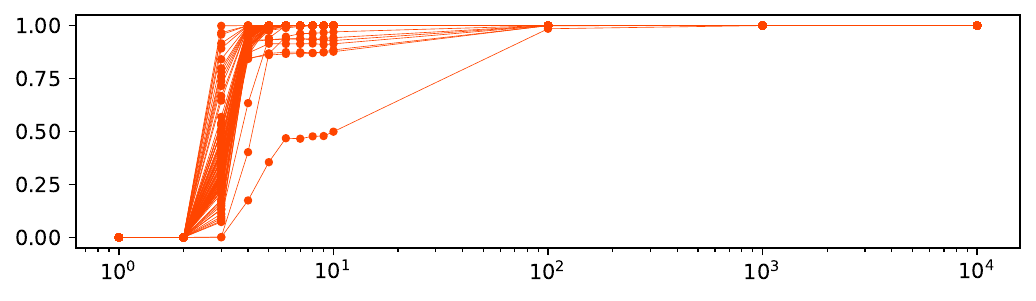}
    \includegraphics[width=0.49\textwidth]{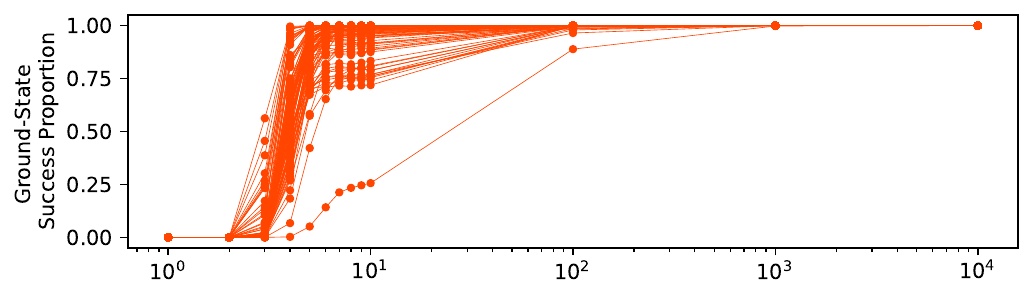}
    \includegraphics[width=0.49\textwidth]{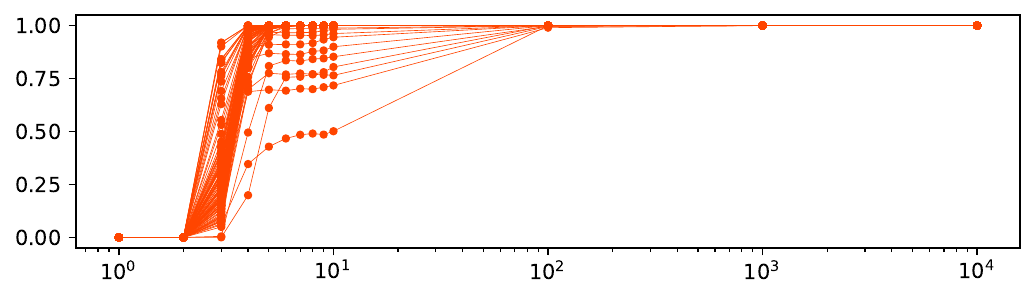}
    \includegraphics[width=0.49\textwidth]{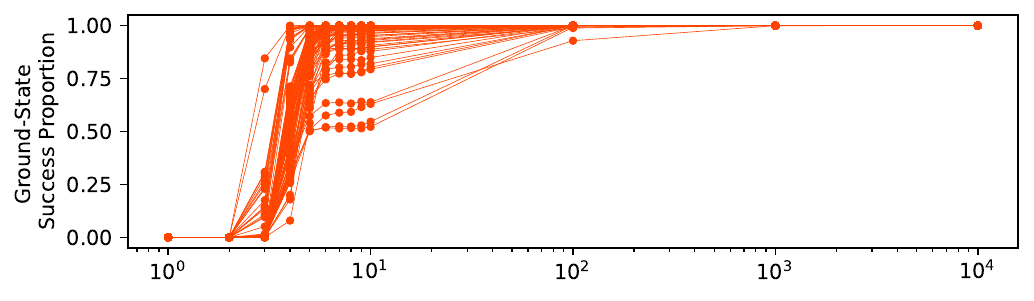}
    \includegraphics[width=0.49\textwidth]{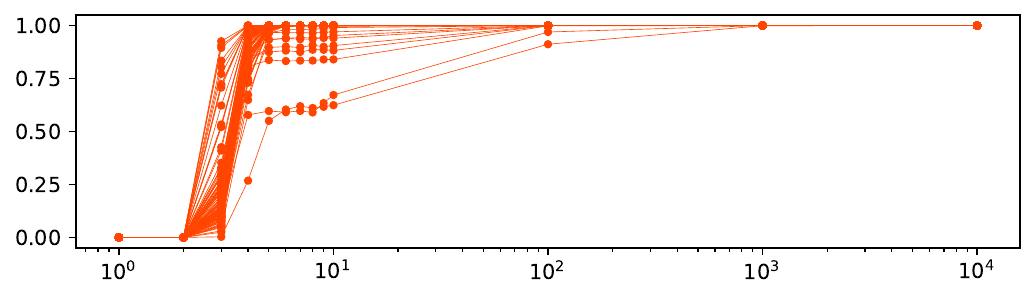}
    \includegraphics[width=0.49\textwidth]{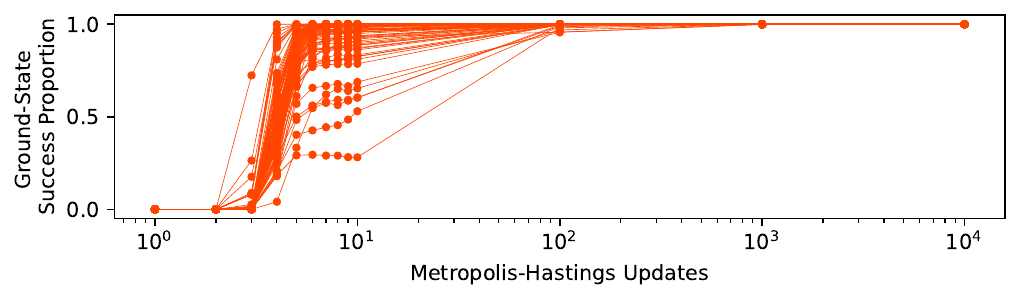}
    \includegraphics[width=0.49\textwidth]{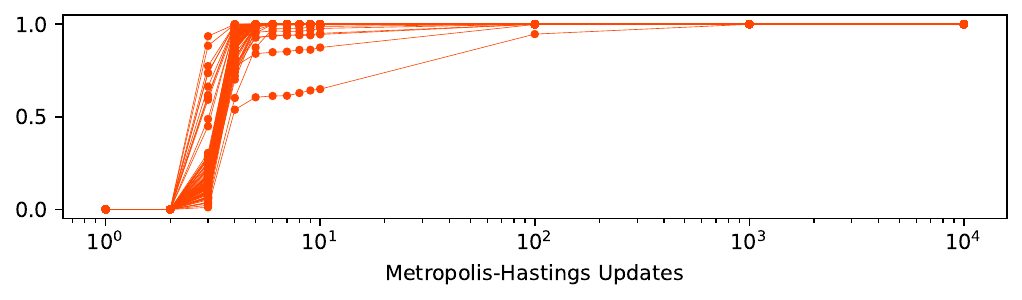}
    \caption{Simulated annealing sampling results. \texttt{Advantage2\_prototype1.1} hardware compatible QUBO results, where each subfigure shows the optimal solution sampling rate for the $100$ different random QUBO instances, as a function of log-scale number of complete sweeps of Metropolis-Hastings updates. Posiform QUBO scaling coefficient is $0.1$. \emph{lin$_{2}$} QUBO instances are shown in the left-hand side column and \emph{lin$_{20}$} QUBO instances are shown in the right-hand side column. }
    \label{fig:SA_success_proportion_Zephyr1.1_0.1}
\end{figure}

Figures \ref{fig:SA_success_proportion_Pegasus4.1_0.1}, \ref{fig:SA_success_proportion_Zephyr1.1_0.01}, \ref{fig:SA_success_proportion_Zephyr1.1_0.1} all show additional simulated annealing ground-state sampling rates.

\clearpage

\setlength\bibitemsep{0pt}
\printbibliography

\end{document}